\documentclass[5p,twocolumn,10pt,times]{elsarticle}
\usepackage{amsmath}
\usepackage{hyperref}
	\makeatletter
	\providecommand{\doi}[1]{%
	  \begingroup
	    \let\bibinfo\@secondoftwo
	    \urlstyle{rm}%
	    \href{http://dx.doi.org/#1}{%
	      doi:\discretionary{}{}{}%
	      \nolinkurl{#1}%
	    }%
	  \endgroup
	}
	\makeatother

	\usepackage[labelformat=parens]{subcaption} 
	\usepackage[export]{adjustbox} 
	
	\usepackage{rotating} 

	\usepackage{wrapfig} 

	\usepackage{multirow} 
	\usepackage{colortbl} 
	\newcolumntype{C}[1]{>{\centering\arraybackslash}m{#1}}   
	\newcolumntype{R}[1]{>{\raggedleft\arraybackslash}m{#1}}  

	\usepackage[capitalise]{cleveref} 

        \usepackage{ amssymb } 
	
	\usepackage[binary-units]{siunitx} 
	\DeclareSIUnit\pixel{px}

	\usepackage{units} 

	\usepackage[inline]{enumitem} 

	\usepackage[toc,page]{appendix} 
	
	\usepackage{pgfplots}
	\usepackage{pgfplotstable} 
	\pgfplotsset{compat=1.14}
	\usetikzlibrary{backgrounds}

	\usepackage{algpseudocode} 
	\usepackage{algorithm} 

    \usepackage{color,soul} 

    \newcommand{\changed}[1]{#1}

\begin{document}
\baselineskip11pt 

\begin{frontmatter} 

\title{CrossFill: Foam Structures with Graded Density for Continuous Material Extrusion}


\author[um,tud]{Tim Kuipers}
\author[tud]{Jun Wu}
\author[cuhk]{Charlie Wang{\corref{cor1}}}
\ead{cwang@mae.cuhk.edu.hk}
\cortext[cor1]{Corresponding author}
\address[um]{Ultimaker, Utrecht, The Netherlands}
\address[tud]{Department of Design Engineering, Delft University of Technology, The Netherlands}
\address[cuhk]{Department of Mechanical and Automation Engineering, The Chinese University of Hong Kong, Hong Kong SAR, China}

\begin{abstract}
The fabrication flexibility of 3D printing has sparked a lot of interest in designing structures with spatially graded material properties. 
In this paper, we propose a new type of density graded structure that is particularly designed for 3D printing systems based on filament extrusion. 
In order to ensure high-quality fabrication results, extrusion-based 3D printing requires not only that the structures are self-supporting, but also that extrusion toolpaths are continuous and free of self-overlap.
The structure proposed in this paper, called \emph{CrossFill}, complies with these requirements.
In particular, CrossFill is a self-supporting foam structure, for which each layer is fabricated by a single, continuous and overlap-free path of material extrusion.
Our method for generating CrossFill is based on a space-filling surface that employs spatially varying subdivision levels.
Dithering of the subdivision levels is performed to accurately reproduce a prescribed density distribution.
We demonstrate the effectiveness of CrossFill on a number of experimental tests and applications.
\end{abstract}

%
%



\end{frontmatter}


\section{Introduction}\label{secIntro}
3D printing enables the fabrication of complex structures with unprecedented geometric detail.
This creates the opportunity to realize 3D shapes with complex internal structures.
Physical properties of these \emph{infill structures} are determined by their geometry and the constitutive material by which they are made. 
Even with a single constitutive material, 3D printing allows to achieve graded physical properties (e.g. density and stiffness) by spatially varying the geometry of infill structures. 
This enables \textit{functionally graded materials} (FGM) at a manufacturable scale. 
Precise realization of graded physical properties can lead to many applications, such as customized insoles, comfort cushioning and medical phantoms. 

\newlength{\voronoiheight}
\setlength{\voronoiheight}{.3\columnwidth}
\begin{figure}
        \centering
    \begin{subfigure}[t]{0.23\columnwidth}
        \centering
         \includegraphics[height=\voronoiheight]{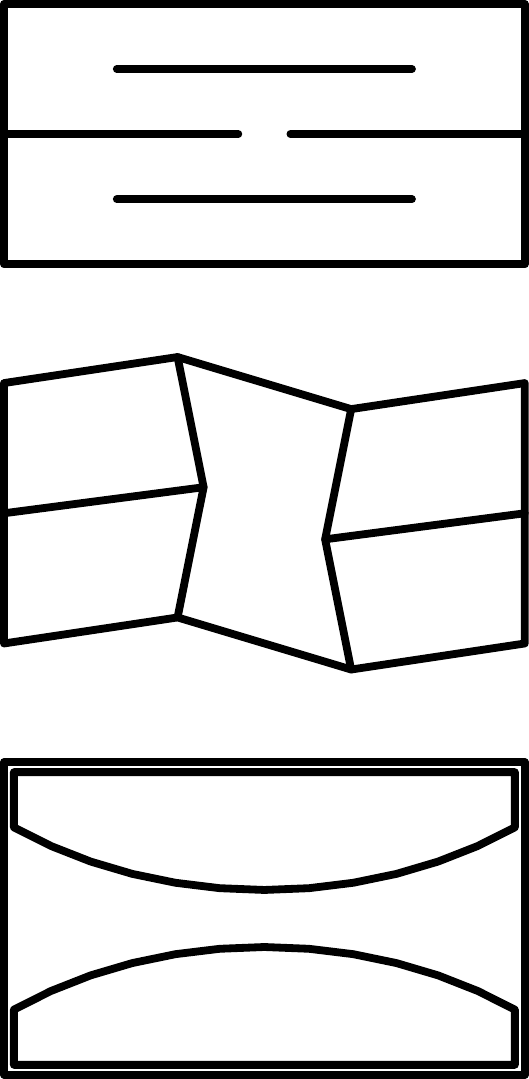}
        \caption{Input}
         \label{fig:discontinuity_print_input}
    \end{subfigure}
    \begin{subfigure}[t]{0.23\columnwidth}
        \centering
         \includegraphics[height=\voronoiheight]{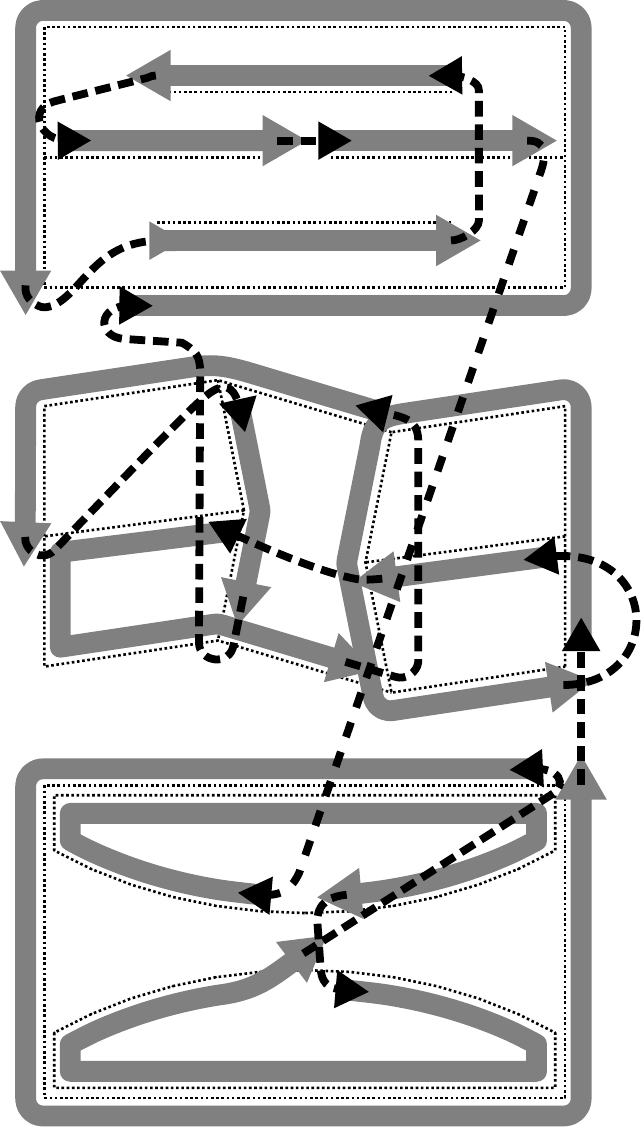}
        \caption{Order}
         \label{fig:discontinuity_print_order}
    \end{subfigure}
    \begin{subfigure}[t]{0.23\columnwidth}
        \centering
         \includegraphics[height=\voronoiheight]{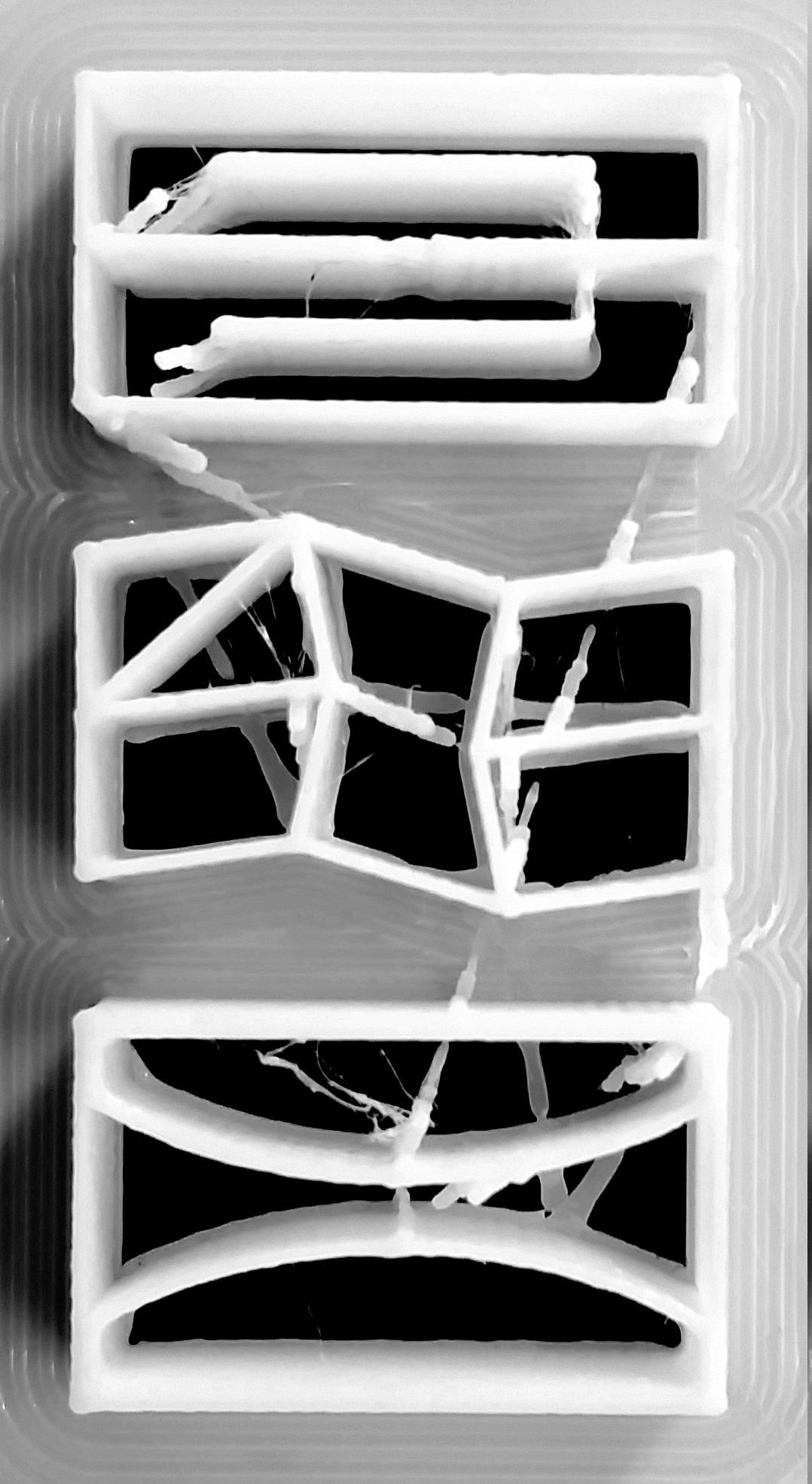}
        \caption{Leakage}
         \label{fig:discontinuity_print_print}
    \end{subfigure}
    \begin{subfigure}[t]{0.23\columnwidth}
        \centering
         \includegraphics[height=\voronoiheight]{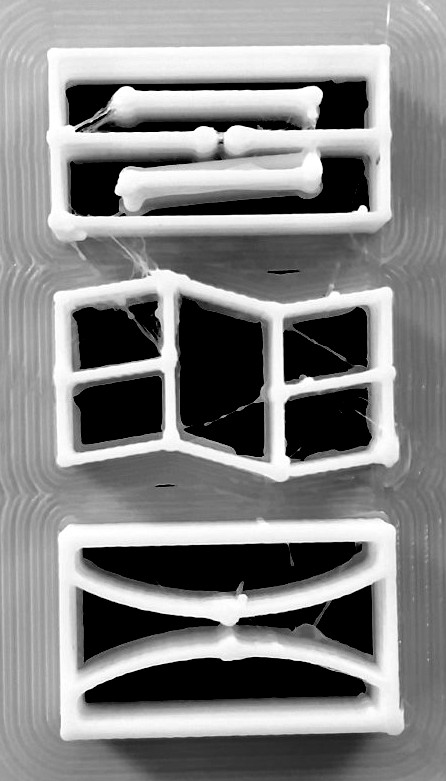}
        \caption{Bulging}
         \label{fig:discontinuity_print_retraction_print}
    \end{subfigure}
    \caption{
\changed{Discontinuity in extrusion paths causes defects.
(TPU, Ultimaker 3)
End points (top), T-junctions (middle) and isolated polygons (bottom) all introduce discontinuity in the extrusion process and rapid travel moves are required.
{\subref{fig:discontinuity_print_print}} When the extrusion motor stops, the material still leaks out during rapid travel moves.
{\subref{fig:discontinuity_print_retraction_print}} Retraction reduces leakage, but instead introduces bulging.} 
    }
    \label{fig:discontinuity_print}
\end{figure}

\begin{figure}
        \centering
    \begin{subfigure}[t]{0.45\columnwidth}
        \centering
         \includegraphics[height=.6\textwidth]{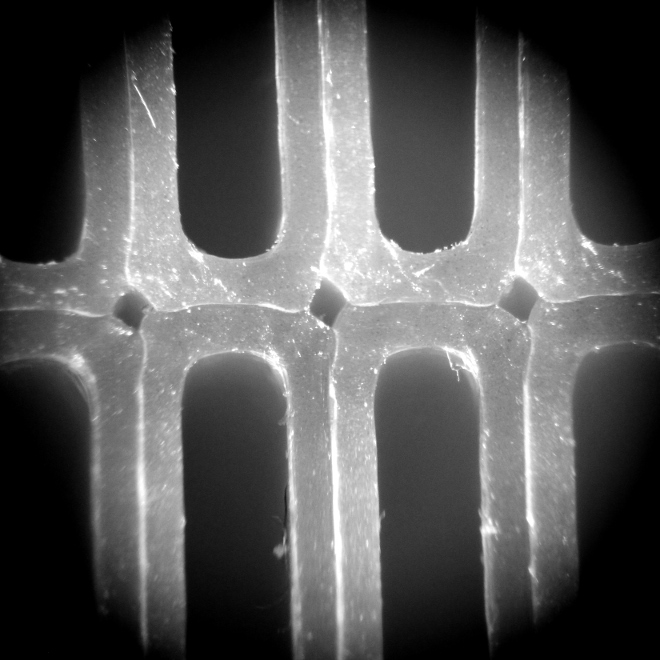}
         \includegraphics[height=.6\textwidth]{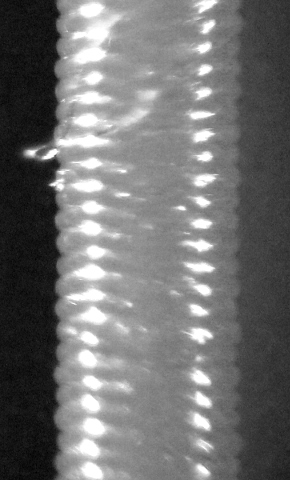}
        \caption{No overlap}
        \label{fig:overlap_constraint_no_overlap}
    \end{subfigure}
    \begin{subfigure}[t]{0.45\columnwidth}
        \centering
         \includegraphics[height=.6\textwidth]{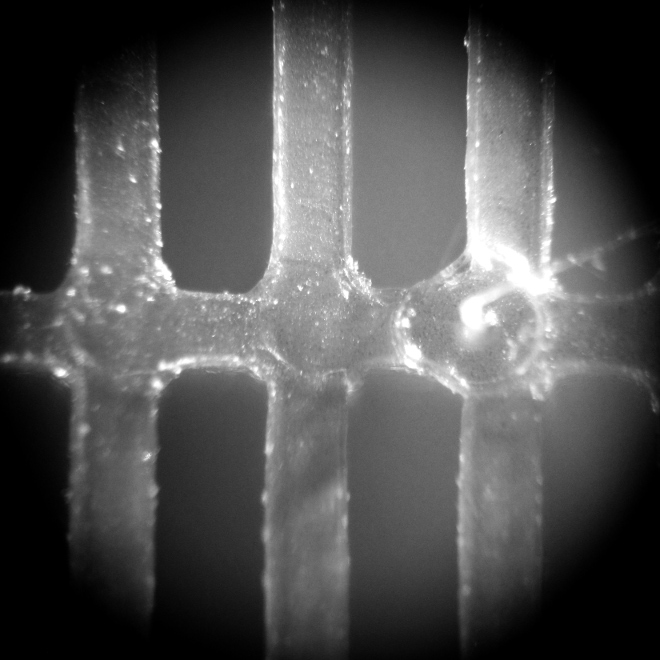}
         \includegraphics[height=.6\textwidth]{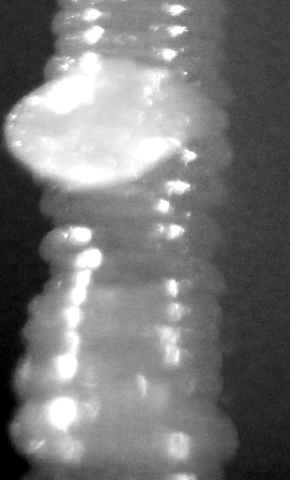}
        \caption{Overlapped}
        \label{fig:overlap_constraint_overlap}
    \end{subfigure}
    \caption{
Microscopic photos of top and side views of printing results with a \SI{0.38}{\milli\meter} wide extrusion path:
 \subref{fig:overlap_constraint_no_overlap} without versus \subref{fig:overlap_constraint_overlap} with overlapping \changed{by} \SI{0.36}{\milli\meter} respectively.
Overlapping extrusion paths exhibit over-extrusion of material at the overlapping region, which results in unwanted blobs on the surface of the print.
    }
    \label{fig:overlap_constraint}
\end{figure}

\emph{Fused deposition modeling} (FDM) is one of the most widely used 3D printing processes as it has a comparatively low running cost and supports a wide variety of materials. 
FDM systems work by extruding melted streams of material from a moving nozzle to form a quickly solidified path. 
However, there is limited study on using FDM to reliably fabricate FGM.
This task is challenging \changed{since} the complicated geometry of a functionally graded infill structure is difficult to meet the different manufacturing constraints required by FDM to ensure printing quality\changed{: }
\begin{itemize}[leftmargin=*]
    \item \textit{Overhanging geometry}: 
If a geometric feature is not properly supported by lower layers, it is said to be \changed{overhanging}. 
While overhanging geometry can be printed using support structures, the complexity of infill structures inside a 3D model does not allow an easy removal of support structures. 
Therefore, infill structures are expected to be self-supporting.

    \item \textit{Discontinuous material extrusion}: 
In extrusion-based fabrication, frequent restarting and stopping extrusion creates defects.
\changed{Simply stopping the extrusion motor} will lead to material leakage.
\changed{One common way to prevent that is to retract the filament backward a bit before starting a rapid travel move,
but that in turn introduces bulges where the extrusion paths start and end.
See} \cref{fig:discontinuity_print}.
In order to fabricate infill structures reliably, it is desired that each layer of the structures is fabricated by continuous extrusion along a single toolpath
\changed{without any interruption.}

    \item \textit{Overlapping extrusion paths}: 
Since material is extruded along a toolpath typically with a constant width, there is an excess of material in a layer when extrusion paths are too close to each other.
The overlap of extrusion paths causes blobs and wider lines, which make it difficult to control the density of infill structures in corresponding regions (see \cref{fig:overlap_constraint}\changed{). }
It is preferred to solve this problem intrinsically by generating infill structures without overlapping toolpaths. 
\end{itemize}
In order to use FDM for fabricating foam structures with graded density inside a given model, a method needs to be developed for generating infill structures according to a user-specified density distribution, which should also avoid the above manufacturing problems.

In this paper, we propose a novel type of foam structure that can achieve the aforementioned objective. 
Specifically, we develop a space-filling surface, called \emph{CrossFill}, \changed{an FDM printable} foam structure as infill for a 3D model. 
Each layer of CrossFill is a space-filling curve that can be continuously extruded along a single overlap-free toolpath. 
The space-filling surface consists of surface patches which are embedded in prism-shaped cells, which can be adaptively subdivided to match the user-specified density distribution.
The adaptive subdivision level results in graded mechanical properties throughout the foam structure.  
Our method consists of a step to determine \changed{a} lower bound for the subdivision levels at each location and a dithering step to refine the local average densities,
so that we can generate CrossFill that closely matches the required density distribution.
A simple and effective algorithm is developed to merge a space-filling curve of CrossFill of a layer into the closed polygonal areas sliced from the input model.
Physical printing tests have been conducted to verify the performance of the CrossFill structures.

\par\vspace{\baselineskip}\noindent
Our approach provides three technical contributions:
\begin{itemize}
\item A novel self-supporting space-filling surface which supports spatially graded density; 

\item A scheme for refining the structure to match a prescribed density distribution;

\item An algorithm for merging the toolpath of an infill structure with the input model's boundary so as to retain continuity.
\end{itemize}

\begin{figure*}
        \centering
    \begin{subfigure}[t]{0.05\textwidth} \centering \includegraphics[height=3\textwidth]{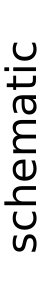}                    \includegraphics[height=3\textwidth]{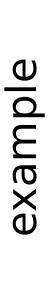}\end{subfigure}
    \begin{subfigure}[t]{0.15\textwidth} \centering \includegraphics[width=\textwidth]{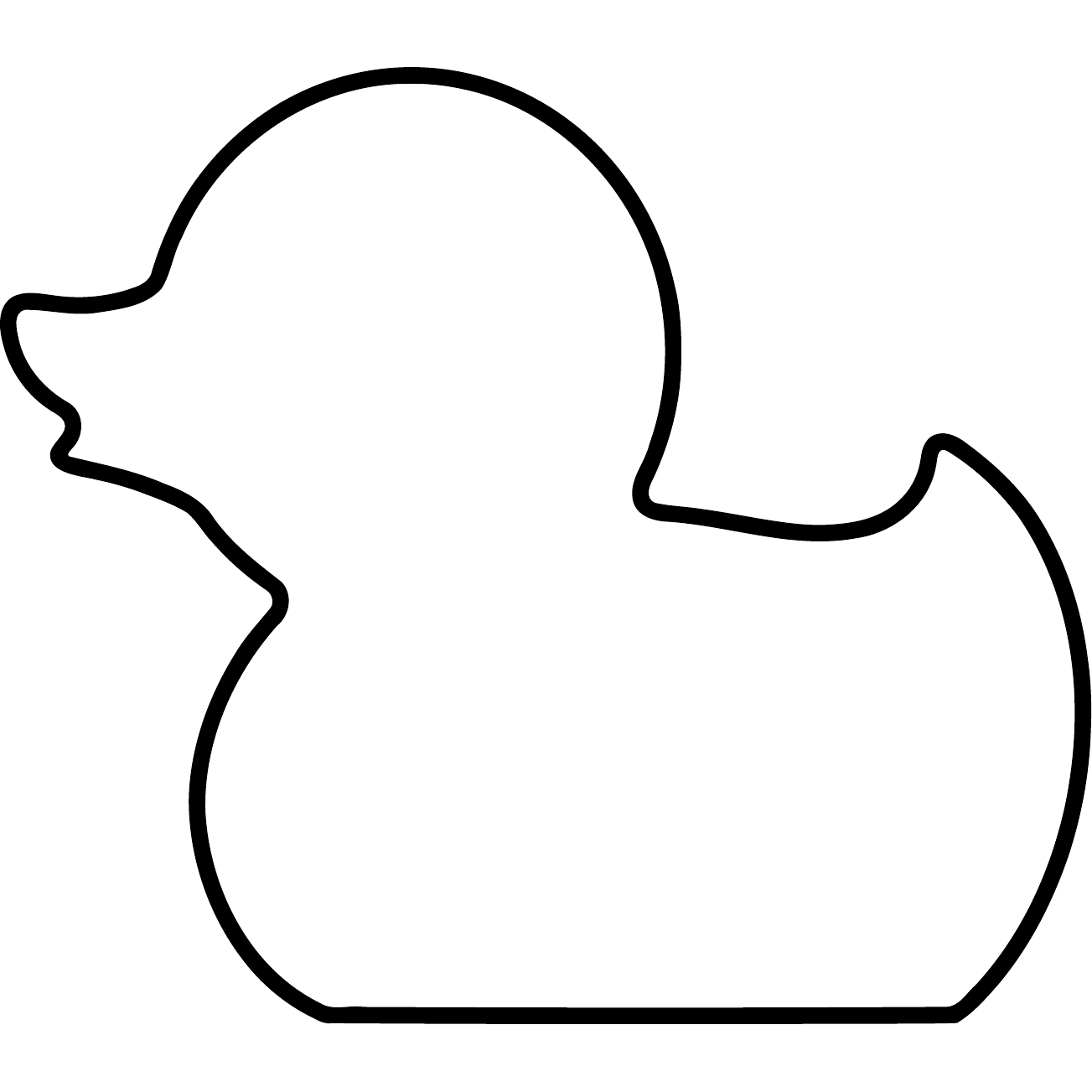}                \includegraphics[height=\textwidth]{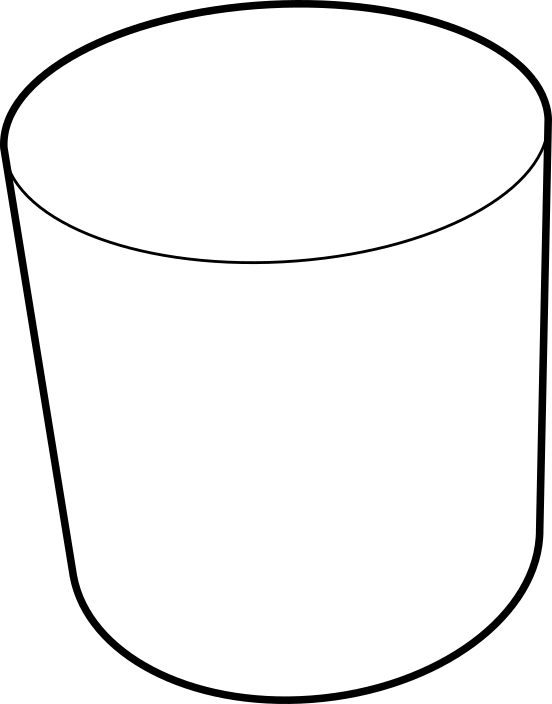} \caption{Boundary mesh} \label{fig:overview_0}\end{subfigure}
    \begin{subfigure}[t]{0.15\textwidth} \centering \includegraphics[width=\textwidth]{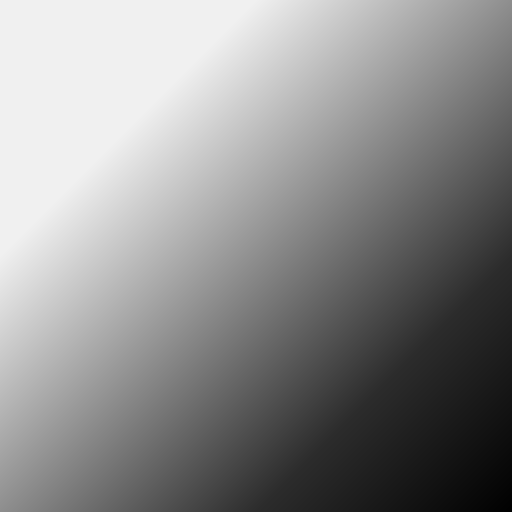}               \includegraphics[height=\textwidth]{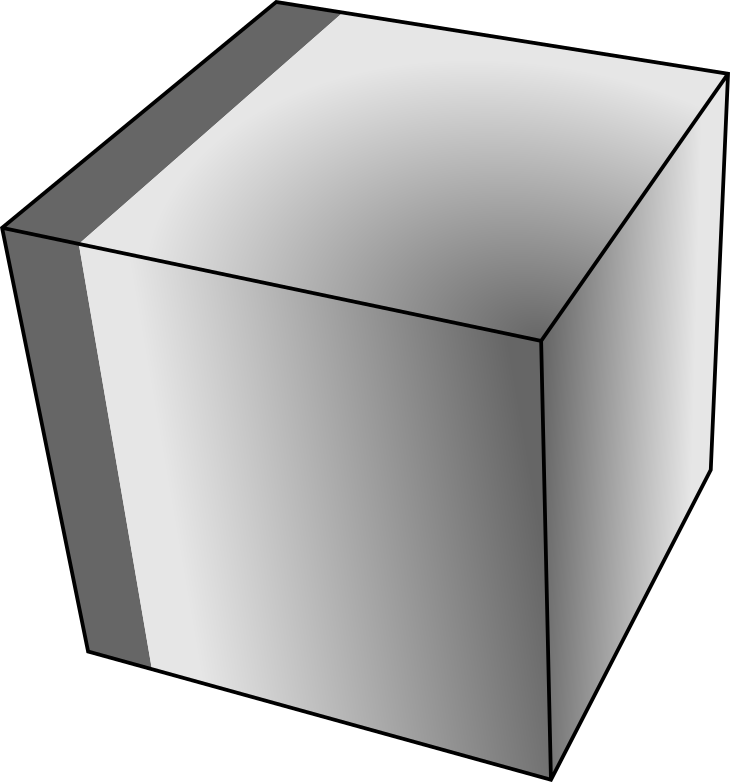} \caption{Density specification} \label{fig:overview_1}\end{subfigure}
    \begin{subfigure}[t]{0.15\textwidth} \centering \includegraphics[width=\textwidth]{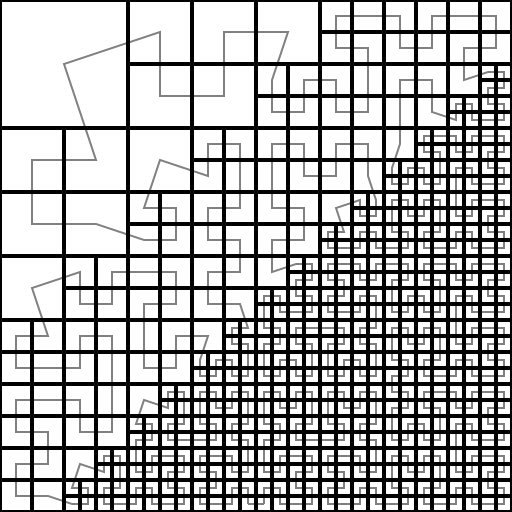}         \includegraphics[height=\textwidth]{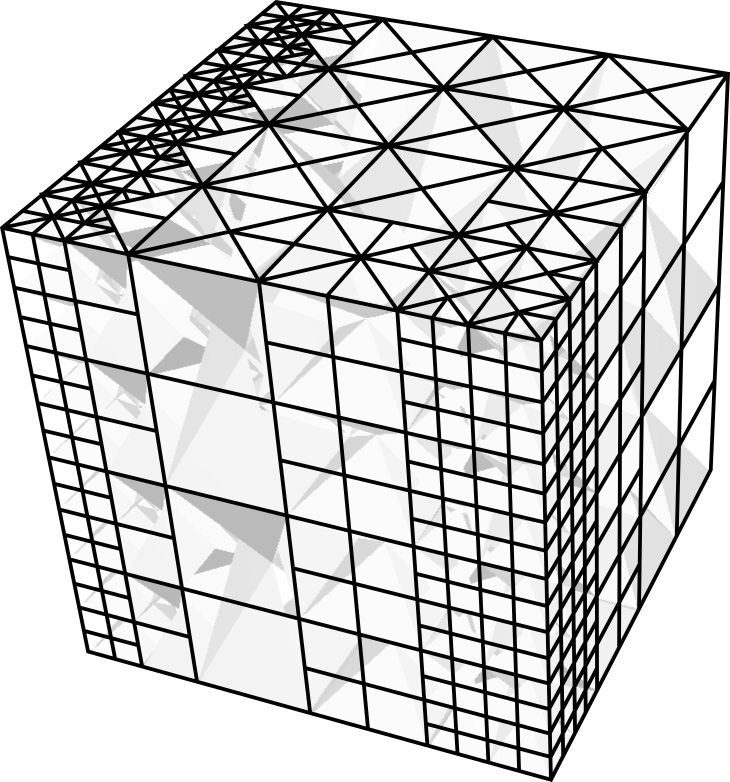} \caption{Lower bound subdivision structure}\label{fig:overview_2}\end{subfigure}
    \begin{subfigure}[t]{0.15\textwidth} \centering \includegraphics[width=\textwidth]{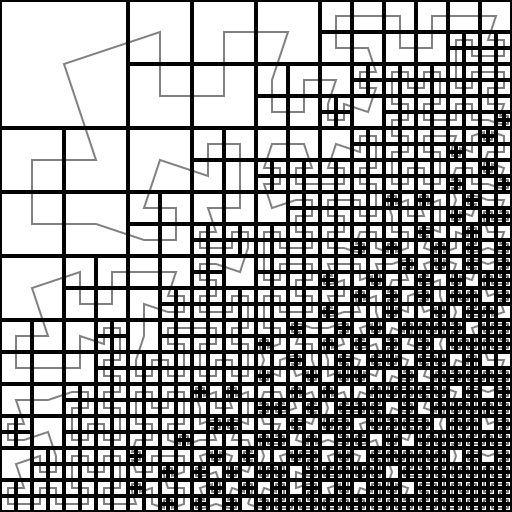}         \includegraphics[height=\textwidth]{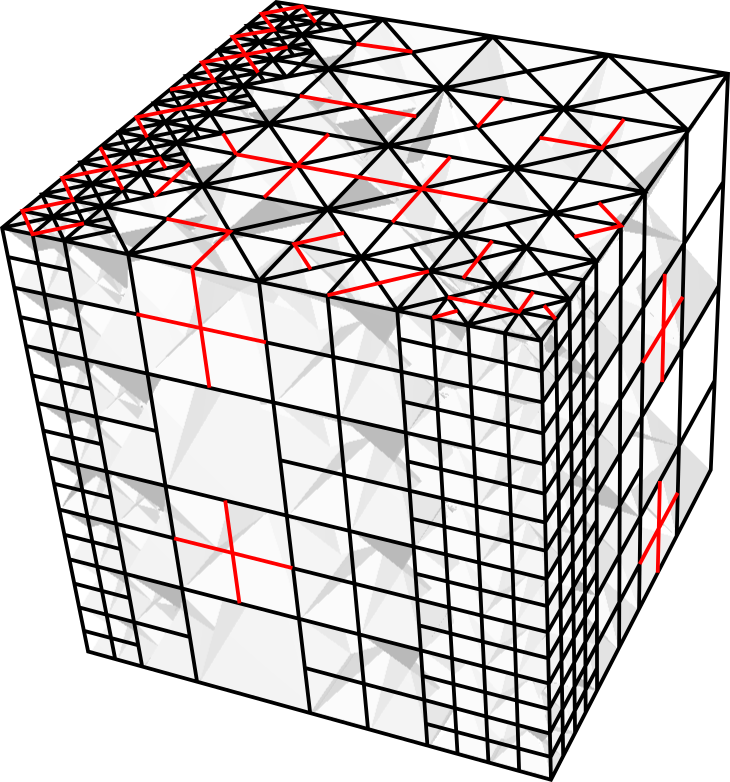} \caption{Dithering}\label{fig:overview_3}\end{subfigure}
    \begin{subfigure}[t]{0.15\textwidth} \centering \includegraphics[width=\textwidth]{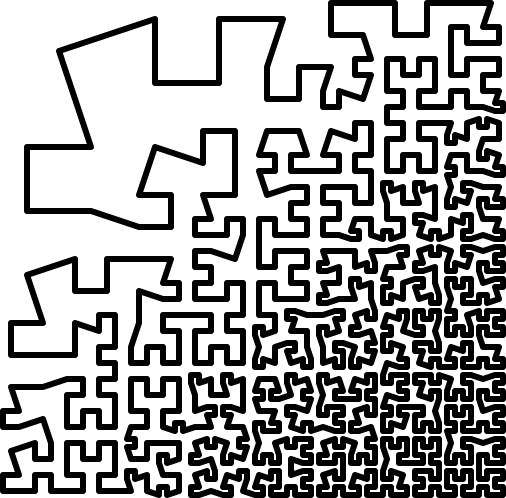}          \includegraphics[height=\textwidth]{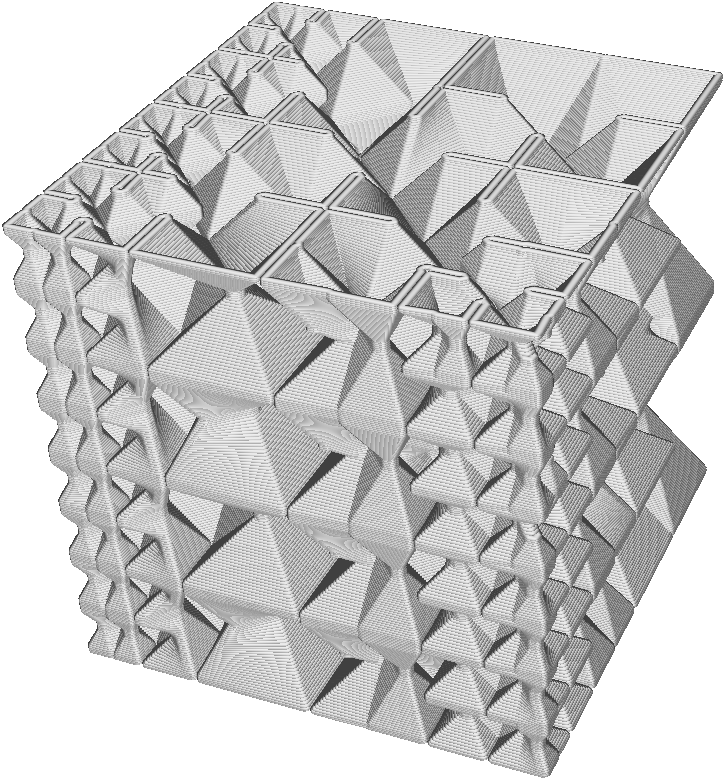} \caption{Space-filling curve extraction}\label{fig:overview_5}\end{subfigure}
    \begin{subfigure}[t]{0.15\textwidth} \centering \includegraphics[width=\textwidth]{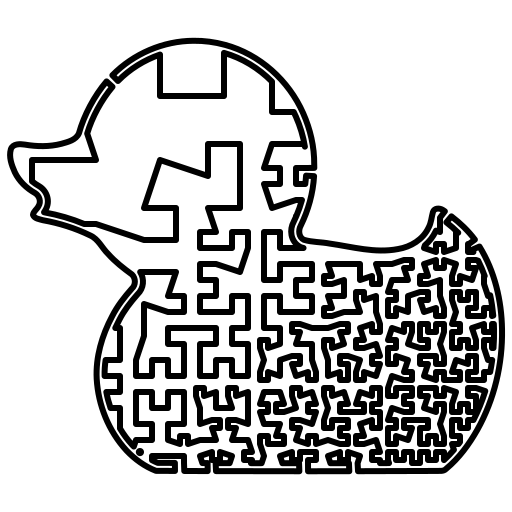}           \includegraphics[height=\textwidth]{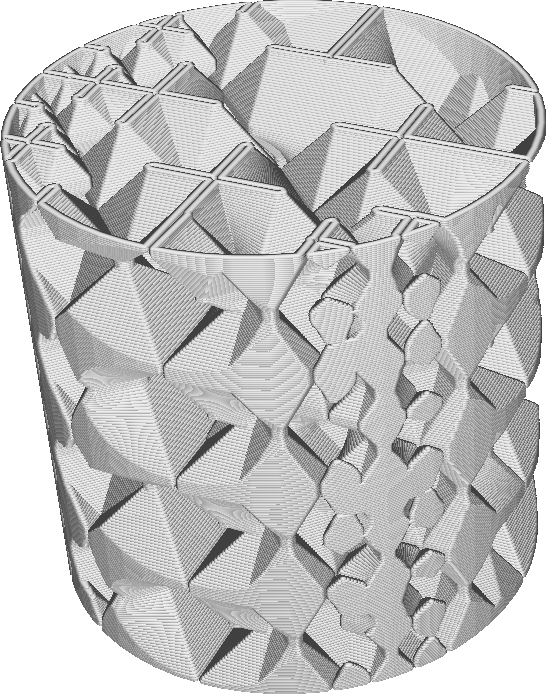} \caption{Fit into infill area}\label{fig:overview_6}\end{subfigure}
    \caption{
Schematic overview of our method.
The top row shows a 2D analogue of our method for clear visualization.
The prism-shaped cells in the bottom row are visualized as semi-opaque solids to keep the visualization uncluttered.
Red lines in the bottom row highlight the local subdivisions performed in the dithering phase.
Note that, the shell of the 3D model is not displayed in (f) for the illustration purpose.
    }
    \label{fig:overview}
\end{figure*}

\section{Related work}\label{secRelatedWork}

\label{section:relatedWork}

For an overview of techniques involved in 3D printing let us refer to the survey articles by \citet{attene2018design} and \citet{Livesu2017}. 
In this section we review the design of microstructures and in particular for extrusion-based 3D printing, as well as the use of space-filling curves and surfaces.


\subsection{Structures with graded properties}

A variety of graded structures have been proposed in recent years,
including lattice structures with varying thickness (e.g.~\cite{Martinez2016,aremu2017voxelbased,bates2018compressive,choy2017compressive,limmahakhun2017stiffness}),
triply periodic minimal surfaces (e.g.~\cite{Li2018,Abueidda2017MD}), free-form microstructures~\cite{massarwi2018hierarchical},
microstructures for expressive deformation (e.g.~\cite{schumacher2015,panetta2015,Coulais2016Nature}),
bone-like microstructures (e.g.~\cite{Liu2015CMS,Wu2018TVCG})
and microstructures optimized by inverse homogenization (e.g.~\cite{Andreassen2014MM,Garner2019AM}).

Most of these complex structures are fabricated with powder-based 3D printing systems such as selective laser sintering (SLS) or with stereolithography (SLA). 
Density gradation is typically achieved by varying the thickness of geometric primitives.
However, reliably fabricating microstructures with varying thickness is challenging for extrusion-based 3D printing. 
When printing beads narrower than the nozzle size, it is difficult to predict at which location exactly the bead will end up; furthermore, when the geometry is wider than the nozzle size the toolpath generation needs to switch from a single bead into several in a controlled manner. 
\changed{Furthermore, lattice structures are sliced into small disconnected components for each layer which violates the continuous extrusion constraint (e.g.~}\cite{Martinez2016,panetta2015,sigmund1995tailoring}).

\subsection{Structures for extrusion-based printing}
3D shapes fabricated by FDM typically comprise uniform infill structures.
Recently \citet{Hornus2018} proposed a method to generate infill structures with spatially graded density by printing the cell membranes of 3D Voronoi diagrams. 
The cells center locations are randomly sampled from a 3D user-specified probability distribution in order to create the spatially graded infill.
Overhang constraints are satisfied by carefully constructing a distance measure which forms the basis of defining the cell membranes.

The generated structures are limited by the following factors.
The extrusion paths are not continuous; the amount of retractions reduces reliability and increases print time.
The density is controlled indirectly through the density of the cell centers.
The actual relation between the two remains unclear.
At high densities the method is likely to generate overlapping extrusion paths, leading to over-extrusion, which causes defects in the print.
These problems are well resolved by our approach.

\citet{wu2016selfsupporting} proposed using a subdivision grid of slanted cubes called \emph{rhombuses} for extrusion-based 3D printing and proposed optimizing the subdivision structure for stiffness~\cite{Wu2018}.
The sides of the rhombuses can then be printed using a single bead. 
However, \changed{the T-junctions} require retractions, which are problematic especially for flexible materials. 
Our method makes use of a subdivision grid as well, but generates a continuous toolpath. 
Moreover, compared to the rigid rhombic structures, the CrossFill structure is more compliant and acts like a foam, which is beneficial for several applications such as an insole.

In this paper we assert the density distribution is prescribed by the user, and present a method to reliably reproduce the distribution using extrusion-based printing. 
The graded density can be specified by the user or by an optimization process.
The latter was exploited for example in~\cite{Liu2015CMS,Martinez2017}.

\subsection{Space-filling curves and surfaces}

Our method makes use of space-filling surfaces, which are analogous to space-filling curves in 2D (e.g. Hilbert curve~\cite{hilbert1891ueber}, Sierpi\'nski curve~\cite{sierpinski,polya1913uber}).
Using space-filling curves with varying degrees of subdivision level has been explored for purposes other than 3D printing, such as robotic exploration tasks~\cite{nair2017hilbert}, finite element analysis~\cite{Gonzaga,Bader2008}, paths for CNC milling~\cite{Griffiths1994}.
In the context of 3D printing, \citet{kumar2009fractal} combined several square based space-filling curves (e.g. Hilbert curve) to generate porous infill structures with spatially graded density and semi-continuous extrusion.
However, this method allows for spatially varying density only in the horizontal plane - not in the vertical direction.

While there is much literature on extending such space-filling curves to \emph{polylines} in 3D, 
a space-filling curve in 2D can also be extended into a space-filling \emph{surface} in 3D.
Space-filling surfaces are first defined by \citet{Ahmed2007}.
Although the space-filling surfaces are continuous, any layer-wise cross section is still discontinuous.
In this paper we propose a new type of space-filling surfaces, which provides cross sections that are continuous.




\section{Overview}\label{secOverview}
CrossFill is a space-filling surface that is constructed using subdivision rules on prism-shaped cells. 
Each cell contains a patch of the surface, which is sliced into a line segment on each layer to be a segment of the extrusion toolpath.
Since the toolpath will be fabricated with a \emph{constant} width, the size of a cell determines the regional fraction of solid material (hereafter referred to as `density').
By adaptively applying the subdivision rules to the prism cells, we create a subdivision structure of cells with a density distribution that closely matches a user-specified input.
Continuity of the space-filling surface across adjacent cells with different subdivision levels -- both horizontally and vertically -- is ensured by the subdivision rules and by post-processing of the surface patches in neighboring cells.

\Cref{fig:overview} provides an overview of our method using a simple 3D example (bottom) and a 2D schematic illustration (top).
From a user-specified 3D density field (\cref{fig:overview_1}),
we first increase the subdivision levels everywhere until one further subdivision would result in an average cell density higher than the average requested density in that region
 (\cref{fig:overview_2}).
The resulting subdivision structure forms the \emph{lower bound} of the final subdivision levels. 
In order to closely match the input density distribution, we develop a dithering method in which we alternate the subdivision level between the lower bound and one level deeper (\cref{fig:overview_3}).
Once the subdivision structure is finalized, we slice it into space-filling curves for the toolpaths on each layer (\cref{fig:overview_5}).
In this step, we adjust the surface patches in the cells such that overlapping toolpaths are prevented.
Lastly, the space-filling curves are trimmed to the infill area and connected to the shell of the input 3D models to form the final toolpaths preserving continuous extrusion (\cref{fig:overview_6}). 

\section{CrossFill}\label{section:crossfill}
CrossFill is an infill structure which consists of a space-filling surface.
CrossFill is so named because the toolpath of this structure resembles crosses (see \cref{fig:simple_crossfill_2D_and_3D}).

\subsection{Initialization and data-structure }\label{section:data_structure}
%
%

\begin{figure}
        \centering
    \begin{subfigure}[t]{0.45\columnwidth}
        \centering
         \includegraphics[width=\textwidth]{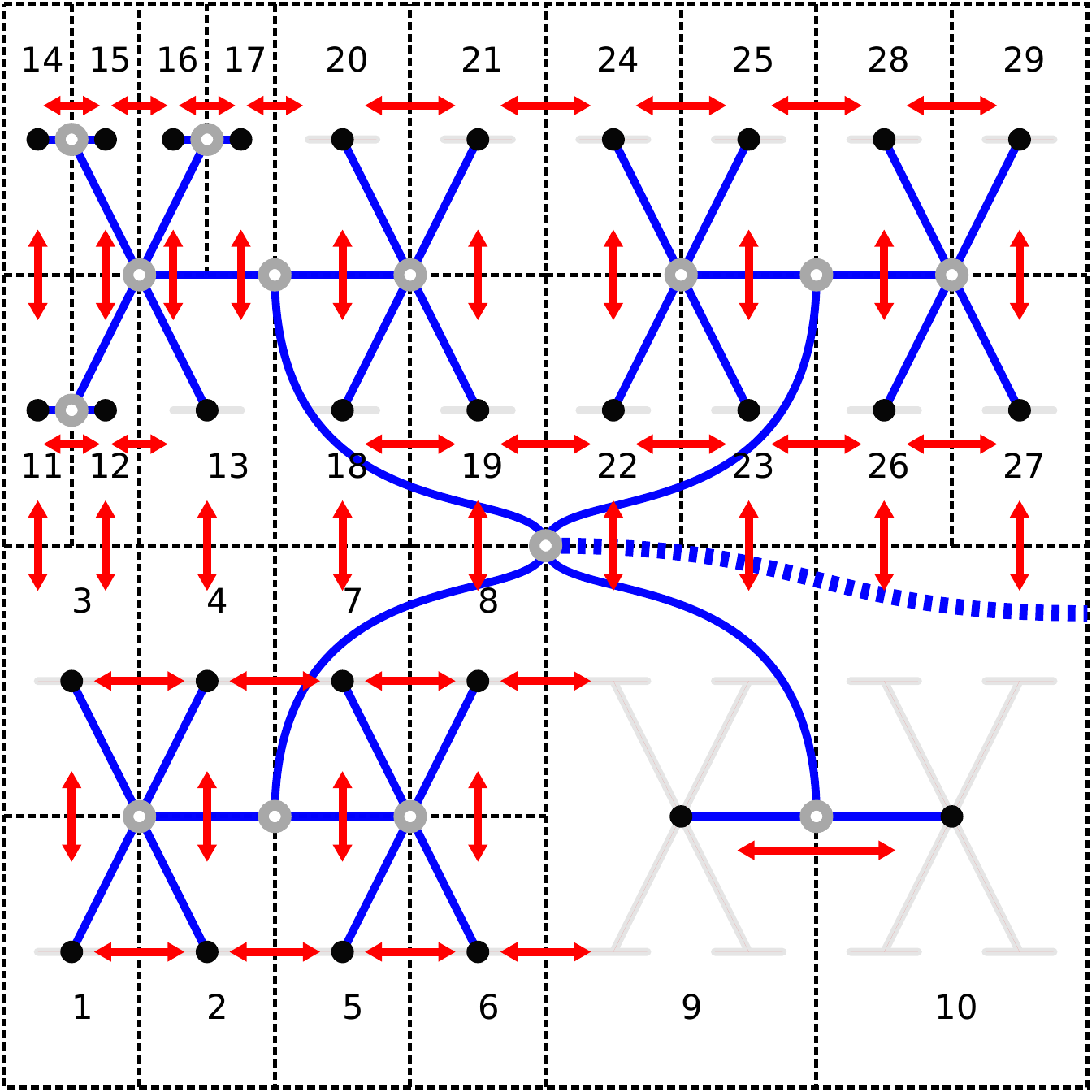}
        \caption{Subdivision tree}
	\label{fig:subdiv_tree_unfolded}
    \end{subfigure}
    \begin{subfigure}[t]{0.45\columnwidth}
        \centering
         \includegraphics[height=\textwidth]{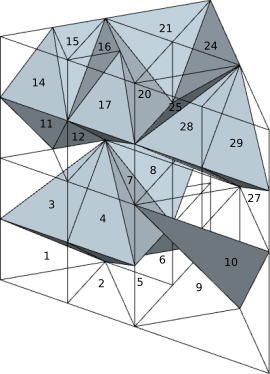}
        \caption{Prism cells}
	\label{fig:subdiv_tree_folded}
    \end{subfigure}
    \caption{
The hybrid tree and graph data-structure employed for CrossFill, where the blue links indicate the subdivision relationship between prism cells and the black dots represent the leaf-nodes on the tree each \changed{of which embeds} a patch of the space-filling surface.
The red arrows denote the connectivity graph for easily traveling from one cell to its neighbors. 
    }
    \label{fig:subdiv_tree_complex}
\end{figure}

The space-filling surface of CrossFill is generated using a subdivision scheme that starts from a cube which encompasses the input 3D model with side lengths $2^{i}w$ for integer values of $i$ and a constant extrusion width $w$.
This cube is divided into four prisms by splitting along the two diagonals of the horizontal faces. 
Starting from these four prisms, the prisms are adaptively subdivided into smaller ones. 
In lieu of the commonly used 1:8 subdivision of cubic cells, we make use of 1:2 and 1:4 subdivision of prism-shaped cells. 
This allows more granularity, which is beneficial for matching the requested density distribution. 
Details on the types of prism-shaped cells and subdivision rules will be presented in \cref{subsec:types} and \ref{subsec:rules}, respectively.

The hierarchy and connectivity of cells are encoded by a combination of a tree and a graph. 
The \emph{subdivision tree} connects a cell to its subdivided constituent cells.
The root node of the tree corresponds to the starting cube, while other nodes correspond to prism cells.
The leaf nodes of the tree constitute the current subdivision structure.
The \emph{connectivity graph} stores the connectivity information among the leaf cells.
Two neighboring leaf cells are linked if their surface patches are connected through the space-filling surface;
however, two cells which share a face which is not crossed by the space-filling surface are not linked.
The links store the relative spatial location of the neighboring cell -- up, down left or right along the space-filling surface.
See \cref{fig:subdiv_tree_complex} for an illustration. 
This connectivity graph facilitates efficient traveling between neighboring cells.
The tree and graph are updated each time a new subdivision is applied to a leaf node.

\subsection{Types of cell}
\label{subsec:types}
\begin{figure}
        \centering
    \begin{subfigure}[t]{0.4\columnwidth}
        \centering
         \includegraphics[height=.5\textwidth]{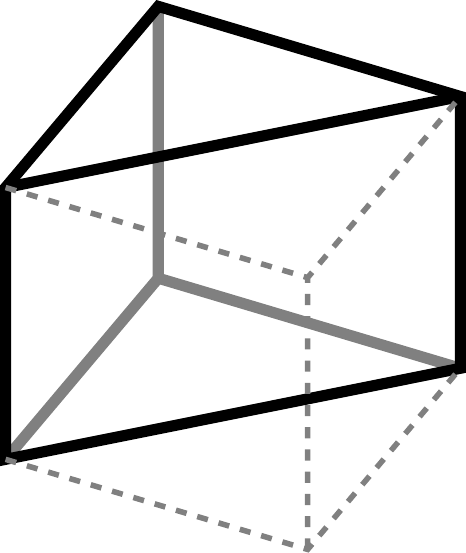}
         \includegraphics[height=.4\textwidth]{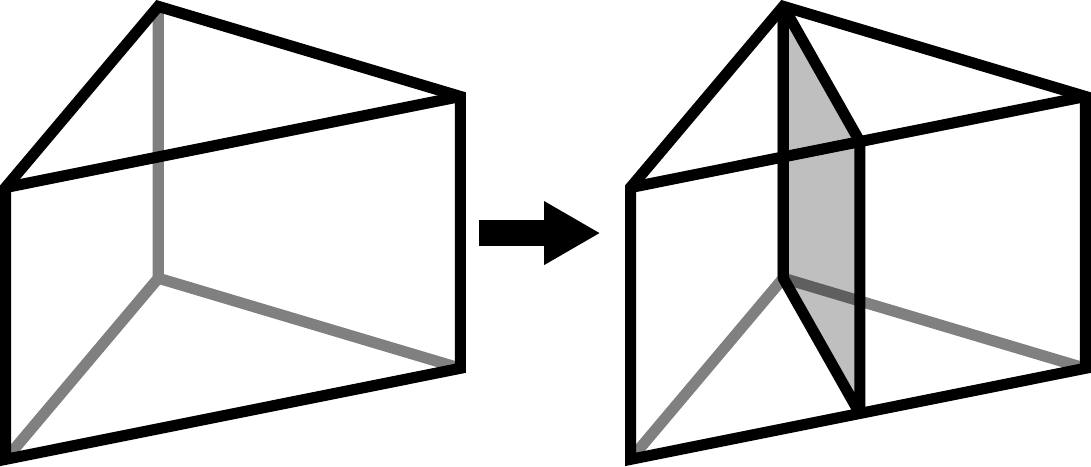}
        \caption{H-prism}
	\label{fig:triangle_subdivision_3d_two_rule}
    \end{subfigure}
    \begin{subfigure}[t]{0.4\columnwidth}
        \centering
         \includegraphics[height=.5\textwidth]{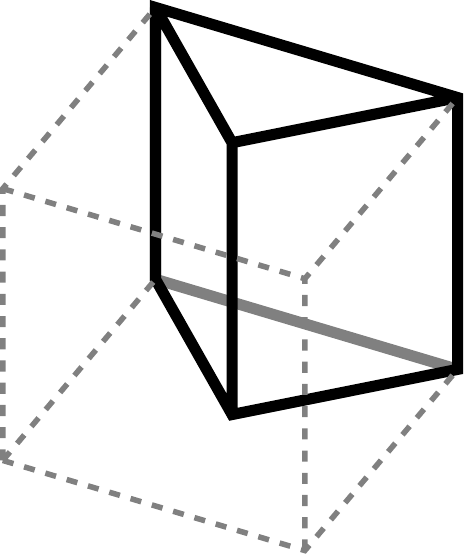}
         \includegraphics[height=.4\textwidth]{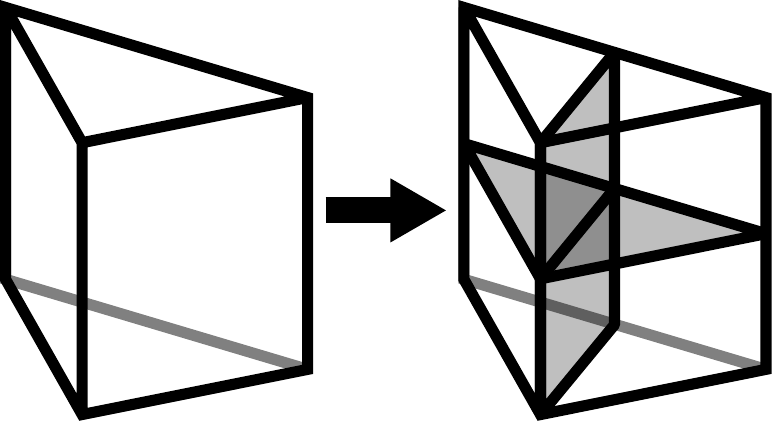}
        \caption{Q-prism}
	\label{fig:triangle_subdivision_3d_four}
    \end{subfigure}
    \caption{
The basic types of prisms in CrossFill: the shape of an H-prism can be obtained by cutting a cube in \emph{half} and the shape of a Q-prism by cutting a cube in \emph{quarters}. A H-prism can be subdivided into two Q-prisms and a Q-prism is always subdivided into four H-prisms in our system. 
    }
    \label{fig:prism_subdivision_rules}
\end{figure}
%
The construction of CrossFill depends on a subdivision tiling consisting of prism-shaped cells.
As shown in \cref{fig:prism_subdivision_rules}, two types of prism are distinguished in our subdivision system:
\begin{description}[align=left]
    \item[H-prism] is constructed by vertically cutting a cube in \emph{half} along a diagonal of \changed{the} horizontal faces.    
    \item[Q-prism] is generated by splitting a cube into \emph{quarters} \changed{along both diagonals of the horizontal faces.}
\end{description}
The subdivision tiling is generated by subdividing an H-prism into two Q-prisms and subdividing a Q-prism into four H-prisms
(see the bottom row of \cref{fig:prism_subdivision_rules}). 

\begin{figure}
        \centering
    \begin{subfigure}[t]{0.155\columnwidth} \includegraphics[height=1.7\textwidth]{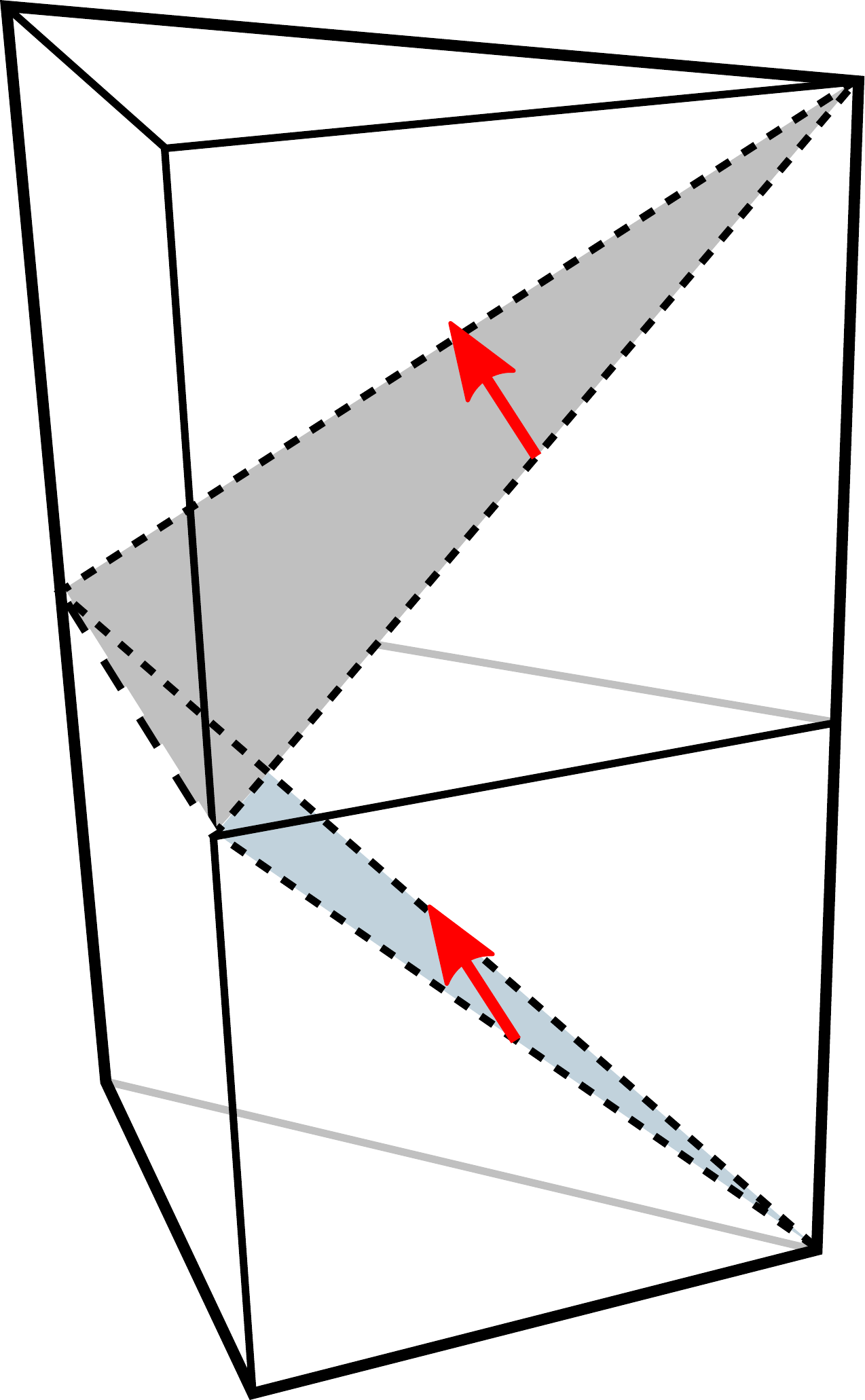}\caption{$\begin{array}{c}\text{HCL}^+\\\text{HEL}^+\end{array}$}\label{fig:basic_cell_HL}\end{subfigure}
    \begin{subfigure}[t]{0.155\columnwidth} \includegraphics[height=1.7\textwidth]{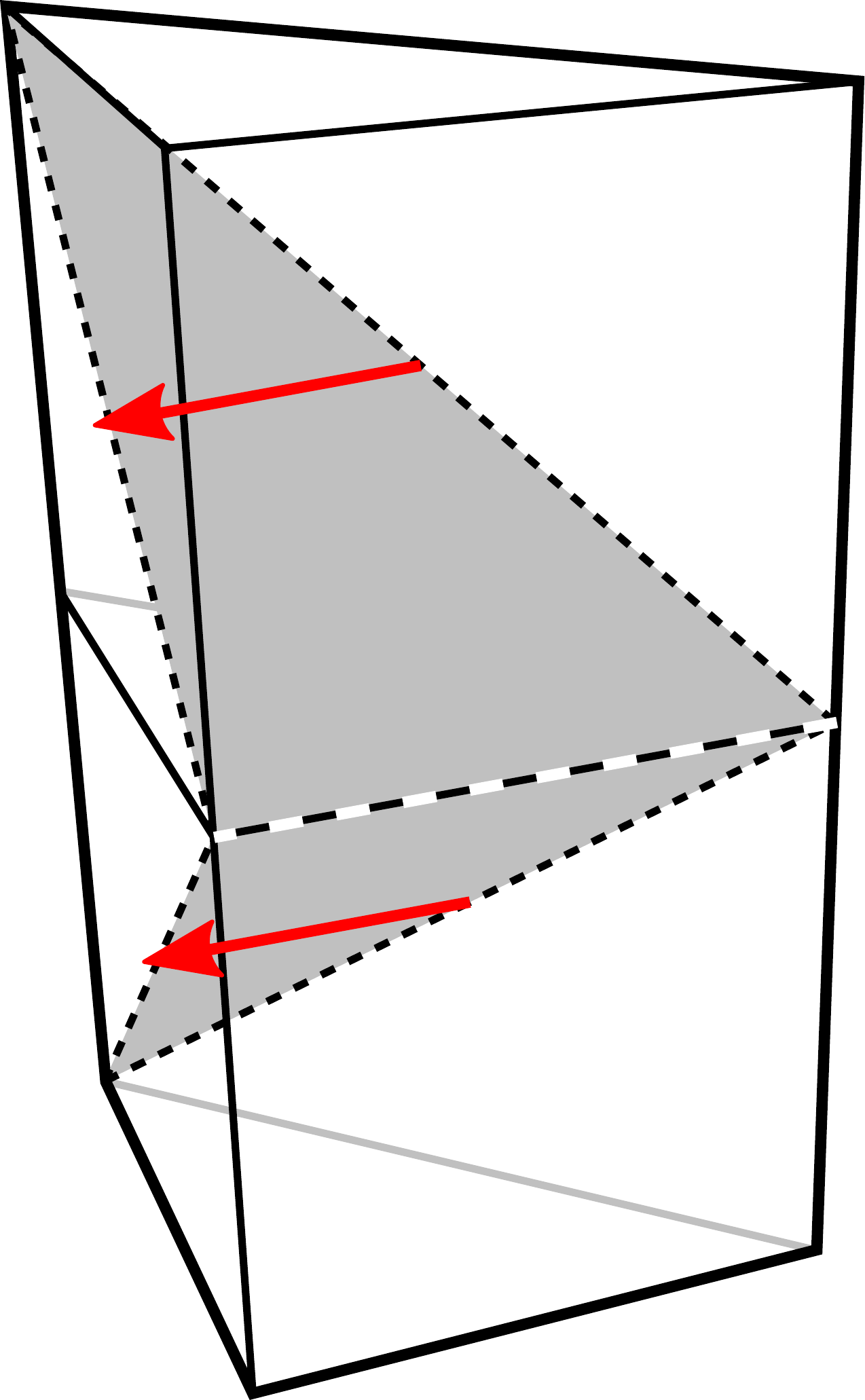}\caption{$\begin{array}{c}\text{HCR}^+\\\text{HER}^+\end{array}$}\label{fig:basic_cell_HR}\end{subfigure}
    \begin{subfigure}[t]{0.155\columnwidth} \includegraphics[height=1.7\textwidth]{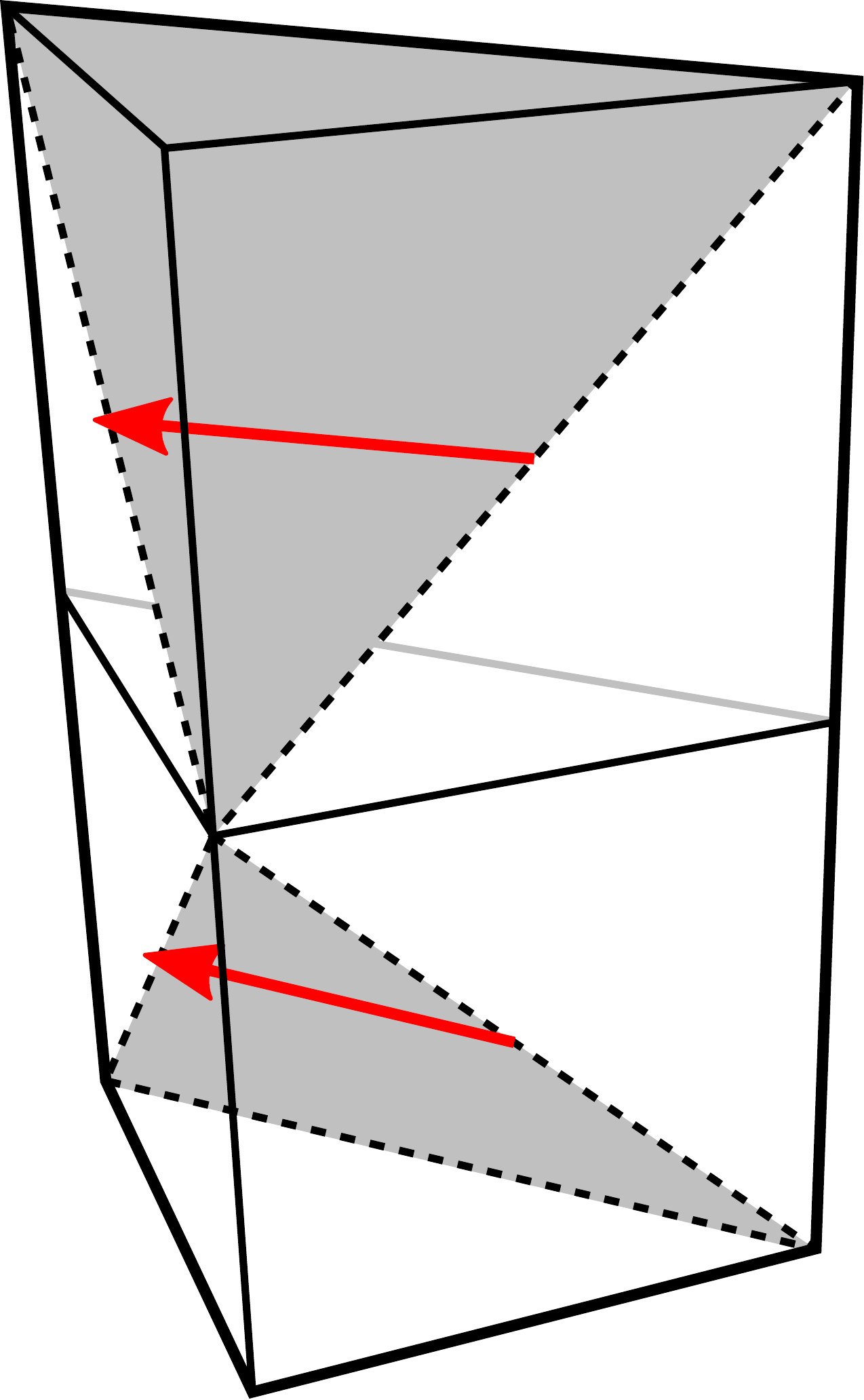}\caption{$\begin{array}{c}\text{HCA}^+\\\text{HEA}^+\end{array}$}\label{fig:basic_cell_HA}\end{subfigure}
    \begin{subfigure}[t]{0.155\columnwidth} \includegraphics[height=1.7\textwidth]{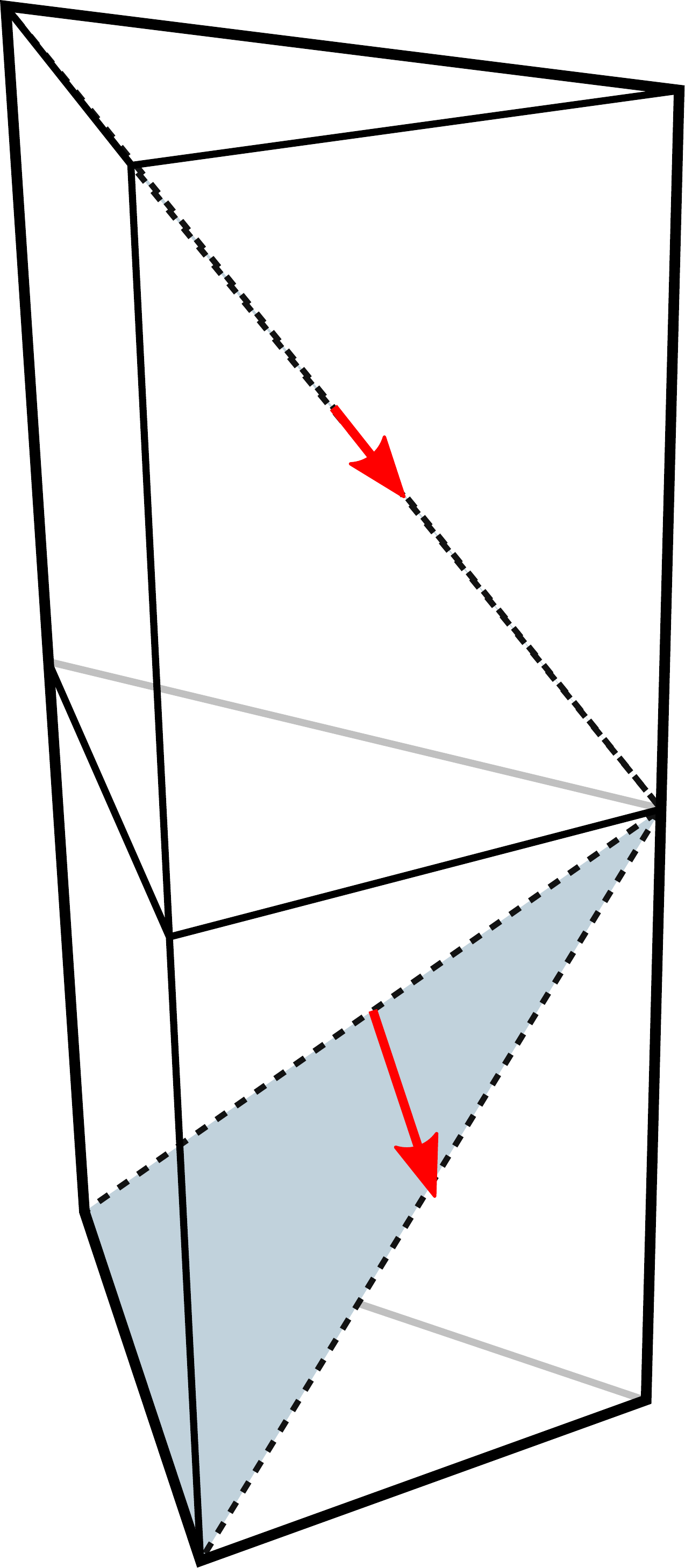}\caption{$\begin{array}{c}\text{QCL}^-\\\text{QEL}^-\end{array}$}\label{fig:basic_cell_QL}\end{subfigure}
    \begin{subfigure}[t]{0.155\columnwidth} \includegraphics[height=1.7\textwidth]{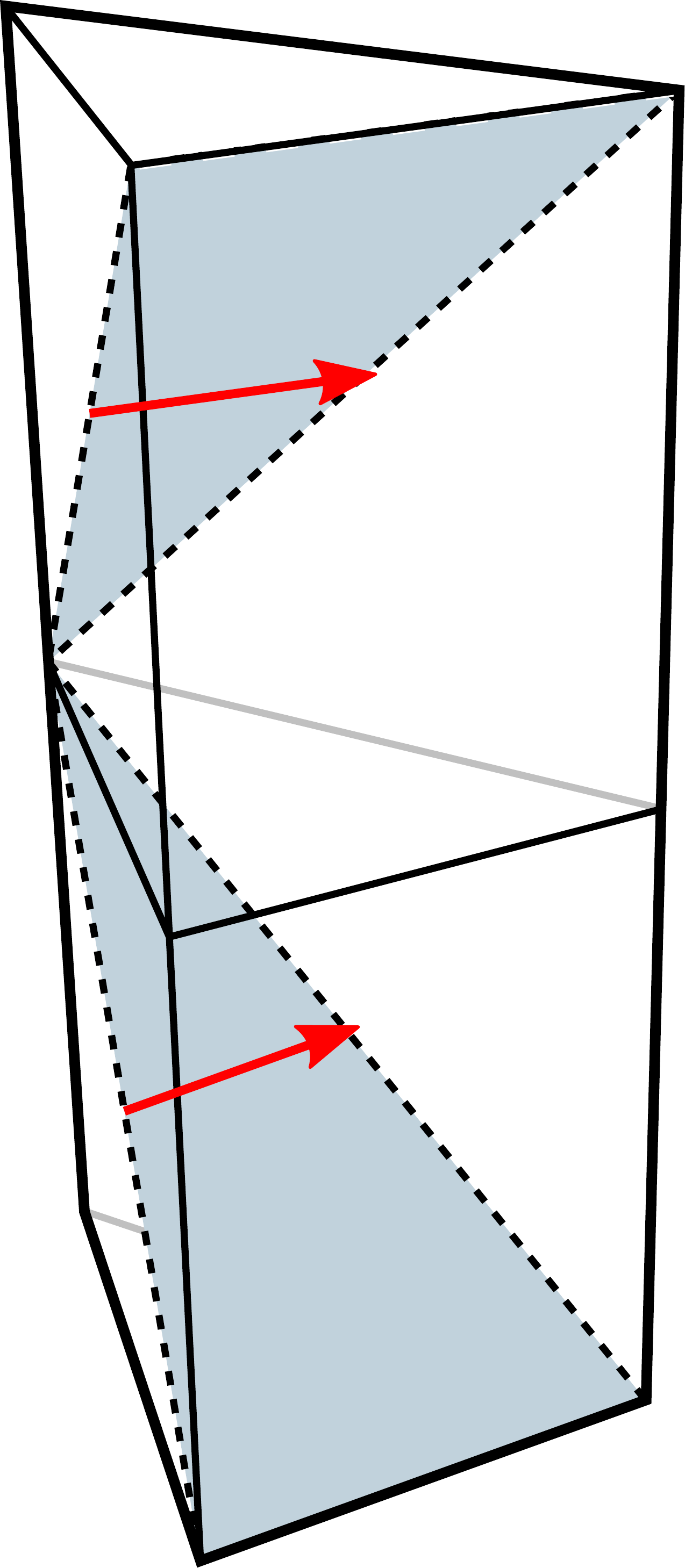}\caption{$\begin{array}{c}\text{QCR}^-\\\text{QER}^-\end{array}$}\label{fig:basic_cell_QR}\end{subfigure}
    \begin{subfigure}[t]{0.155\columnwidth} \includegraphics[height=1.7\textwidth]{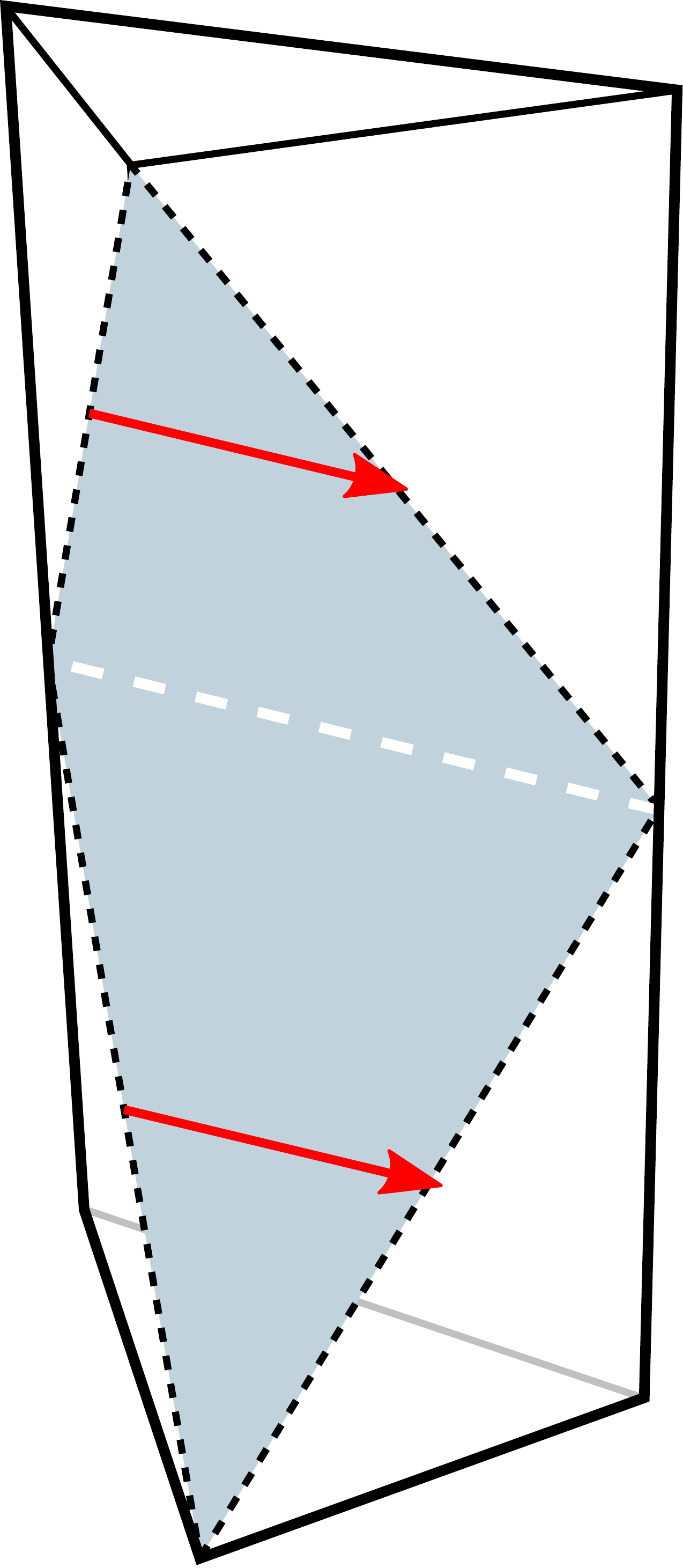}\caption{$\begin{array}{c}\text{QCA}^-\\\text{QEA}^-\end{array}$}\label{fig:basic_cell_QA}\end{subfigure}
     \caption{
The types of prism cells by considering the embedded space-filling surfaces.
     }
    \label{fig:basic_cells_with_triangles}
\end{figure}

\begin{figure}
        \centering
    \begin{subfigure}[t]{\columnwidth}
        \centering
         \includegraphics[width=.5\textwidth]{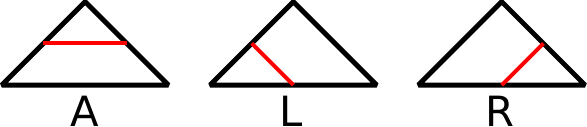}
        \caption{Three different types of route}
	\label{fig:subdivision_rules_types}
    \end{subfigure}
    \begin{subfigure}[t]{\columnwidth}
        \centering
         \includegraphics[width=\textwidth]{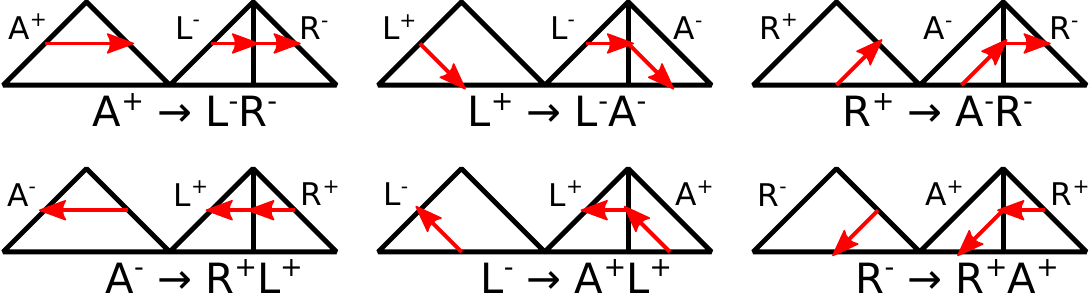}
        \caption{Subrules for `$x \to a | b$' production}
	\label{fig:subdivision_rules_rules}
    \end{subfigure}
    \caption{
Types of surface patch distinguished by the route (which sides of the cell the segment crosses) and the direction.
When a cell is subdivided, the type of route $x$ is substituted with different routes in the two newly constructed cells as $a$ and $b$ respectively.
    }
    \label{fig:subdivision_rules_route}
\end{figure}

Each prism cell encompasses a triangular patch of the space-filling surface.
The prism cells are categorized according to which side faces of the prism are crossed by the triangle surface patch (see the horizontal cross sections visualized in \cref{fig:subdivision_rules_types}):
\begin{description}[align=left]
    \item[A-route] the surface is spanned across the faces connected to the two catheti of the right triangle at the base of the prism.
    \item[L-route] the surface crosses between the face connected to the hypotenuse and the left cathetus.
    \item[R-route] the surface crosses between the face connected to the hypotenuse and the right cathetus.
\end{description}

In order to keep track of the spatial ordering between cells when subdividing and when updating the connectivity graph, we introduce the horizontal  direction of traversal (see  \cref{fig:basic_cells_with_triangles}):
\begin{description}[align=left]
\item[$+$ direction] travel from left to right, from left to hypotenuse or from hypotenuse to right.
\item[$-$ direction] travel from right to left, from right to hypotenuse or from hypotenuse to left.
\end{description}

In order to fully characterize a cell, we also distinguish between the two fashions in which the 3D surface patch is embedded in the prism
(compare the \changed{bottom and top} in \cref{fig:basic_cells_with_triangles}): 

\begin{description}[align=left]
\item[E-embedding] when vertically exploring a cell from bottom to top by horizontal cross sections, the embedded surface is said to be \emph{expanding} if it is moving from the right of the traveling direction to the left.
\item[C-embedding] the embedded surface is \emph{contracting} \changed{otherwise.} 
\changed{(Note that the embedding is not defined in terms of whether the triangle surface patch points up or down.)}
\end{description}

In total we consider 12 different types of prism cell; see \cref{fig:basic_cells_with_triangles}.
It can be easily verified that the overhanging angle of embedded surfaces in all types of prism cell is always less than \SI{45}{\degree} -- i.e., the \emph{self-supporting} constraint is satisfied.

\begin{figure}
        \centering
    \begin{subfigure}[t]{0.45\columnwidth}
        \centering
         \includegraphics[width=\textwidth]{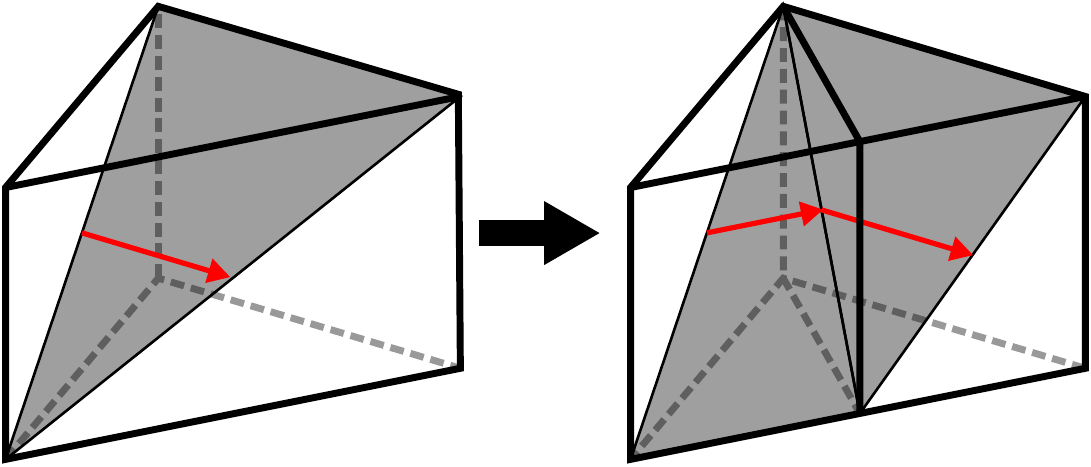}
        \caption{$\text{HCL}^+ \to \begin{array}{c|c} \text{QCL}^- & \text{QCA}^-\end{array} $}
	\label{fig:triangle_subdivision_3d_two_rule_example}
    \end{subfigure}
    \begin{subfigure}[t]{0.45\columnwidth}
        \centering
         \includegraphics[width=.8\textwidth]{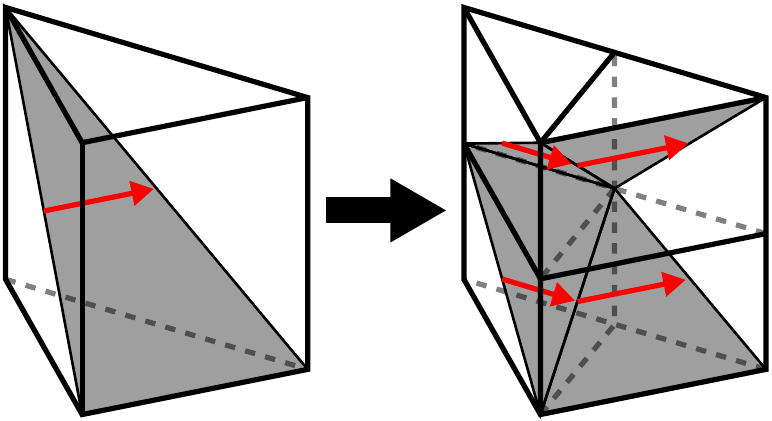}
        \caption{$\text{QCR}^- \to \begin{array}{c|c} \text{HER}^+& \text{HEA}^+ \\ \hline \text{HCR}^+& \text{HCA}^+ \end{array}$}
	\label{fig:triangle_subdivision_3d_four_rule_example}
    \end{subfigure}
    \caption{
Examples of combined subdivision rules.
    }
    \label{fig:prism_subdivision_examples}
\end{figure}

\subsection{Subdivision rules}
\label{subsec:rules}

We now define the subdivision rules on the prism cells, which depend on the type of prism (H- or Q-prism), the route type (A-, L- or R-route), the direction of traversal ($+$ or $-$) and the type of embedding (C- or E-embedding):
\begin{equation}
\begin{split}
\text{HC}x \to \begin{array}{c|c} \text{QC}a & \text{QC}b \end{array} & \quad 
\text{HE}x \to \begin{array}{c|c} \text{QE}a & \text{QE}b \end{array}
\\
\text{QC}x \to \begin{array}{c|c} \text{HE}a & \text{HE}b \\ \hline \text{HC}a & \text{HC}b \end{array} & \quad 
\text{QE}x \to \begin{array}{c|c} \text{HC}a & \text{HC}b \\ \hline \text{HE}a & \text{HE}b \end{array}
\end{split}
\label{eq:prism_subdivision_rules}
\end{equation}
where the pattern `$x \to a | b$' can be filled with any of the subrules for route type and traveling direction as given in \changed{{\cref{fig:subdivision_rules_rules}}}.
The spatial ordering on the right hand side of the rule indicates the spatial ordering of the cells: on top of each other and horizontally next to each other when following the route along its direction.
Two examples of filled subdivision rules described by \cref{eq:prism_subdivision_rules} are illustrated in \cref{fig:prism_subdivision_examples}.

We initialize the CrossFill fractal with four QCA$^-$ cells such that the surface patches form a pyramid.
Together with the subdivision rules this forms a system closely related to an L-system.

\begin{figure}
        \centering
    \begin{subfigure}[t]{0.35\columnwidth}
        \centering
         \includegraphics[width=.9\textwidth]{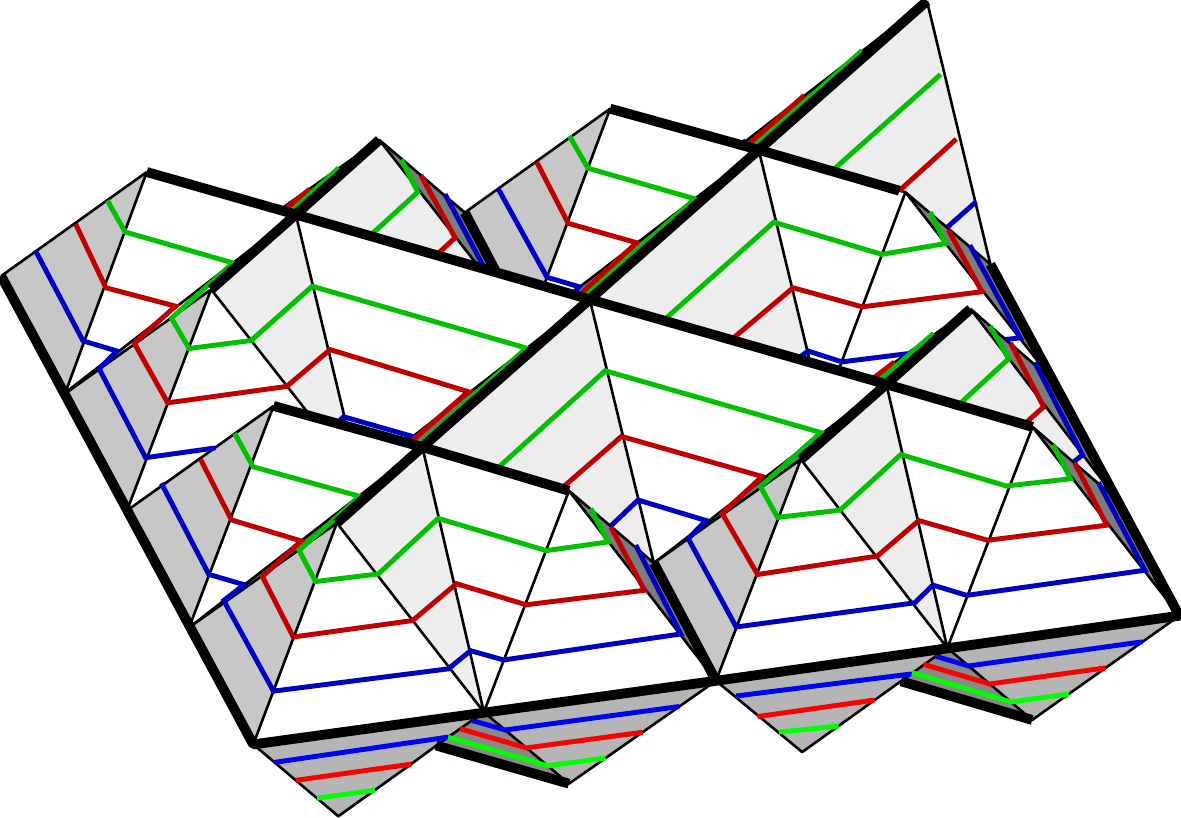}
        \caption{CrossFill 3D}
	\label{fig:cross_oscillating_extrusion}
    \end{subfigure}
    \begin{subfigure}[t]{0.25\columnwidth}
        \centering
         \includegraphics[width=.9\textwidth]{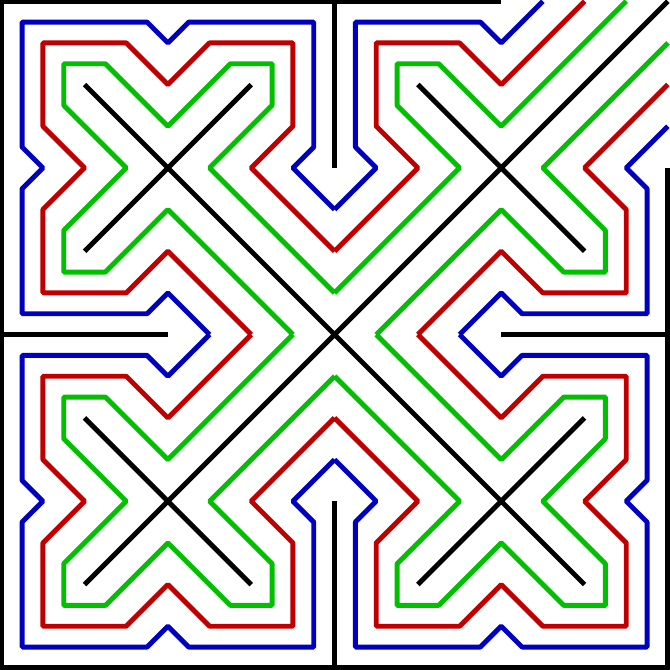}
        \caption{Layers}
	\label{fig:cross_oscillating_extrusion_crosssection}
    \end{subfigure}
    \begin{subfigure}[t]{0.25\columnwidth}
        \centering
         \includegraphics[width=.9\textwidth]{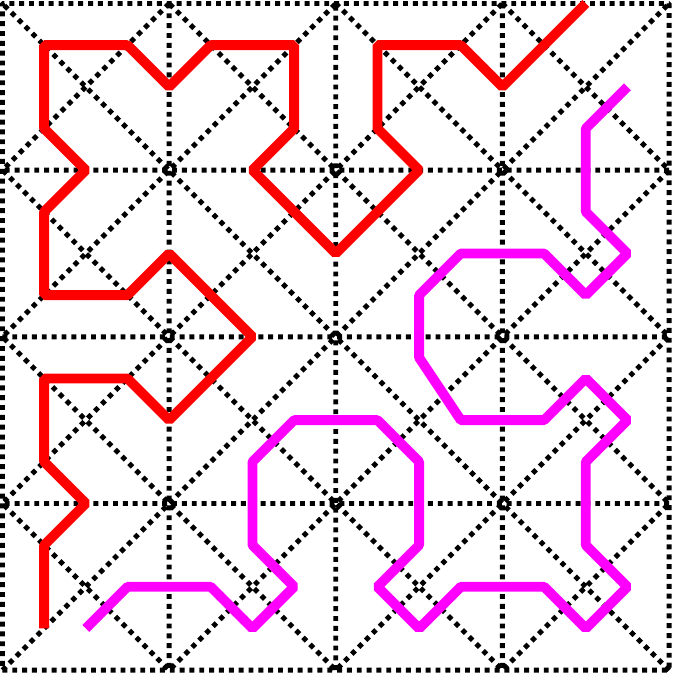}
        \caption{2D L-systems}
	\label{fig:2D_L_systems}
    \end{subfigure}
    \caption{
Intersecting the uniformly subdivided CrossFill \subref{fig:cross_oscillating_extrusion} by horizontal planes results in space-filling curves \subref{fig:cross_oscillating_extrusion_crosssection}.
The red space-filling curve can be obtained by applying the subrules of \cref{fig:triangle_subdivision_3d_four_rule_example}.
The Sierpi\'nski curve is visualized in purple in \subref{fig:2D_L_systems}.
    }
    \label{fig:simple_crossfill_2D_and_3D}
\end{figure}

\begin{figure}
\includegraphics[width=\columnwidth]{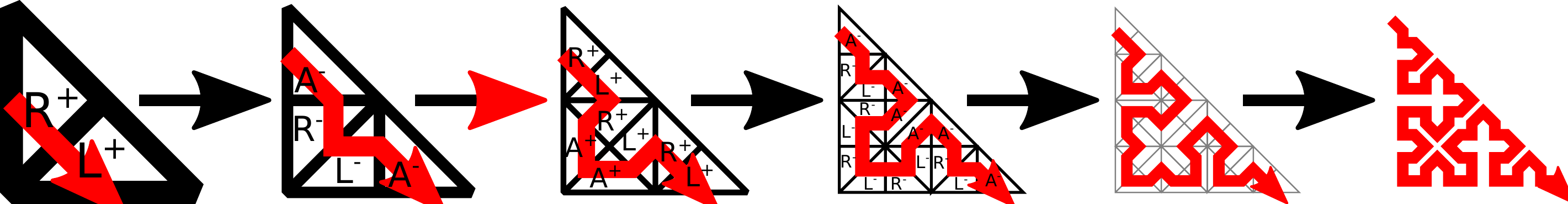}
\caption{Repeated application of the subdivision rules defined in \cref{fig:subdivision_rules_rules}.
Denser and denser space-filling curves can be generated from the subdivided space tiling.}
    \label{fig:subdivision_rules_reiteration}
\end{figure}

\paragraph{Relation to 2D L-systems}
As illustrated in \cref{fig:cross_oscillating_extrusion} and \subref{fig:cross_oscillating_extrusion_crosssection}, a subdivision tiling of right triangles and an embedded 2D space-filling curve can be obtained when intersecting a uniformly subdivided CrossFill structure with a horizontal\changed{plane} at an altitude of half the prism's height.
The rules used above with integrated traversal information can help generate a 2D space-filling curve similar to the Sierpi\'nski curve (see the purple curve in \cref{fig:2D_L_systems}). 
Whereas the Sierpi\'nski curve is generated by connecting the centers of the triangle cells, the CrossFill pattern is generated by connecting the vertices located on the edges which are crossed by the curve.
Compared to the 1:4 subdivision of square cells (e.g. for constructing the Hilbert curve), the 1:2 subdivision of triangles allows more granularity.
Similarly, our 3D subdivision rules provide more granularity than a cubic 1:8 subdivision.
As illustrated in \cref{fig:subdivision_rules_reiteration}, repeatedly denser space-filling curves can be obtained when reapplying the subdivision rules defined in \changed{{\cref{fig:subdivision_rules_rules}}}.

\subsection{Compatibility and continuity}\label{section:continuity_enforcement}
%
%
%
The surface patches embedded in a subdivision structure with uniform subdivision level form a continuous space-filling surface.
With adaptive subdivision levels, neighboring cells can have different sizes.
The boundary of the surface patch embedded in a big cell does not always match with the boundary of surface patches in neighboring linked smaller cells (e.g. \changed{prisms 10 and 27 in {\cref{fig:subdiv_tree_complex}}}). 

\begin{figure}
        \centering
         \includegraphics[width=\columnwidth]{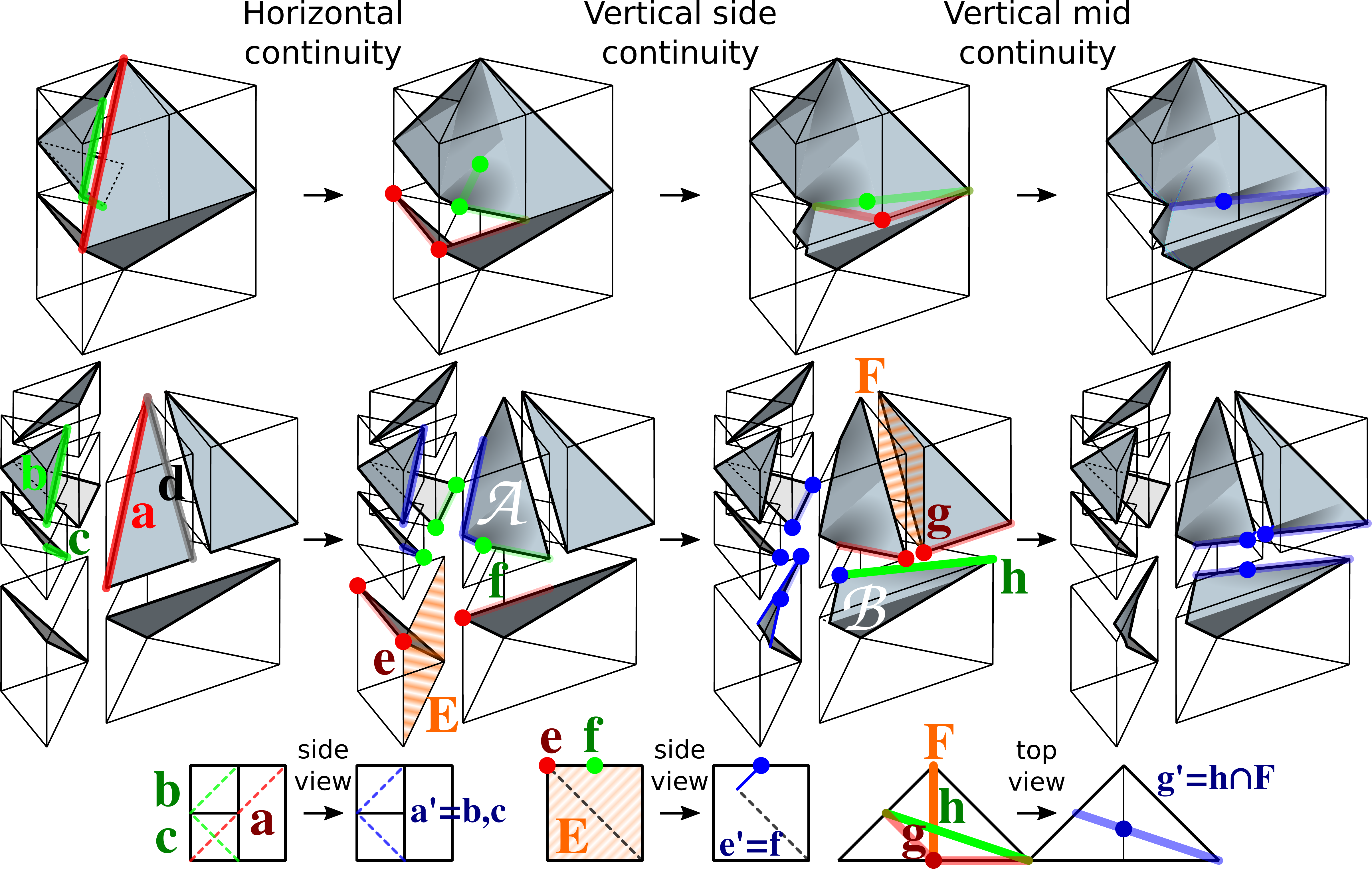}
    \caption{
Three steps for adjusting the space-filling surface to enforce the continuity between neighboring linked cells.
\changed{Adjustments are made to edges on the sides of the surface patches and the patches are thereby transformed into ruled surfaces
Red elements (vertices and edges) are to be changed into the corresponding blue elements in order to match the green elements in this figure.
(top) Assembled view. (middle) Exploded view. (bottom) Closeups of the interface between two cells.}
    }
    \label{fig:disconnect}
\end{figure}

To solve this problem in general is difficult;
however, it is solvable when neighboring linked cells have only one level difference in the \changed{subdivision} tree, because there is only a limited set of configurations. 
The continuity can be enforced on such a structure with heterogeneous subdivision level in three steps.

\paragraph{Step 1: Horizontal continuity enforcement}
\changed{When horizontally neighboring cells have a different subdivision level their embedded surface patches may not match at the sides where the cells meet.
For example the edge $\mathbf{a}$ in {\cref{fig:disconnect}} doesn't align with $\mathbf{b}$ and $\mathbf{c}$.
In such a case the edge of the lower subdivision level cell is transformed such that it matches the higher subdivision level cells and its surface is converted into a ruled surface $\mathcal{A}$:
on each layer we connect the one edge ($\mathbf{d}$) of the surface patch with either of the two edges ($\mathbf{b}$ or $\mathbf{c}$) of the smaller surface patches using a horizontally straight line segment.
See {\cref{fig:disconnect}} and {\cref{fig:continuity_enforcement_result_ruled_surface}}.}

\paragraph{Step 2: Vertical side continuity enforcement}
\changed{When vertically neighboring cells have a different subdivision level the edges of their surfaces may not end in the same location at the interface where the cells meet.
For example the vertex $\mathbf{e}$ doesn't coincide with $\mathbf{f}$ in {\cref{fig:disconnect}}.
In such a case the edge(s) of the lower subdivision level cell are transformed:
part of the edge is flipped horizontally in the plane $\mathbf{E}$ along the cell side where the edge resides.
This adjustment is also performed if the discontinuity was introduced because of the horizontal continuity enforcement,
as is illustrated by surface patch $\mathcal{B}$ in the middle of {\cref{fig:disconnect}}.}

\paragraph{Step 3: Vertical mid continuity enforcement}
\changed{When vertically neighboring cells have different subdivision level and have ruled surfaces the horizontal edges may not align on the horizontal side where the cells meet.
For example vertex $\mathbf{g}$ doesn't lie on the edge $\mathbf{h}$ in {\cref{fig:disconnect}}.
In such a case the vertices of the higher subdivision level cells are adjusted to lie on the intersection between edge $\mathbf{h}$ and the side $\mathbf{F}$ where the two higher subdivision cells meet horizontally.
Similar to vertical side continuity enforcement, we flip part of the edge to which the adjusted vertex belongs and introduce ruled surfaces.}

\bigskip

\noindent
The space-filling surface in linked cells with subdivision level differences can be effectively enforced to be continuous by this approach. 
\changed{Examples of CrossFill structures with continuity enforcement are shown in {\cref{fig:continuity_enforcement_results}}.
For more extensive examples we refer to the video in the supplementary material.}
Note that in the implementation we do not actually construct the surface;
instead, we compute the vertex locations of the edge segments of the surface patches, slice those at the height of a printing layer and connect the resulting locations using straight line segments.

\newlength{\contenfheight}
\setlength{\contenfheight}{.25\columnwidth}
\begin{figure}
        \centering
    \begin{subfigure}[t]{0.20\columnwidth}
        \centering
         \includegraphics[height=\contenfheight]{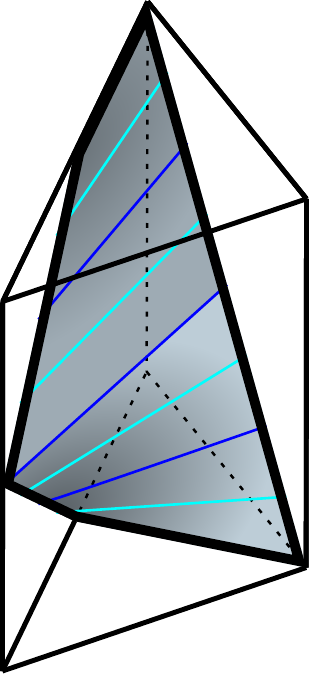}
        \caption{Ruled surface}
	\label{fig:continuity_enforcement_result_ruled_surface}
    \end{subfigure}
    \begin{subfigure}[t]{0.25\columnwidth}
        \centering
         \includegraphics[height=\contenfheight]{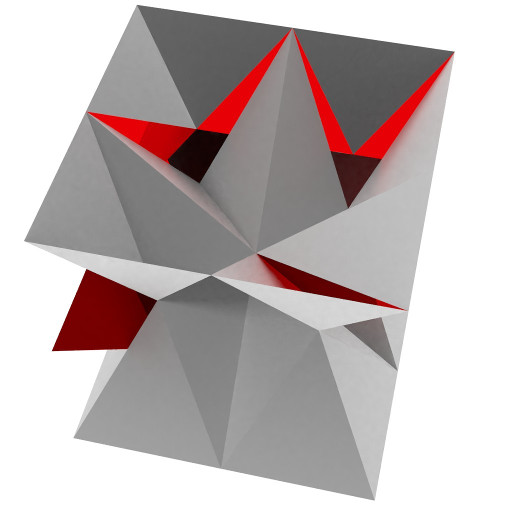}
        \caption{Before enforcement}
	\label{fig:continuity_enforcement_result_before}
    \end{subfigure}
    \begin{subfigure}[t]{0.25\columnwidth}
        \centering
         \includegraphics[height=\contenfheight]{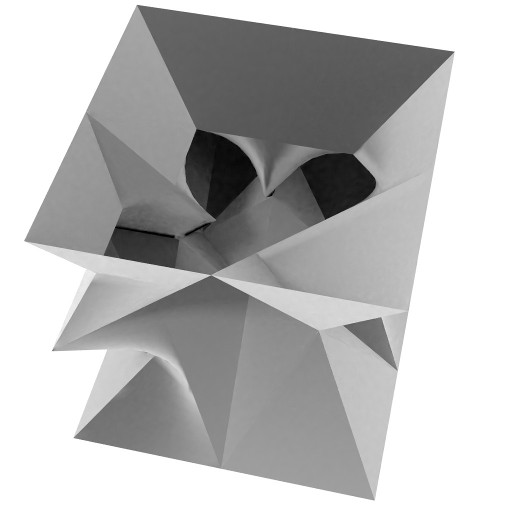}
        \caption{After enforcement}
	\label{fig:continuity_enforcement_result_after}
    \end{subfigure}
    \begin{subfigure}[t]{0.25\columnwidth}
        \centering
         \includegraphics[height=\contenfheight]{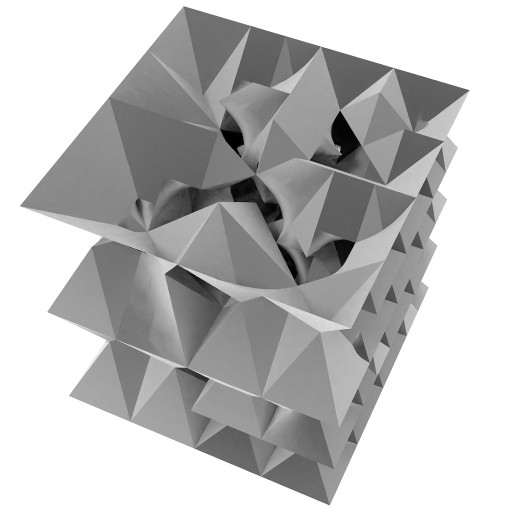}
        \caption{Another example}
	\label{fig:continuity_enforcement_result_after_2}
    \end{subfigure}
    \caption{
\changed{The impact of continuity enforcement on CrossFill structures.
The enforcement causes surface patches to become ruled surfaces.
Blue lines show the isolines which are straight at each Z height.
Red patches show discontinuities in the space filling surface.}
    }
    \label{fig:continuity_enforcement_results}
\end{figure}

\paragraph{Self-supporting}
It should be noted that the curved surfaces introduced by enforcing the continuity of neighboring linked prism cells will not violate the self-supporting property of the space-filling surface.
With the help of the carefully designed continuity enforcement algorithm, we generate surfaces that have overhang $\leq$ \SI{45}{\degree} in most places.
There is only one exceptional case for the side enforcement in an H-prism, where the overhang can be increased to $\tan^{-1}\sqrt{2} \approx$ \SI{55}{\degree}.
However, \changed{geometry overhanging with an angle of} \SI{55}{\degree} is not a problem for most FDM 3D printers, so the self-supporting constraint is not violated.

\paragraph{Density}
Sudden jumps in density are hard to be realized on our infill structure. 
For example, a density distribution which is \SI{10}{\percent} in the bottom half and \SI{80}{\percent} in the top half is not easily realized while satisfying the overhang constraint.
Our space-filling surface requires some distance to change from the low to the very high density along the vertical direction. 
Also, the surface patch with enforced continuity would be considerably different from the original triangular surface patches, which might have a large influence on the physical properties associated with a given density.
The situation is controlled by imposing the constraint that cells linked to each other only allow to differ by a single subdivision level at most.

\changed{The distance required to change from a low density to a high density depends heavily on the size of the cell associated with the lower density.
For a simple square subdivision grid the distance between the side of a cell with subdivision level $n$ and height $h$ to a cell with subdivision level $m$ is minimally $h \cdot \left( \frac{1}{2^1} + \frac{1}{2^2} + ... + \frac{1}{2^{m-n}} \right)$,
which converges to $h$ for $m \to \infty$. 
For our prism based subdivision approach the distance converges to $2h$ vertically and $0.97h$ horizontally.
However, depending on the positioning of the most dense cell w.r.t. the grid of the least dense cell, the required distance can increase to $4h$ vertically and $2.75h$ horizontally.
The horizontal distance is measured along the space filling surface in terms of the average length of segments in the 2D L-system.
This means that two cells which are spatially next to each other can have a large difference in density so long as the space filling surface takes a large detour between the two.}

\section{Adaptive subdivision}\label{secGradedDensity}
This section presents our approach for generating a subdivision structure with subdivision levels which closely matches the requested density distribution.
Our approach consists of two steps: the first decides a lower bound subdivision level at each location and the second fine-tunes the local density distribution by dithering \changed{between the lower bound and a higher bound.}
Before presenting this approach, we introduce the methods for evaluating the target and the current density in a cell.


\subsection{Target cell density}\label{section:input_density_distribution}
\changed{Common ways of specifying a density distribution are as a scalar field defined on a tetrahedral mesh or a voxel model or as a procedural function. 
In order to make our program compatible with several commercial software packages, we construct a voxel model from a sequence of gray-scale image files.}
The relationship between the gray-scale value and the physical density is specified by the user.
The target density $\rho_T$ of a cell $\mathcal{P}$ is computed as the average density of voxels $\{ v_i \}$ covered by $\mathcal{P}$:
\begin{equation}\label{eq:targetDensity}
    \rho_T(\mathcal{P}) = \sum_{v_i} \text{Vol}(v_i \cap \mathcal{P}) \rho(v_i) \, \left/ \,  \sum_{v_i} \text{Vol}(v_i \cap \mathcal{P}) \right.
\end{equation}
where the function $\text{Vol}(v_i \cap \mathcal{P})$ computes the volume of the common region of a voxel $v_i$ and a cell $\mathcal{P}$.
We define the \emph{target mass} $M_T$ as the requested amount of volume to be filled with solid material in a cell:
\begin{equation}
    M_T(\mathcal{P}) = \rho_T(\mathcal{P})\text{Vol}(\mathcal{P}).
\end{equation}
\changed{Because the size of starting cube of CrossFill is a power of $2$ times the extrusion width $w$, the fractal can start with a volume which is considerably larger than the input model and its density distribution.}
For a cell lying completely outside the voxel set, we use the density of its nearest voxel as its density $\rho_T(\cdot)$.


\subsection{Current cell density}\label{section:cell_density}
%
For a cell $\mathcal{P}_n$ located at level $n$ in the subdivision tree, we calculate the current amount of material in the surface patch according to the size of $\mathcal{P}$ and the type of route (L, R or A).
By considering the configurations of embedded triangles in a prism cell (\cref{fig:basic_cells_with_triangles}), we calculate the current density estimate $\rho_C$ of $\mathcal{P}_n$  and the corresponding current mass $M_C$ as follows:
\begin{wrapfigure}[3]{l}{0.2\linewidth}
\begin{center}
\includegraphics[width=\linewidth]{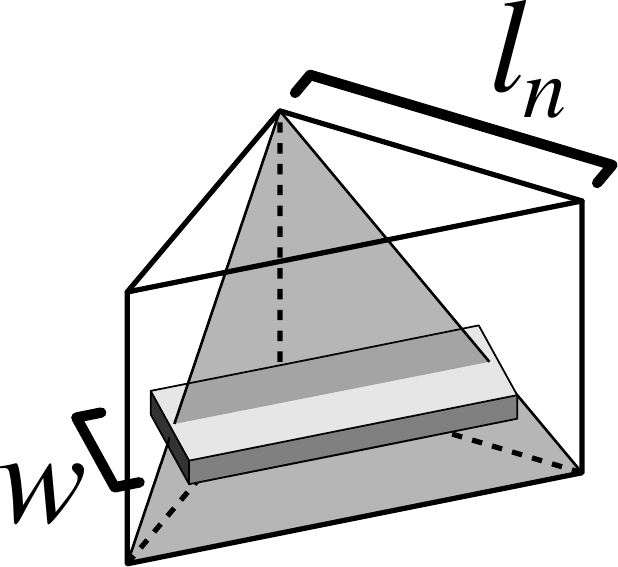}
\end{center}
\end{wrapfigure}
\begin{equation}
\begin{split}
\rho_C(\mathcal{P}_n) &= \frac{w}{l_n} \cdot
\begin{cases}
\sqrt{2} & \quad \text{for A}  \\
1  & \quad \text{for L and R} 
\end{cases}
\\
l_n &= l_\text{init} 2^{-n/2}
\\
M_C(\mathcal{P}) &= \rho_C(\mathcal{P}_n)\text{Vol}(\mathcal{P}_n)
\end{split}
\label{eq:cell_density}
\end{equation}
where $w$ is the constant width of material extrusion, 
$l_n$ is the length of the cathetus at the \changed{top} of the prism
and $l_\text{init}$ is the side length of the starting cube.
Note that $w$ is the horizontal width which differs from the thickness in the direction normal to the surface.
Note also that the density, i.e., the fraction of solid material, is independent of the height of the prism, the embedding of the surface patch and the direction;
it is determined by the average of horizontal segment length and the extrusion width $w$, which are unaffected by these factors.
In this analysis, for the sake of simplicity, we neglect the effect that linked cells influence the density of a given cell due to the continuity enforcement.
We provide a method for compensating for the errors induced by continuity enforcement in \cref{sec:simplified_density_compensation}.

From \cref{eq:cell_density} and the subdivision rules in \cref{fig:subdivision_rules_route} we can derive the increment in mass when performing a subdivision, which is used to supervise the generation of an adaptive subdivision structure.
The new mass the children cells would have after a subdivision $M_N$ is as follows:
\begin{equation}
M_N(\mathcal{P}) = M_C(\mathcal{P}) \cdot
\begin{cases}
1 & \quad \text{for A}  \\
1 + \frac12 \sqrt{2} & \quad \text{for L and R} 
\end{cases}
\end{equation}

The ratio of A-, L- and R-route cells in a CrossFill structure with uniform subdivision level quickly converges to \nicefrac{1}{3} after a depth of only 5 subdivisions.
If the current density of A-route cells is \SI{100}{\percent} then the other \nicefrac{2}{3} of all cells is still at a lower density,
so the maximum attainable density in the whole structure is $1/3 + 1/3 \sqrt{2} \approx$  \SI{80}{\percent}.

From the above we can derive that the average density increment factor is $\sqrt{2}$.
This provides more granularity than other fractal structures, e.g. the Hilbert curve, which has a density increment factor of $2$ for each subdivision.
\changed{One step of the Hilbert curve quadruples the amount of cells, while it only doubles the distance between connected cells, so the total length of the curve is doubled, which means that the curve with a constant line width comes to cover double the area.}

\subsection{Lower bound subdivision levels}\label{section:lower_bound}
\begin{algorithm}[t]
\caption{Lower bound subdivision structure generation}
\label{alg:lower_bound}
\begin{algorithmic}
\Function{LowerBoundSubd}{cell $\mathcal{P}$}
    \If{$\mathcal{P}$ is a \emph{leaf-node}}
        \State Compute $M_C$,  $M_N$ and  $M_T$;
        
		
		\If {$M_N<M_T$} \Comment{$\mathcal{P}$ needs to be subdivided}
		    \State \Call{Subdivide}{$\mathcal{P}$};
		\EndIf
    \EndIf
    \ForAll{$c \in$  $\mathcal{P}$.children} 
        \State \Call{LowerBoundSubd}{$c$};
    \EndFor
\EndFunction

\Function{Subdivide}{cell $\mathcal{P}$}
    \ForAll{$c \in$  $\mathcal{P}$.links} \Comment{For linked neighbors of $\mathcal{P}$}
        \If{$c$.depth $< \mathcal{P}$.depth} \Comment{For level constraint}
            \State \Call{Subdivide}{$c$};
        \EndIf
    \EndFor
    
    \State Subdivide $\mathcal{P}$ according to the rules;
    
    \State Update the corresponding links of neighbors;
\EndFunction
\end{algorithmic}
\end{algorithm}
%

Given the specified density distribution, we adaptively refine the subdivision structure such that the current density of each cell approaches, but does not exceed, the average target density (i.e., \cref{eq:targetDensity}).
This is achieved by a top-down pass on the subdivision tree.
In order to accomplish that we subdivide a cell $\mathcal{P}$ if it satisfies the following condition: $M_N(\mathcal{P}) < M_T(\mathcal{P})$.
To restrict the subdivision level difference between linked cells to be at most one, before subdividing a candidate cell $\mathcal{P}$ that satisfies the condition, we subdivide its linked cells with a shallower subdivision level first.

The pseudo-code of our algorithm is presented in \cref{alg:lower_bound}.
By calling the function \caps{LowerBoundSubd}($\cdot$) on the root of the tree, the subdivision level is decided in each location.
This constitutes the lower bound subdivision levels.
The following subsection presents a dithering approach to further reduce the approximation error.

\begin{figure}
    \centering
    \includegraphics[width=\columnwidth]{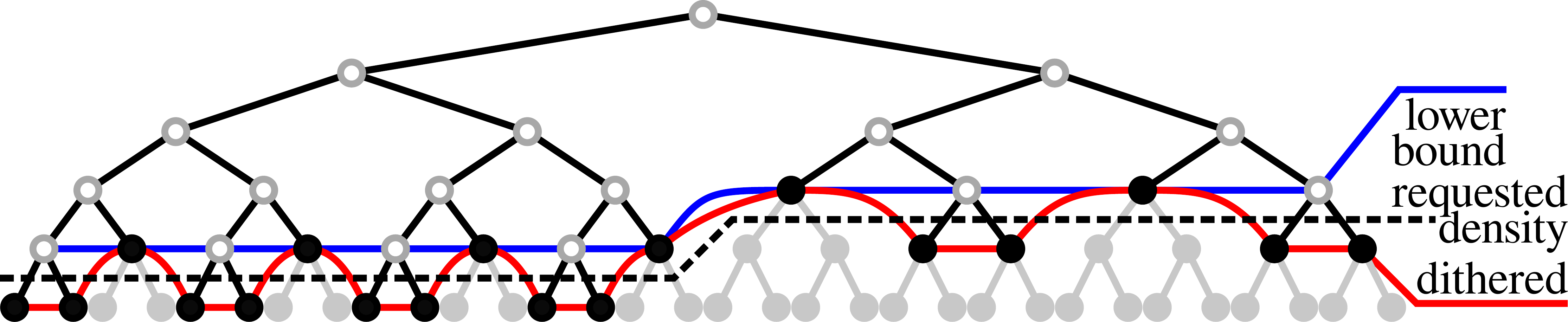}
    \caption{
Schematic overview of dithering from a subdivision tree with lower bound subdivision levels.
    }
    \label{fig:subdiv_tree_dithering}
\end{figure}

\subsection{Dithering}\label{section:dithering}
%
Because the input is a continuous density distribution while the output only admits a limited set of subdivision levels, choosing a subdivision level of a CrossFill cell always induces a discretization error a.k.a. \emph{quantization error}.
The idea is to \emph{diffuse} this quantization error to linked cells in the neighborhood so as to influence the chosen subdivision level there.
This causes the subdivision levels to oscillate between the subdivision levels closest to the target density (see \cref{fig:subdiv_tree_dithering}).
This is akin to the widely employed \emph{dithering} technique in multimedia processing.

\begin{figure}
        \centering
    \begin{subfigure}[t]{0.322\columnwidth}
        \centering
         \includegraphics[width=\textwidth]{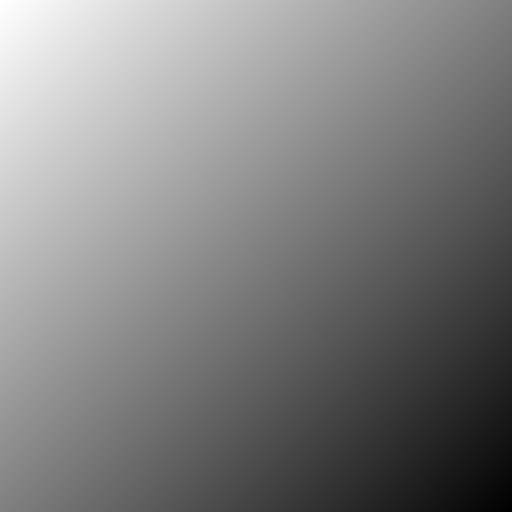}
        \caption{Input density distribution}
        \label{fig:gradient_input}
    \end{subfigure}
    \begin{subfigure}[t]{0.322\columnwidth}
        \centering
         \includegraphics[width=\textwidth]{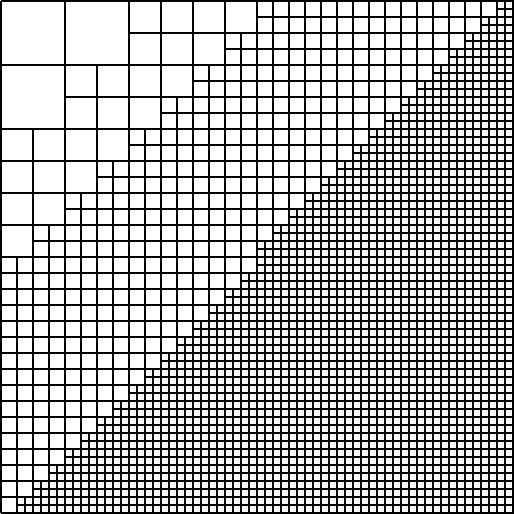}
        \caption{Lower bound subdivision levels}
        \label{fig:gradient_no_dither}
    \end{subfigure}
    \begin{subfigure}[t]{0.322\columnwidth}
        \centering
         \includegraphics[width=\textwidth]{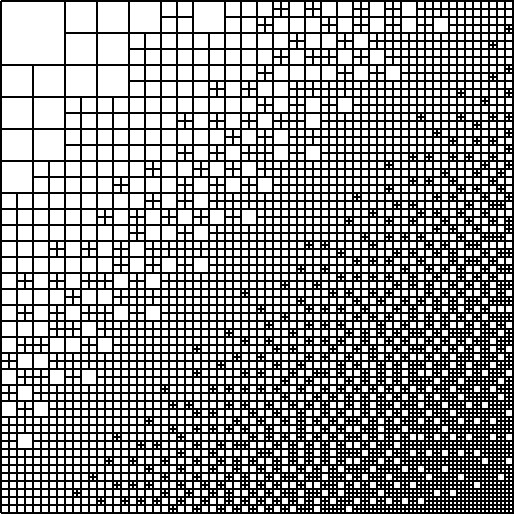}
        \caption{After dithering}
        \label{fig:gradient_dither}
    \end{subfigure}
    \caption{
A square subdivision tiling fitted to an input density distribution image with a diagonal gradient from \SI{0}{\percent} to \SI{100}{\percent} density.
The dithering step eliminates banding artifacts from the lower bound subdivision structure and creates a smooth density distribution. 
    }
    \label{fig:gradient}
\end{figure}

We define the quantization error as the difference between target mass and current mass.
Diffusing this error to dither the subdivision level leads to a CrossFill structure with densities better matching the target distribution regionally.
See \cref{fig:gradient} for an example on a simple square subdivision tiling.
The lower bound subdivision structure exhibits strong banding artifacts; these artifacts are eliminated by dithering.

\paragraph{Dithering order}
\setlength\intextsep{0pt}
\begin{wrapfigure}[6]{r}{0.12\linewidth}
\begin{center}
\includegraphics[width=\linewidth]{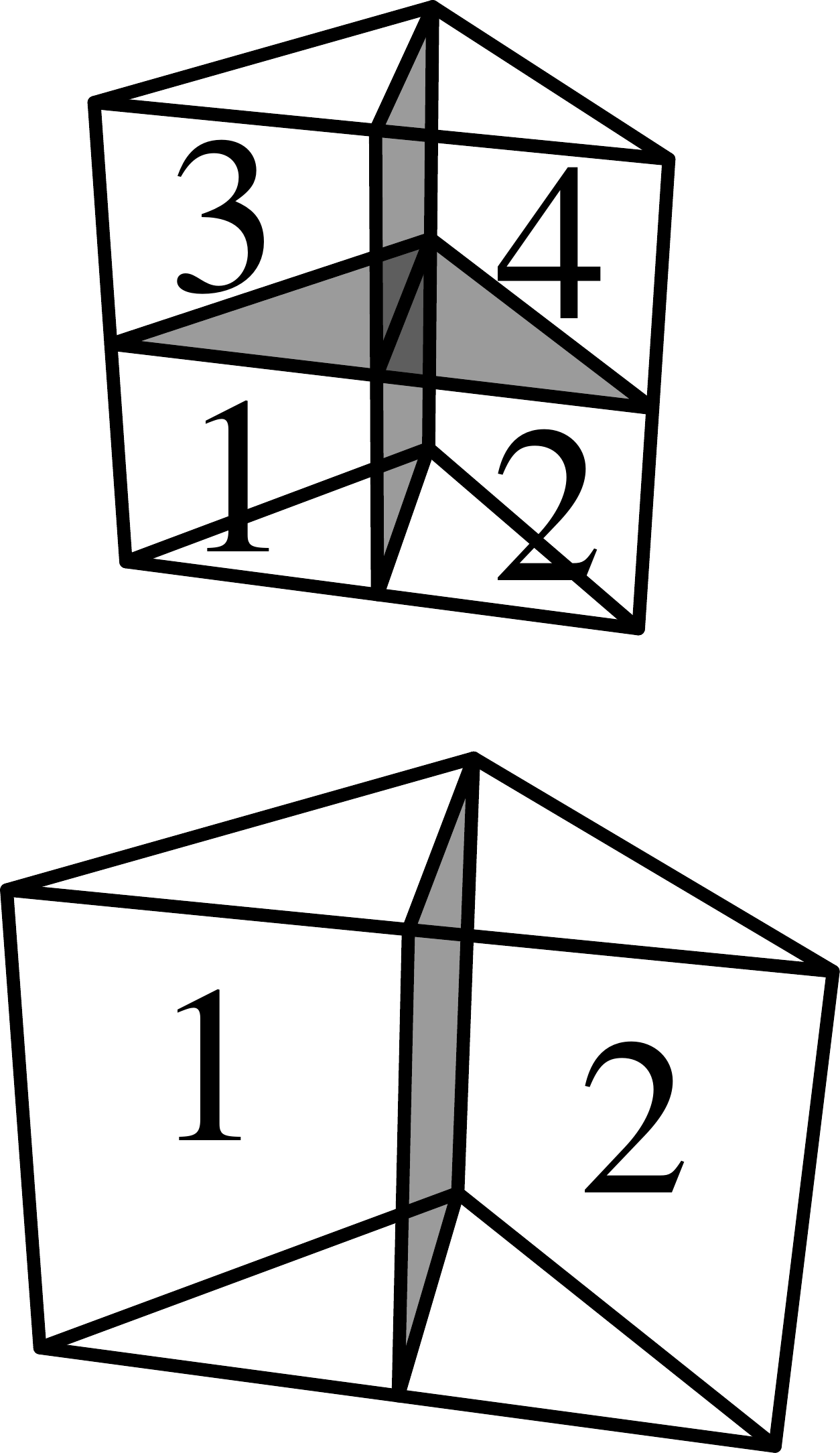}
\end{center}
\end{wrapfigure}
When processing cells for dithering we consider the leaf nodes
\changed{in a sequence analogous to the \emph{Morton order}~{\cite{morton1966computer}}:
We traverse the tree in depth-first order and at each non-leaf cell we recurse the children in the following order: first the bottom left, then the bottom right and, if it is a Q-prism, then the top left and the top right.}
Here `right' refers to linked cells along the direction of the horizontal traversal of the surface patches, which is clockwise around the space-filling polygon of  each layer.
Quantization error of the current cell $\mathcal{P}$ is then redistributed to those cells in the neighborhood of $\mathcal{P}$ which have not yet been processed in dithering.

\paragraph{Neighborhood}
Dithering along a manifold with a lower dimension makes the implementation simpler while retaining the advantages of dithering.
\citeauthor{VelhoDigitalHalftoning}~\cite{VelhoDigitalHalftoning} have shown that propagating quantization error along the directions of a space-filling curve (as 1D manifold) for 2D images can yield a halftoning technique with appealing properties.
Similarly, rather than considering all geometrically neighboring cells of a prism cell $\mathcal{P}$, we only consider the cells which are neighboring $\mathcal{P}$ in the connectivity graph.
As such we only disperse quantization error along the 2D manifold of the space-filling surface within 3D space. 
In order to visualize the dithering process effectively, we `unfold' the space-filling surface and consider the resulting 2D topology (see \cref{fig:subdiv_tree_unfolded} and the supplementary video).


\begin{figure}
        \centering
    \begin{subfigure}[t]{0.45\columnwidth}
        \centering
         \includegraphics[height=.6\textwidth]{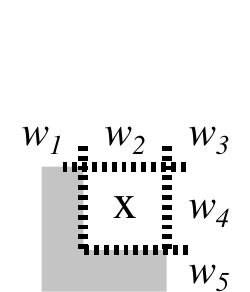}
	\caption{Positioned weights}
        \label{fig:dithering_weights_general}
    \end{subfigure}
    \begin{subfigure}[t]{0.45\columnwidth}
        \centering
         \includegraphics[height=.6\textwidth]{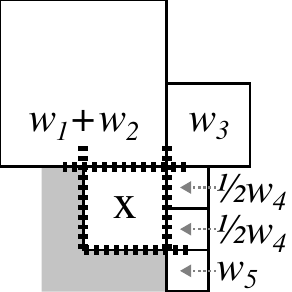}
	\caption{Application example}
        \label{fig:dithering_weights_complex}
    \end{subfigure}
    \caption{
Quantization error propagation weights in various scenarios.
Connected cells may cover an area at multiple positions relative to the checking cell marked by `x'.
The horizontal dimension depicted here is along the traversal of the space-filling surface (i.e., two-manifold). 
The gray area indicates locations which have already been processed.
    }
        \label{fig:dithering_weights}
\end{figure}
%
\paragraph{Weights}
The amount of the quantization error distributed to each (yet unprocessed) cell in the neighborhood depends on its relative position with respect to the current cell $\mathcal{P}$ being checked. 
Connected cells to the left and bottom have always already been processed because of the Morton order. 
For a subdivision structure with uniform subdivision, the configuration is therefore as shown in \cref{fig:dithering_weights_general}. 
The diagonal cells are obtained by accessing the links of directly linked neighbors.
There are various error diffusion schemes with different weights and different configurations (ref.~\cite{floyd1976adaptive,Jarvis1976,Stucki1981}). 
A comprehensive study about their performance is beyond the scope of this paper.
We employ a simple error diffusion scheme with the following weights: $w_1 = w_3 = 1.0$, $w_2 = w_4 = 2.0$ and $w_5 = 0.0$. 
Because not all positions relative to the checking cell are always occupied by (unprocessed) cells,
the weights are normalized to compute the quantization error to be diffused to unprocessed linked neighbors:
\changed{$
\hat{w}_i = w_i  / \sum_{c \in \mathcal{P}.\text{neighborhood}} c.w
$}

Because linked cells can have a size different from $\mathcal{P}$, the weights need to be adjusted to account for the change of configuration. 
A linked cell can occupy more space than $\mathcal{P}$ or only a portion of the space occupied by $\mathcal{P}$. 
In the former case multiple weights are added together, while in the latter case the weight is split equally.
The diagonal positions always retain the same weight.
An illustration of such a configuration can be found in \cref{fig:dithering_weights_complex}. 

\paragraph{Algorithm}
\changed{The dithering algorithm decides on the final subdivision level by choosing between the lower bound subdivision level and a higher bound of one subdivision level deeper by comparing the quantization error of those two subdivision levels.}
The quantization error $\mathcal{E}$ of a prism cell $\mathcal{P}$ before subdivision ($\mathcal{E}_C$) and after subdivision ($\mathcal{E}_N$) are calculated by
\begin{equation}
\begin{split}
\mathcal{E}_C(\mathcal{P}) &= M_C(\mathcal{P}) + M_\mathcal{E}(\mathcal{P}) - M_T(\mathcal{P}) 
\\
\mathcal{E}_N(\mathcal{P}) &=  M_N(\mathcal{P}) + M_\mathcal{E}(\mathcal{P})  - M_T(\mathcal{P})
\end{split}
\end{equation}
where 
$M_\mathcal{E}(\mathcal{P})$ is the quantization error diffused to $\mathcal{P}$ from already processed cells.
$\mathcal{P}$ is subdivided if the absolute value of $\mathcal{E}_N(\mathcal{P})$ is smaller than the absolute value of $\mathcal{E}_C(\mathcal{P})$.
An exception to this rule is if $\mathcal{P}$ has linked neighbors with a lower subdivision level in order to comply with the constraint that linked cell can only differ by a single subdivision level.
After the subdivision decision the corresponding quantization error is diffused to the unprocessed linked neighbors according to the weighing scheme described above.
Pseudo-code of this dithering approach is given in \cref{alg:dithering}.
The diffused error $\mathcal{P}.M_\mathcal{E}$ is initialized as \emph{zero} on each cell $\mathcal{P}$ before calling \caps{Dither}$(\cdot)$ on the root of the subdivision tree.
\Cref{fig:gradient_dither} shows the result on a 2D example. 


\begin{algorithm}[t]
\caption{Dithering}
\label{alg:dithering}
\begin{algorithmic}
\Function{Dither}{cell $\mathcal{P}$} 
    \If{$\mathcal{P}$ is NOT a \emph{leaf-node}}
        \ForAll{$c \in$  $\mathcal{P}$.children} \Comment{using the Morton order}
            \State \Call{Dither}{$c$};
        \EndFor
    \Else
        
        \State Compute $M_C$,  $M_N$ and  $M_T$;
        
        
        \State Define $M = M_C$;
        
		\If {$\nicefrac12 (M_C + M_N) +  \mathcal{P}.M_\mathcal{E} < M_T$} 
		    
   			\State Subdivide $\mathcal{P}$ according to the rules;
   			
    
    			\State Update the corresponding links to neighbors;
		    
		    \State Update mass as $M = M_N$;
		\EndIf
        \State $\mathcal{E} = M + \mathcal{P}.M_\mathcal{E} - M_T$;
		    
		\State Re-distribute $\mathcal{E}$ to $\mathcal{P}$'s unprocessed neighbors; 
		    
		\Comment{The $M_\mathcal{E}$ of cells in the neighborhood is updated.}

    \EndIf
\EndFunction
\end{algorithmic}
\end{algorithm}

\section{Toolpath generation}\label{section:toolpath_extraction}
%
We now have a method for creating an infill structure with spatially graded density according to a user-specified density distribution. 
The infill structure is defined in a cubic region and can be fabricated by continuous material extrusion. 
%
In this section, we will first explain how to effectively slice the structure into a continuous 2D polygonal curve for each layer. 
After that, we will fit the 2D polygonal curve of a layer into the region of an input 3D model.

\subsection{Slicing}\label{section:slicing}
The first step toward generating the toolpath of a layer for 3D printing is to generate the space-filling curve which lies on the intersection between the surface of CrossFill and the horizontal plane at the height $z$ of the printing layer. 
As mentioned above, the space-filling surface only exists conceptually in our implementation. 
We directly generate the space-filling curves from the type of prism cells and their linkage in the connectivity graph. 
Given a height $z$, we first find the sequence of cells covering this height in the subdivision structure.
The cell which is closest to the last point on the toolpath of the previous layer, is chosen as the first cell for exploring the horizontally linked cells. 
The whole sequence of cells can be traced out by following the links in the connectivity graph which are pointing to the right, i.e., by following the cells along the horizontal traversal direction (see \cref{fig:basic_cells_with_triangles}).
When appending a cell to the sequence we take take the right linked cell which intersects with the $z$ height.
After employing the continuity enforcement rules from \cref{section:continuity_enforcement} we are left with the edges of the surface patches, which are sliced at $z$ to serve as the vertices of the space-filling curve of that layer.

\begin{figure}
    \begin{subfigure}[t]{0.32\columnwidth}
        \centering
         \includegraphics[width=.8\textwidth]{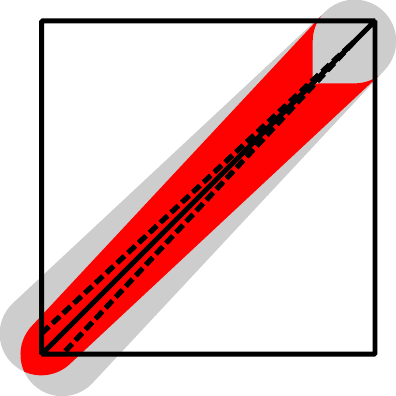}
        \caption{Overlap problem}
	\label{fig:overlap_problem}
    \end{subfigure}
    \begin{subfigure}[t]{0.32\columnwidth}
        \centering
         \includegraphics[width=.8\textwidth]{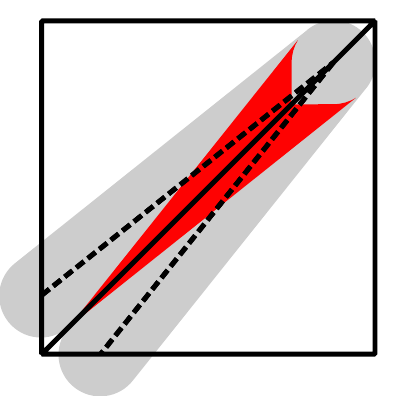}
        \caption{Clamping}
	\label{fig:overlap_clamping}
    \end{subfigure}
    \begin{subfigure}[t]{0.32\columnwidth}
        \centering
         \includegraphics[width=.8\textwidth]{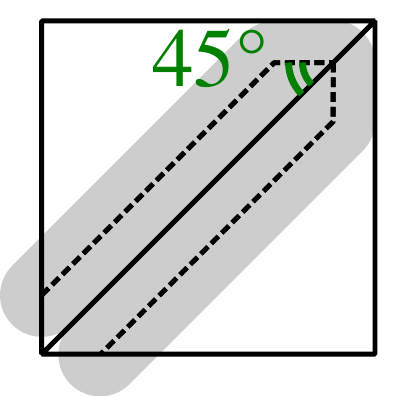}
        \caption{Detouring}
	\label{fig:overlap_add_turn}
    \end{subfigure}
    \caption{
Dealing with overlapping paths in neighboring prisms (the red color indicates overlapping regions) -- 
the figures illustrate a top-view of two prisms crossed by extrusion paths (dashed black lines) which initially lie close to each other (i.e., overlap occurs in \subref{fig:overlap_problem}).
By clamping the endpoints to a position that is at least \nicefrac{1}{2} the extrusion width away from the other prism \subref{fig:overlap_clamping} and introducing an additional \SI{45}{\degree} turn on the line segments \subref{fig:overlap_add_turn}, 
the overlap can be avoided.
    }
    \label{fig:overlap}
\end{figure}

\subsection{Overlap prevention}\label{sec:overlap_prevention}
These space-filling curves can have overlap near the boundary of a prism when a surface patch segment is too close to the neighboring prism (see \cref{fig:overlap_problem} for an example). 
This occurs when the distance \changed{between the two segments is less than the horizontal width $w$ of extrusion toolpaths}. 
In order to prevent path overlap, we \emph{clamp} the endpoints of sliced line segments to a position that is $w/2$ away from the neighboring prism (see \cref{fig:overlap_clamping}). 

However, only applying the clamping step cannot completely avoid overlap in situations where the toolpath makes a sharp turn. 
When the \changed{\emph{turning angle} between the segment and the side of the cell} is sharper than \SI{45}{\degree}, we introduce an additional vertex at a position with distance $w \cdot \nicefrac{1}{2} \sqrt{2}$ away from the \changed{vertex on the side of the cell} -- called \emph{detouring}.
See \cref{fig:overlap_add_turn}.
It should be noted that in a uniformly subdivided CrossFill structure, the turning angle on a space-filling curve generated by slicing is either \SI{45}{\degree} or \SI{90}{\degree} (see \cref{fig:cross_oscillating_extrusion_crosssection} for an example), which means that detouring is not needed in such a context.

\changed{Detouring does not violate the overhang constraint.
The introduced vertex is supported either by a detoured vertex below or a segment with a turning angle just above {\SI{45}{\degree}}.
Moreover, detouring does not reduce any area covered by the extrusion path, so a detoured layer still supports the layer above.
The increased density caused by detouring can be compensated for using a method introduced in} \cref{sec:simplified_density_compensation}.

\subsection{Conversion into infill structure}\label{section:infill_extraction}
\begin{figure}
        \centering
    \begin{subfigure}[t]{0.24\columnwidth}
        \centering
         \includegraphics[width=\textwidth]{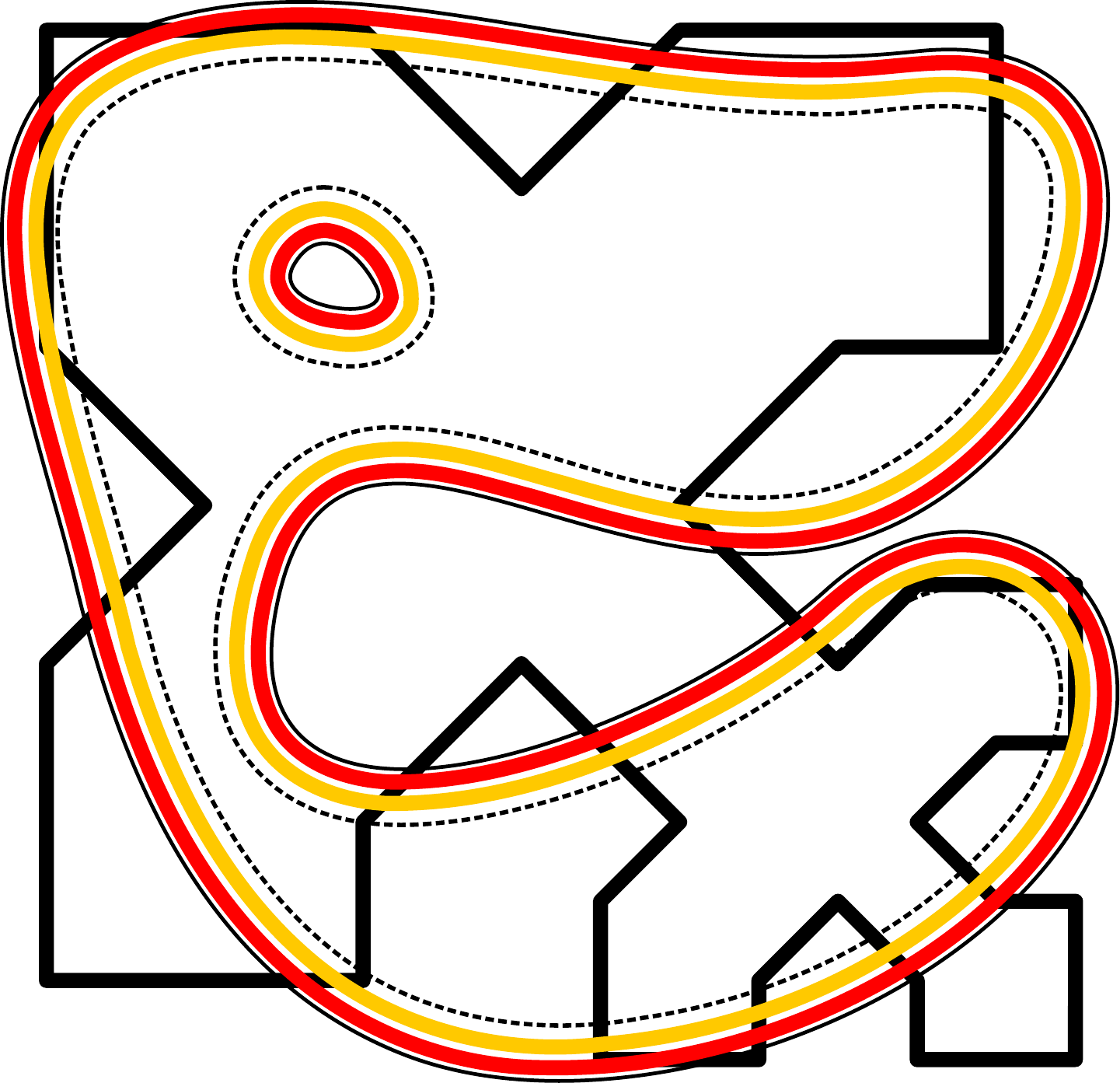}
        \caption{Raw}
	\label{fig:fitting_unfit}
    \end{subfigure}
    \begin{subfigure}[t]{0.24\columnwidth}
        \centering
         \includegraphics[width=\textwidth]{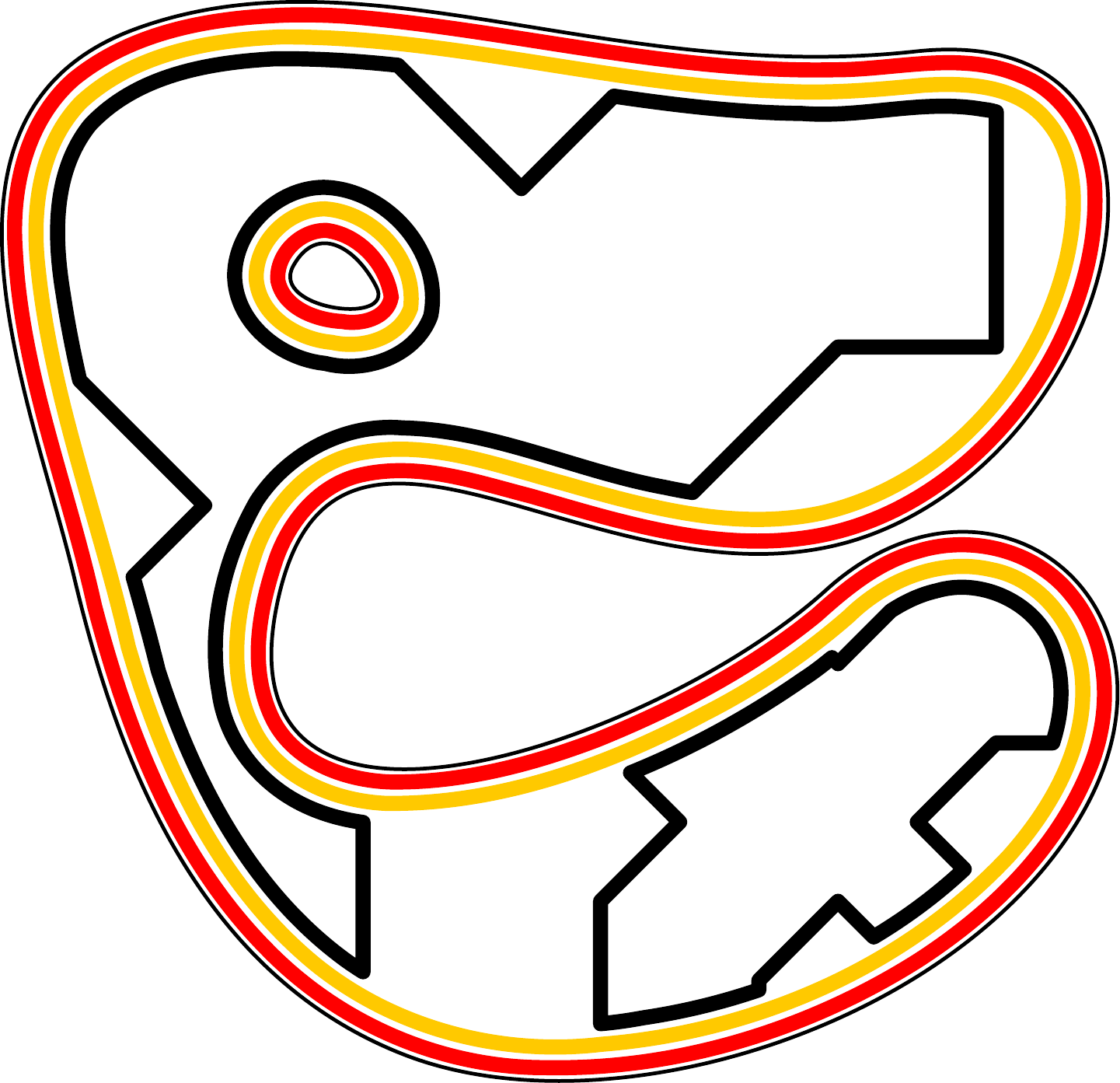}
        \caption{Intersected}
	\label{fig:fitting_intersect}
    \end{subfigure}
    \begin{subfigure}[t]{0.24\columnwidth}
        \centering
         \includegraphics[width=\textwidth]{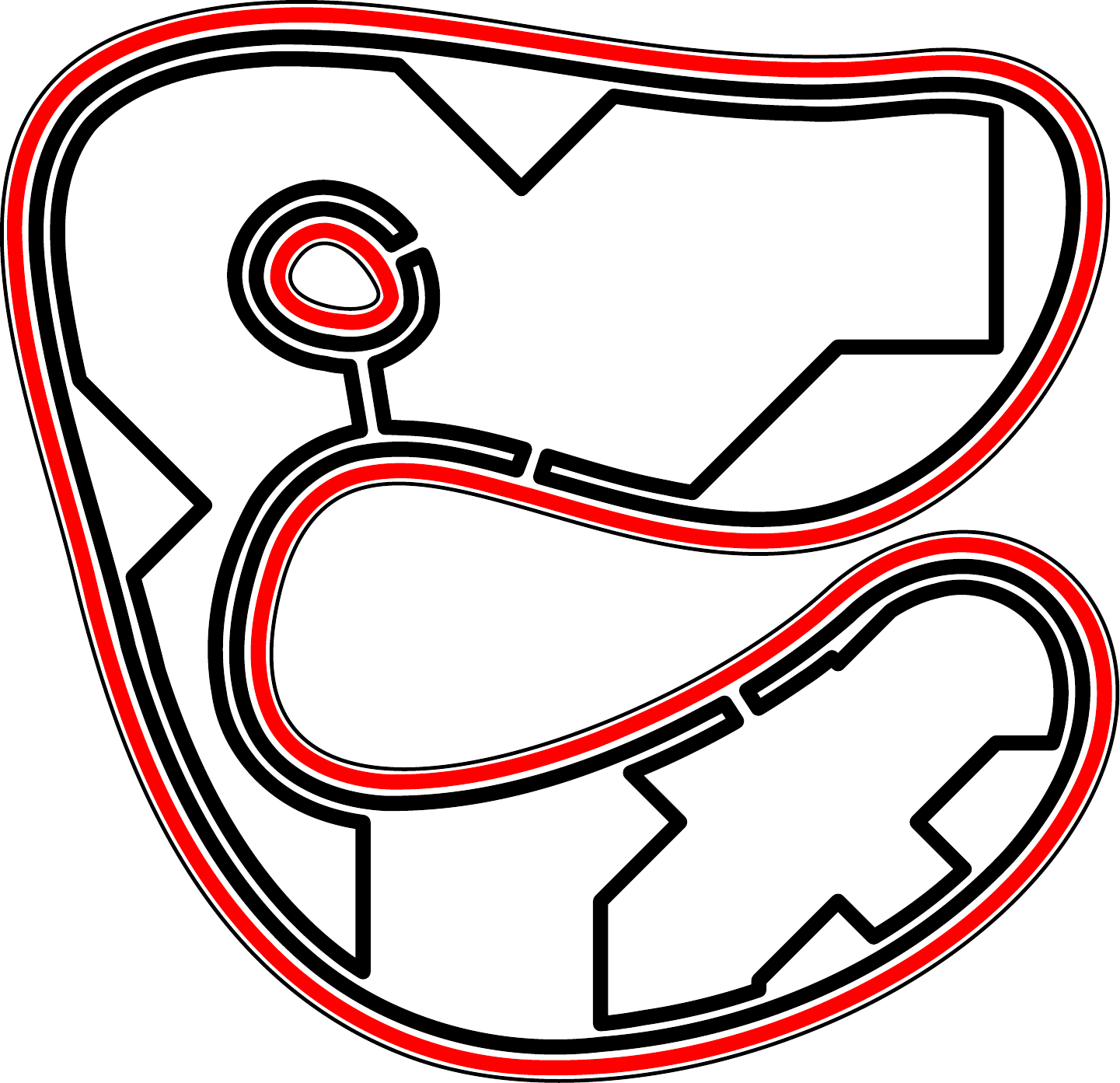}
        \caption{Single line}
	\label{fig:fitting_intersect_connect}
    \end{subfigure}
    \begin{subfigure}[t]{0.24\columnwidth}
        \centering
         \includegraphics[width=\textwidth]{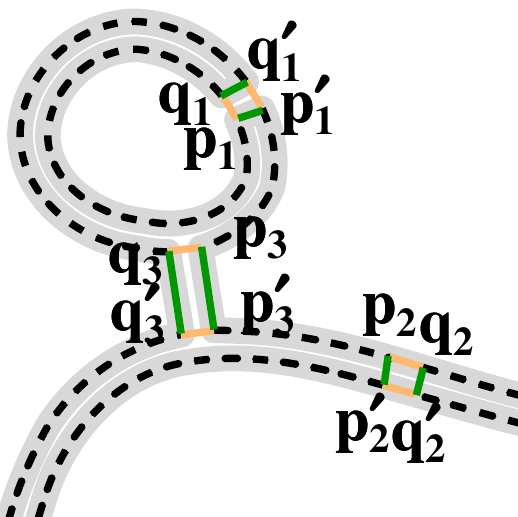}
        \caption{Closeup}
	\label{fig:fitting_connecting}
    \end{subfigure}
    \caption{
Inserting the cross-filling curve into to an infill area, where the cross-filling curve is displayed in black color, the outer wall is in red color and the second wall is shown in orange color. 
\subref{fig:fitting_unfit} The original cross-filling curve. 
\subref{fig:fitting_intersect} The intersected space-filling structure. 
\subref{fig:fitting_intersect_connect} The finally connected toolpaths of multiple walls and the infill structure.
\subref{fig:fitting_connecting} A zoom-view of \subref{fig:fitting_intersect_connect}.
    }
    \label{fig:fitting}
\end{figure}

%
Now that we have a non-overlapping space-filling curve as toolpath for fabricating the CrossFill structure of each layer,
We limit it to the interior area of an input 3D model while retaining the continuity of the toolpaths. 
Specifically, we first intersect the space-filling curve \changed{with} the infill area shrunk by $w/2$ (see \cref{fig:fitting_intersect_connect}). 
This operation makes the trimmed space-filling curve connected to itself via the perimeters of that layer.

Performing an intersection between the space-filling curve and the infill area could result in a single polygon.
However, there are cases in which additional polygons are generated:
\begin{enumerate}
\item the infill area splits the space-filling curve into multiple parts (in the bottom right of \cref{fig:fitting_intersect});
\item the infill area contains a polygon which doesn't touch the space-filling curve (in the top left of \cref{fig:fitting_intersect}).
\end{enumerate}
We tackle \changed{both} these problems by connecting all polygons to the innermost perimeter \changed{of the shell of the print} and to each other afterwards. 
The problem to be solved here for generating continuous toolpaths is different from \cite{Zhao2016}, in which spiraling toolpaths are generated to completely fill a given region. 
By contrast, we are tackling the problem of connecting multiple polygons into a single polygonal curve.

Two polygons are connected into a single polygon by building a bridge between them as follows.
First, two points $\mathbf{p}$ and $\mathbf{q}$ with $\|\mathbf{p} - \mathbf{q}\|=w$ on a polygon are considered.
We consider the point $\mathbf{p}'$ on the other polygons closest to $\mathbf{p}$ and find a point $\mathbf{q}'$ on the other polygon such that $\mathbf{qq}'$ is parallel to $\mathbf{pp}'$.
We search for such pairs of points until we find a bridge for which the length is at most $\nicefrac{3}{2}w$.
New line segments $\mathbf{p}\mathbf{p}'$ and $\mathbf{q}\mathbf{q}'$ are then added
and the line segments $\mathbf{p}\mathbf{q}$ and $\mathbf{p}'\mathbf{q}'$ are removed.
Examples for building bridges can be found in \cref{fig:fitting_connecting}.
Repeatedly building bridges between polygons can connect all into a single polygon.
In order to minimize the number of sub-optimal bridges, we start by connecting the smallest polygons to suitable neighbors and work our way outward.

Our method has several advantages.
When there are many possible candidate locations for building a bridge, we can select the `optimal' position according to various criteria.
For example, we can promote bridges at regions with low curvature in order to minimize the influence on the extruded amount of material.
Another option is to build bridges that are closer to interior regions so as to minimize the visual surface impact.

Connecting polygons which have distance more than $2w$ introduces new line \changed{segments} hanging in the air.
Strictly this conflicts with the overhang constraint; however, this often is not a problem for FDM printing.
Such distances are rarely long, and these lines of bridges do not have to support any material above.
In practice, the extruded beads of FDM can stay in mid-air because of the high viscosity of the melted plastic.

\section{Results}\label{secResult}

\subsection{Experiments}\label{subsecExperiments}
%

Experiments were performed on an Intel Core i7-7500U CPU @ \SI{2.70}{\giga\hertz} using a single core and \SI{16.3}{\giga\byte} memory.
We have printed test structures on several Ultimaker 3 machines, loaded with white Ultimaker TPU 95A in AA \SI{0.4}{\milli\meter} Print Cores.
The basic print path settings were taken from the default Cura 4.0 profile of \SI{0.1}{\milli\meter} layer thickness for this setup.
qMost notably the setting for Infill Line Width was \changed{$w = $} \SI{0.38}{\milli\meter} and the Speed was \SI{25}{\milli\meter\per\second}.
In order to enable the connect polygons functionality in some of our tests, we set the Extra Skin Wall Count to zero and instead set the Extra Infill Wall Count to one,
so that we can enable the setting Connect Infill Polygons.
The Infill Line Distance setting was set to \SI{0.76}{\milli\meter} meaning that the smallest possible prism has sides of length $l_\text{max} = 2 w \sqrt{2}$, which corresponds to a density of \SI{40}{\percent} in order to save computation time.

For various tests we have used a test cube with side lengths of \SI{48.64}{\milli\meter}, which is \changed{$2^7 w$},
so that the starting cube of the subdivision tree matches the 3D model.
\changed{In order to isolate the infill structure, the settings Top/Bottom Thickness and the Wall Thickness have been set to {\SI{0}{\milli\meter}}.}

For some of the models, the requested infill densities are too low to support the dense top skin.
In order to overcome this problem, we enforced a minimal subdivision level for prism cells which overlap with top skin after the dithering stage,
by iteratively calling \caps{Subdivide} from \cref{alg:lower_bound}.
By modifying the subdivision structure in this way, we guarantee a required percentage of infill for supporting top skin regions.


\subsubsection{Simplified density measure compensation}\label{sec:simplified_density_compensation}
The method we propose uses a simplified density estimate based on cell size and route type.
Because this doesn't take into account effects from continuity enforcement (\cref{section:continuity_enforcement}) or from clamping and detouring (\cref{sec:overlap_prevention}) the output density is different from the simplified density estimates.
Once we evaluate the discrepancy, we can compensate for it.

\begin{figure}
        \centering
        \includegraphics[width=.8\columnwidth]{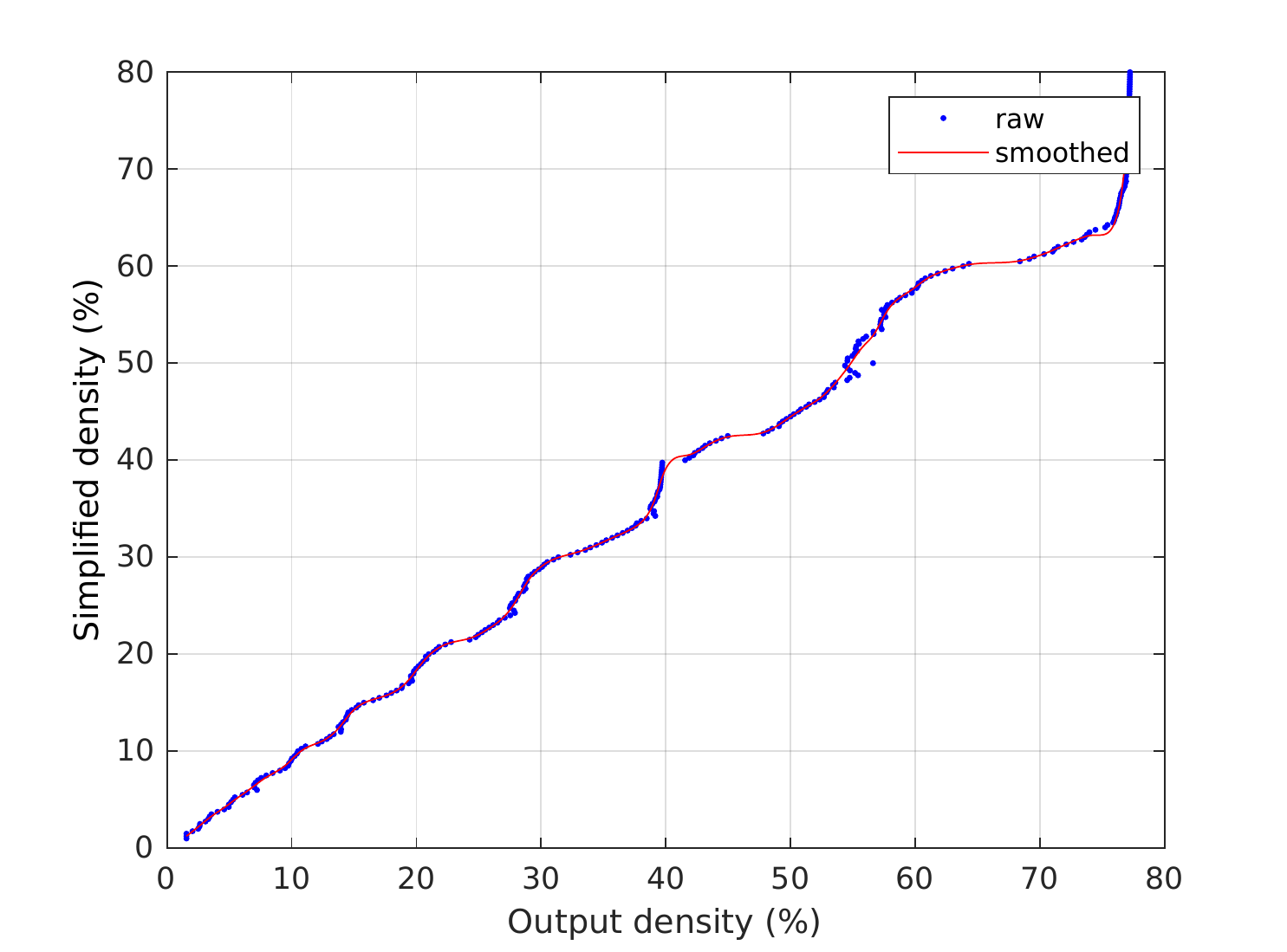}
\caption{
Compensating for inaccuracies of the simplified density measure.
For several simplified density values the actual total amount of volume is recorded in blue,
so that we can map required output densities to the corresponding simplified input densities
using a smoothing spline fitted to that data in red.
}
\label{fig:inoutdensity}
\end{figure}

We have generated test cubes with a homogeneous simplified density \changed{({\cref{eq:cell_density}})} within the range of \SI{1}{\percent} to \SI{80}{\percent}
and analyzed the total extruded volume compared to the total volume of the cube.
These results are shown in blue in \cref{fig:inoutdensity}.
We fit a \caps{Matlab} smoothing spline to the data with a smoothing parameter of \SI{0.75}{}.
The resulting curve is then used to compensate for the disparity between the simplified density estimates and the actual densities
by mapping the density requirements to the corresponding simplified density estimates prior to applying our algorithms.

\subsubsection{Accuracy}
\begin{figure}
        \centering
    \begin{subfigure}[t]{0.4\columnwidth}
        \centering
        \includegraphics[height=.7\columnwidth]{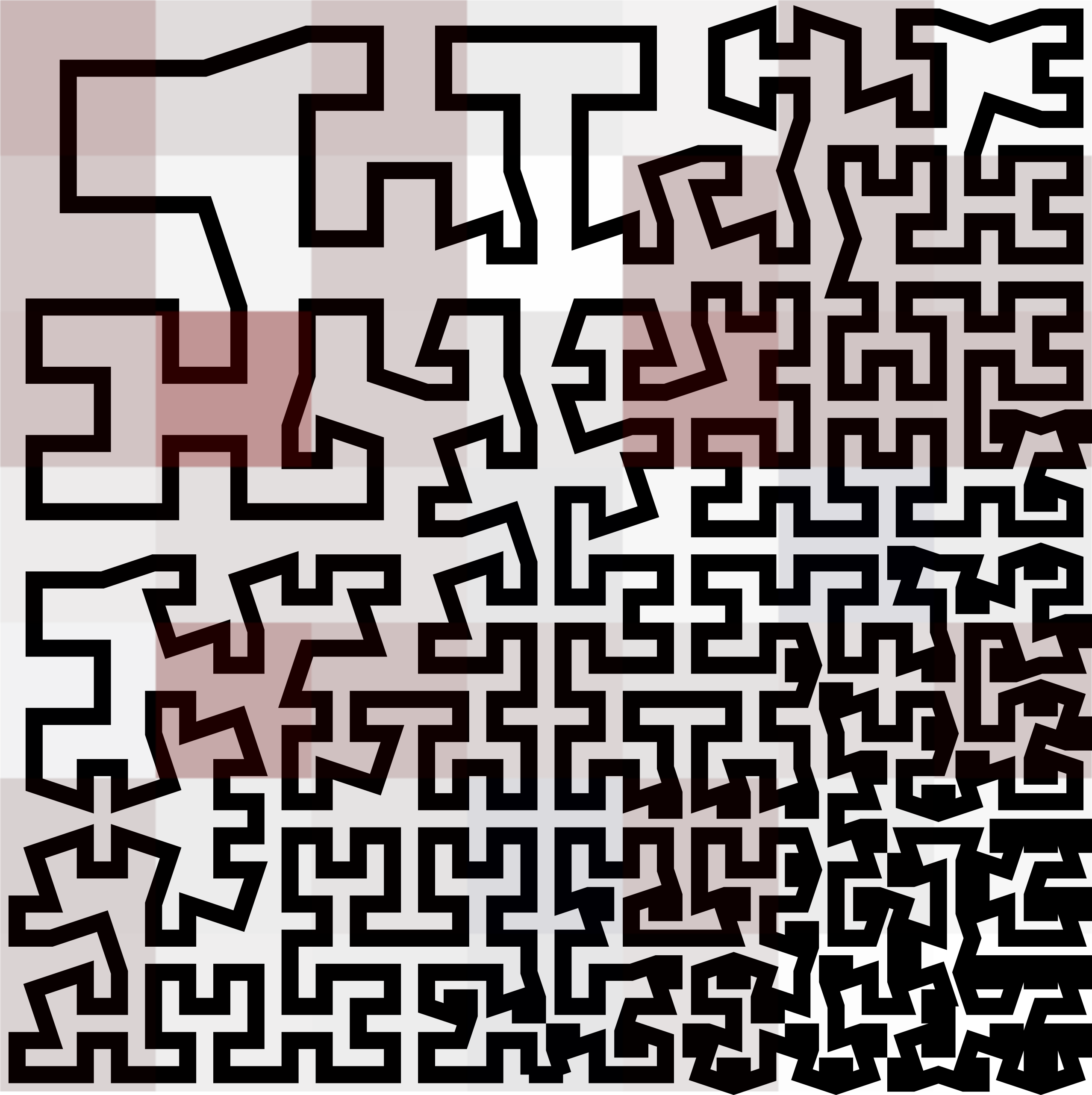}
        \caption{
Kernel size $w\cdot64/7$
        }
    \end{subfigure}
    \begin{subfigure}[t]{0.4\columnwidth}
        \centering
        \includegraphics[height=.7\columnwidth]{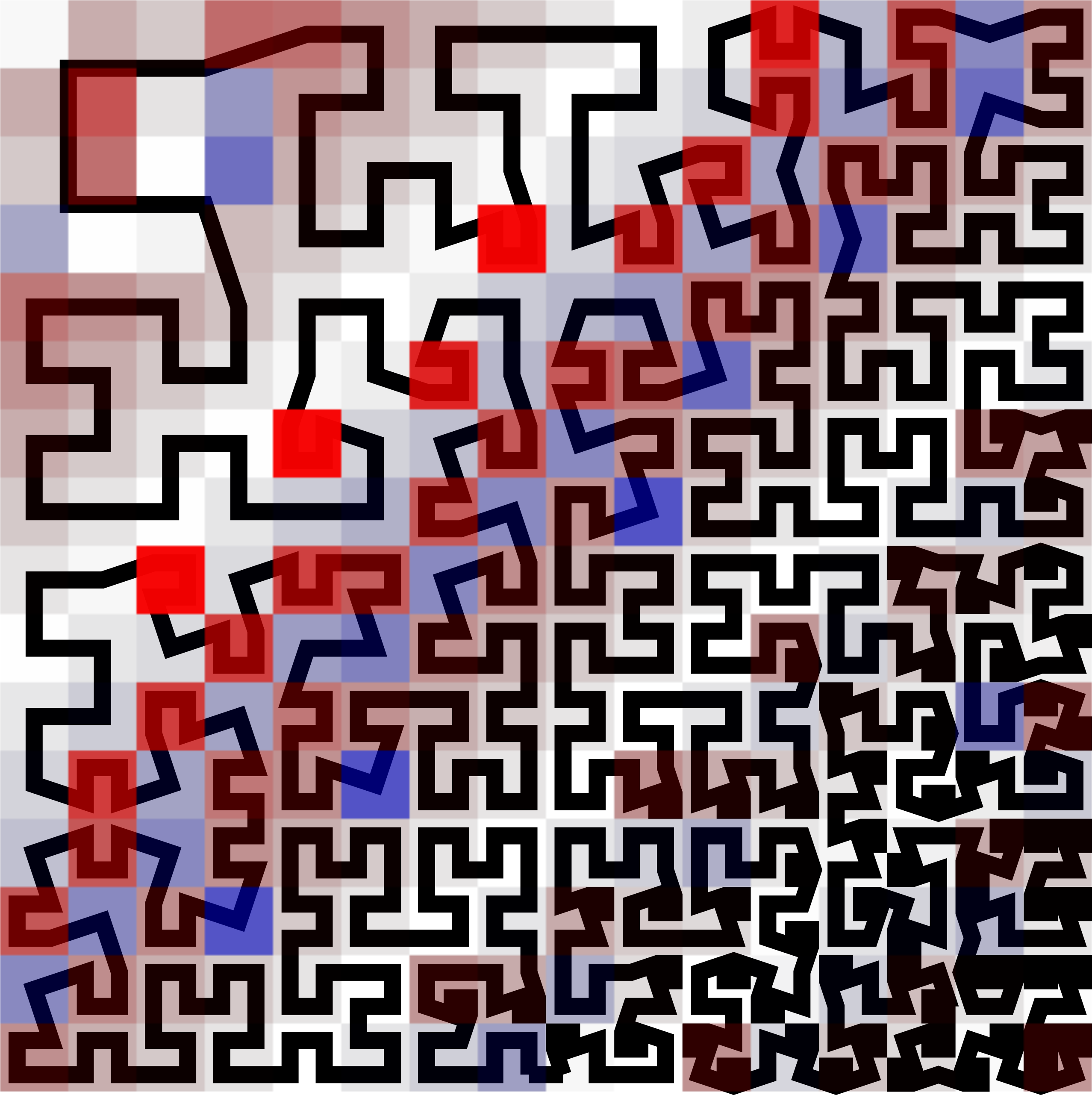}
        \caption{
Kernel size $w\cdot64/16$
        }
    \end{subfigure}
    \begin{subfigure}[t]{0.16\columnwidth}
        \centering
        \includegraphics[height=1.75\columnwidth]{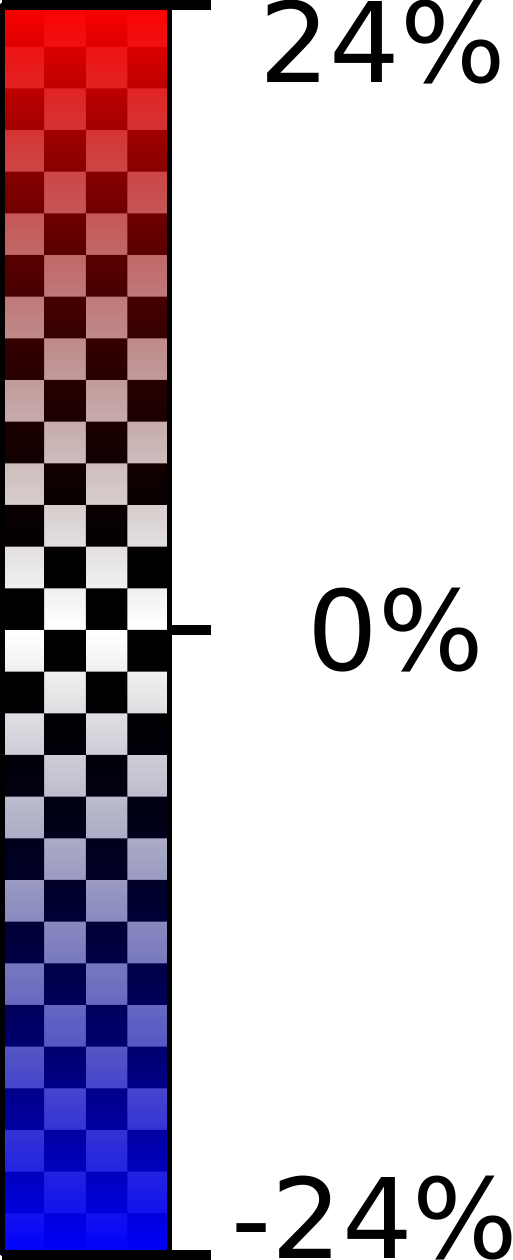}
        \caption{Legend}
    \end{subfigure}
\caption{
Example of local errors on a 2D space-filling curve generated from a diagonal gradient from \SI{10}{\percent} to \SI{80}{\percent} density.
The errors are translated to opacity and overlaid with the space-filling curve.
$w$ is the line width.
When analyzing at lower resolution the errors are higher.
The overall error is positive because the diagonal line segments introduced at locations where consecutive cells are a different subdivision level have a higher density than the simplified density estimate used.
}
\label{fig:local_error_2d}
\end{figure}

The accuracy of a functionally graded material is inherently related \changed{to a viewing resolution.
When viewing any print at a resolution close to the printer resolution, the density is either {\SI{100}{\percent}} or {\SI{0}{\percent}} regardless of the user specified density at each location, which means the accuracy at that resolution is low.}
In order to evaluate the accuracy of our functionally graded material, we evaluate the average local error at a range of resolutions.
For each resolution we divide the specification and the generated infill structure into smaller cubes and
compute the local error as the absolute difference between the average specification density throughout that cube and the average realized density throughout that cube.
The accuracy is then given by the average local error across all subcubes for that resolution (see \cref{fig:local_error_2d}).
Note that at kernel sizes of powers of two the subcubes align with the prism-shaped cells, which lowers the error measure.

\begin{figure}
        \centering
        \includegraphics[width=.8\columnwidth]{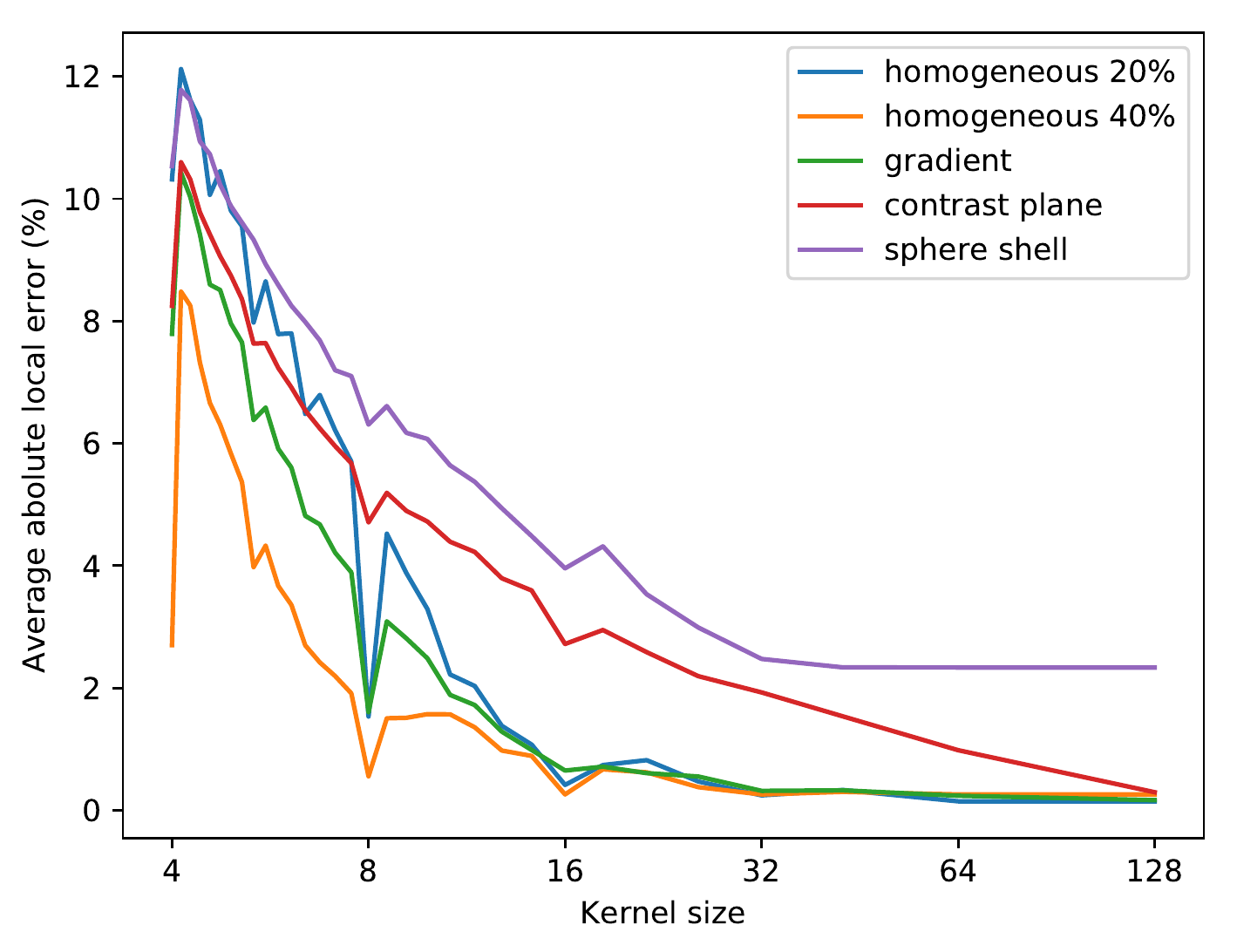}
\caption{
Local average error for a range of kernel sizes on several test specifications.
Kernel size is in multiples of the line width, while the average local error is measured in terms of infill density percentage.
When viewing the structures at lower resolutions, the accuracy is higher.
High-frequency specifications such as the sphere shell perform worse.
}
\label{fig:accuracy}
\end{figure}

We define several density specifications for our test cube on which we evaluate the accuracy.
\begin{enumerate*}[label={\alph*)}]
\item Homogeneous at \SI{20}{\percent}.
\item Homogeneous at \SI{40}{\percent}.
\item Gradient: a smooth linear gradient from \SI{10}{\percent} in \changed{one} bottom corner to \SI{40}{\percent} in the diametrically opposite corner.
\item Contrast plane: half of the cube is \SI{10}{\percent} infill density and the other half is \SI{40}{\percent}.
The plane which separates these two halves makes an angle of \SI{22.5}{\degree} with the X axis in the horizontal plane
and has an overhang angle of \SI{45}{\degree}.
\item Sphere shell: a sphere with a radius of half the cube side length and a shell thickness of 1/7 the size of the side lengths of the cube.
The density of the shell is  \SI{40}{\percent} and the density inside and outside is  \SI{10}{\percent}.
\end{enumerate*}
All these are specifications consisting of $512$ images of \SI{512 x 512}{\pixel}.
The accuracy results are shown in \cref{fig:accuracy}.

\subsubsection{Computation time}
In order to evaluate the running time of our algorithms, we consider four of the application models discussed in \cref{section:applications}.
We consider the test models and corresponding settings displayed in \cref{table:example_model_settings}.
The computation times are shown in \cref{table:computation_time}.

\begin{table}
  \caption{Example model settings.}
  \label{table:example_model_settings}
  \begingroup
\setlength{\tabcolsep}{0.25\tabcolsep}
  \begin{tabular}{ l | c  c  c  c c }
        & white & black & top & spec size (\si{\pixel}) & phys. size (\si{\milli\meter}) \\
    \hline
    Sole & \SI{5}{\percent} & \SI{40}{\percent}  & \SI{0}{\percent}  & \SI{1456 x 564 x 1}{}       & \SI{155 x 58 x 14}{}\\
    Bunny     & \SI{10}{\percent} & \SI{80}{\percent}  & \SI{20}{\percent} & \SI{45 x 35 x 2785}{}    & \SI{129 x 100 x 126}{} \\
    Phantom   & \SI{0}{\percent}  & \SI{100}{\percent} & \SI{20}{\percent} & \SI{417 x 412 x 146}{} & \SI{123 x 121 x 87}{}\\
    Saddle & \SI{40}{\percent} & \SI{10}{\percent}  & \SI{0}{\percent}  & \SI{60 x 44 x 47}{}              &  \SI{250x188x63}{}\\
  \end{tabular}
    \endgroup
\end{table}
\medskip

\begin{table}
  \caption{Computation time in seconds.}
  \label{table:computation_time}
  \begin{tabular}{ l | c  c  c c }
	&	Sole	&	Bunny	&	Phantom & Saddle	\\
\hline                                  
    Lower bound        &   6.2    &    11.2    &    4.3    &   150.9   \\
    Dithering          &   0.5    &    2.3    &    0.4    &   10.5   \\
    Extract polygon    &   0.4    &    15.7   &    3.4    &   12.8   \\
    Limit polygon      &   0.3    &    11.3   &    3.8    &   12.0   \\
    Reconnecting       &   4.6    &    98.9   &    65.4   &   135    \\
\hline
    Total gcode        &   14     &    182    &    105    &   346    \\
  \end{tabular}
\end{table}


\subsubsection{Elastic behavior}
\label{section:experiments}
\changed{Because the generated structures are similar to foams, which are often used in a compressive context, it would be interesting to find out their compressive behavior.
Because the non-linear material properties at high strain values when compressing a foam are difficult to capture when using a finite elements method, we have performed actual physical tests instead.}

We have printed samples with homogeneous subdivision level, and we have printed samples using dithering to approximate several homogeneous density specifications with simplified densities between \SI{10}{\percent} and \SI{30}{\percent}.
Compressions were performed in both the vertical and the horizontal direction.
Because our structures are rotationally symmetric around the Z axis we only need to test 2 of the 3 dimensions.
In total 78 prints were made in 42 test configurations (some configurations were tested multiple times): 2 testing directions, 4 homogeneous subdivision levels \changed{({\SI{10.1}{\percent}}, {\SI{14.0}{\percent}}, {\SI{20.1}{\percent}}, {\SI{28.5}{\percent}})} and 17 heterogeneous subdivision levels \changed{in the same range}.

\begin{figure}
        \centering
    \begin{subfigure}[t]{0.48\columnwidth}
        \centering
        \includegraphics[width=\columnwidth]{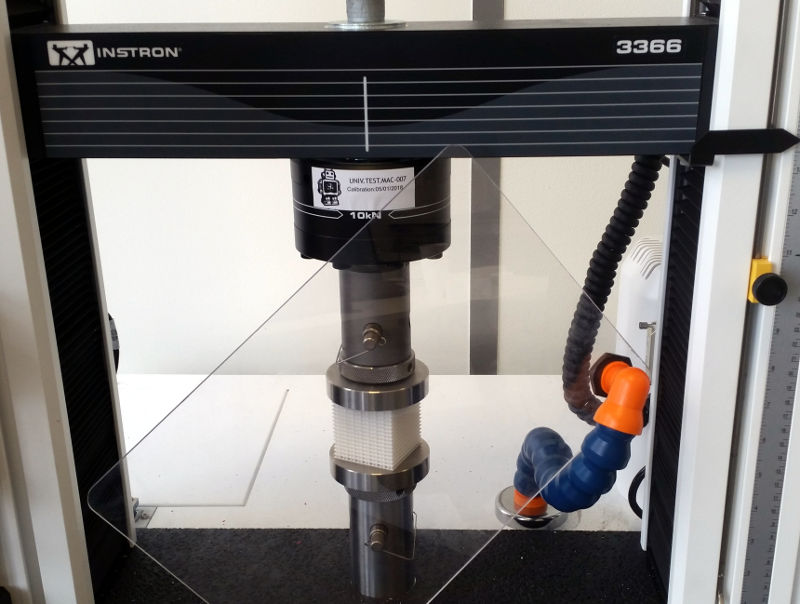}
        \caption{
The compression testing setup.
        }
        \label{fig:instron}
    \end{subfigure}
    \begin{subfigure}[t]{0.48\columnwidth}
        \centering
        \includegraphics[width=\columnwidth]{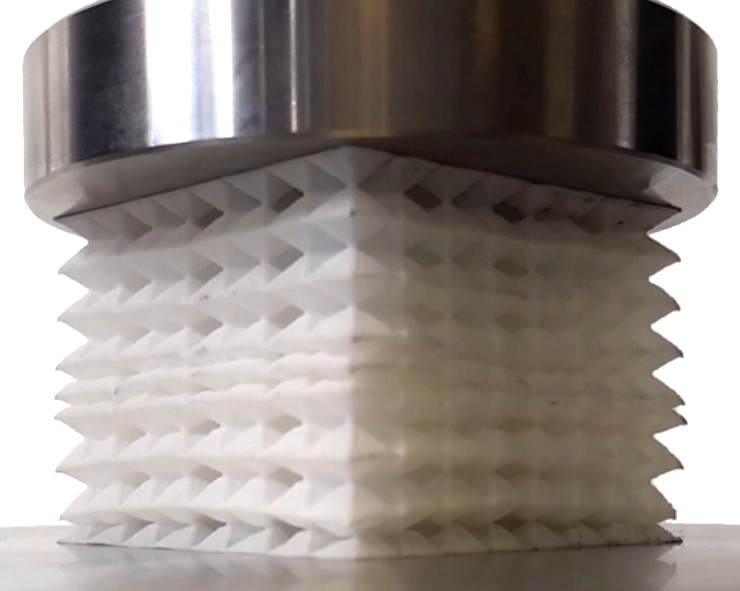}
        \caption{
        Partial collapse.
        }
        \label{fig:partial_collapse}
    \end{subfigure}
\caption{
Compression testing.
Under stress some cells in the structure start to collapse. Cells tend to collapse in groups on the same heights.
}
\end{figure}

We have performed \changed{compression} tests on the Instron 3366, fitted with compression plates.
See \cref{fig:instron}.
Compressions were performed at a speed of \SI{0.5}{\milli\meter\per\second} up to a maximum force of \SI{2}{\kilo\newton} after which the sample was decompressed.

\begin{figure}
        \centering
    \begin{subfigure}[t]{0.9\columnwidth}
        \centering
         \includegraphics[width=\columnwidth]{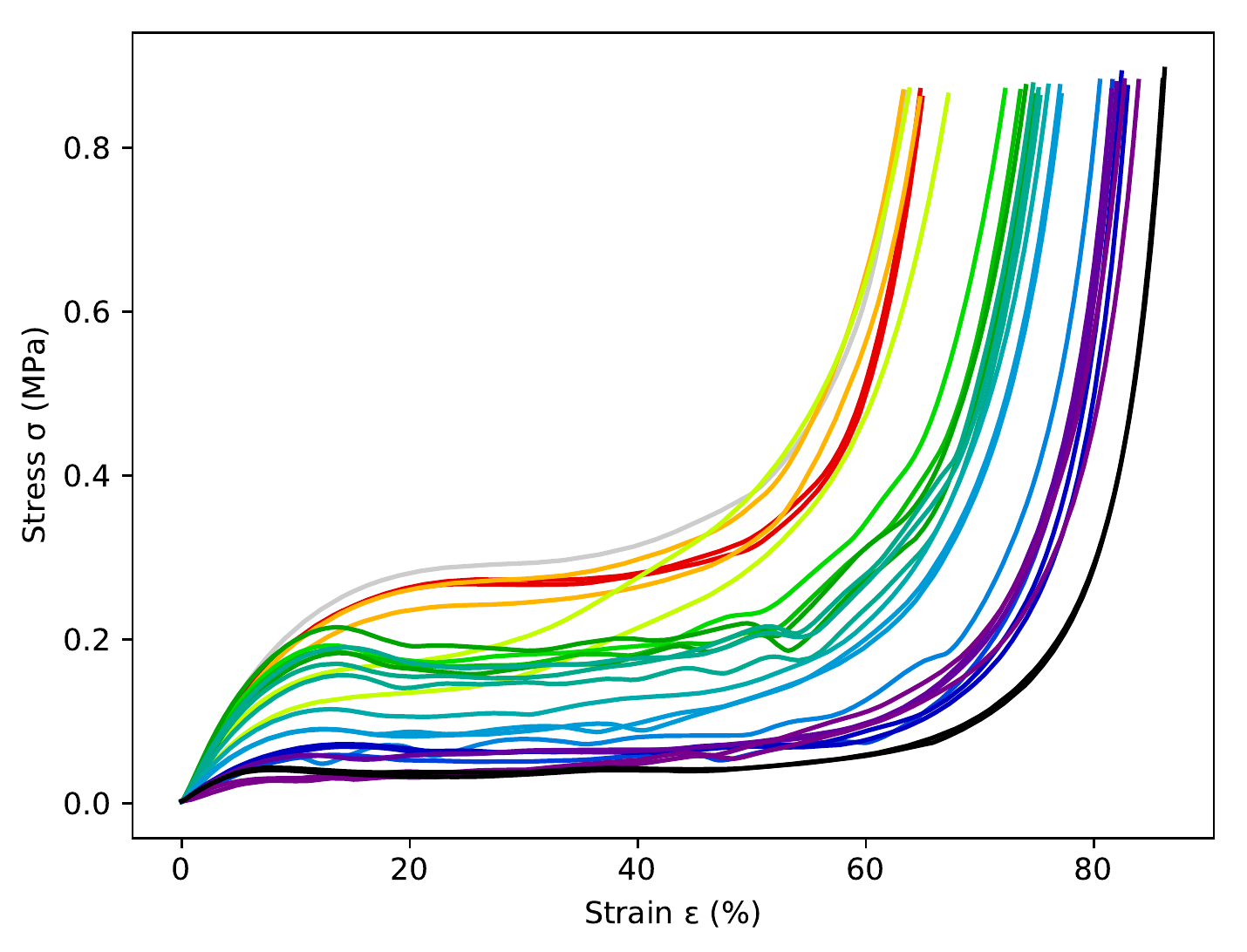}
        \caption{Vertical compression}
        \label{fig:stress_strain_results_top}
    \end{subfigure}
    \begin{subfigure}[t]{0.9\columnwidth}
        \centering
         \includegraphics[width=\columnwidth]{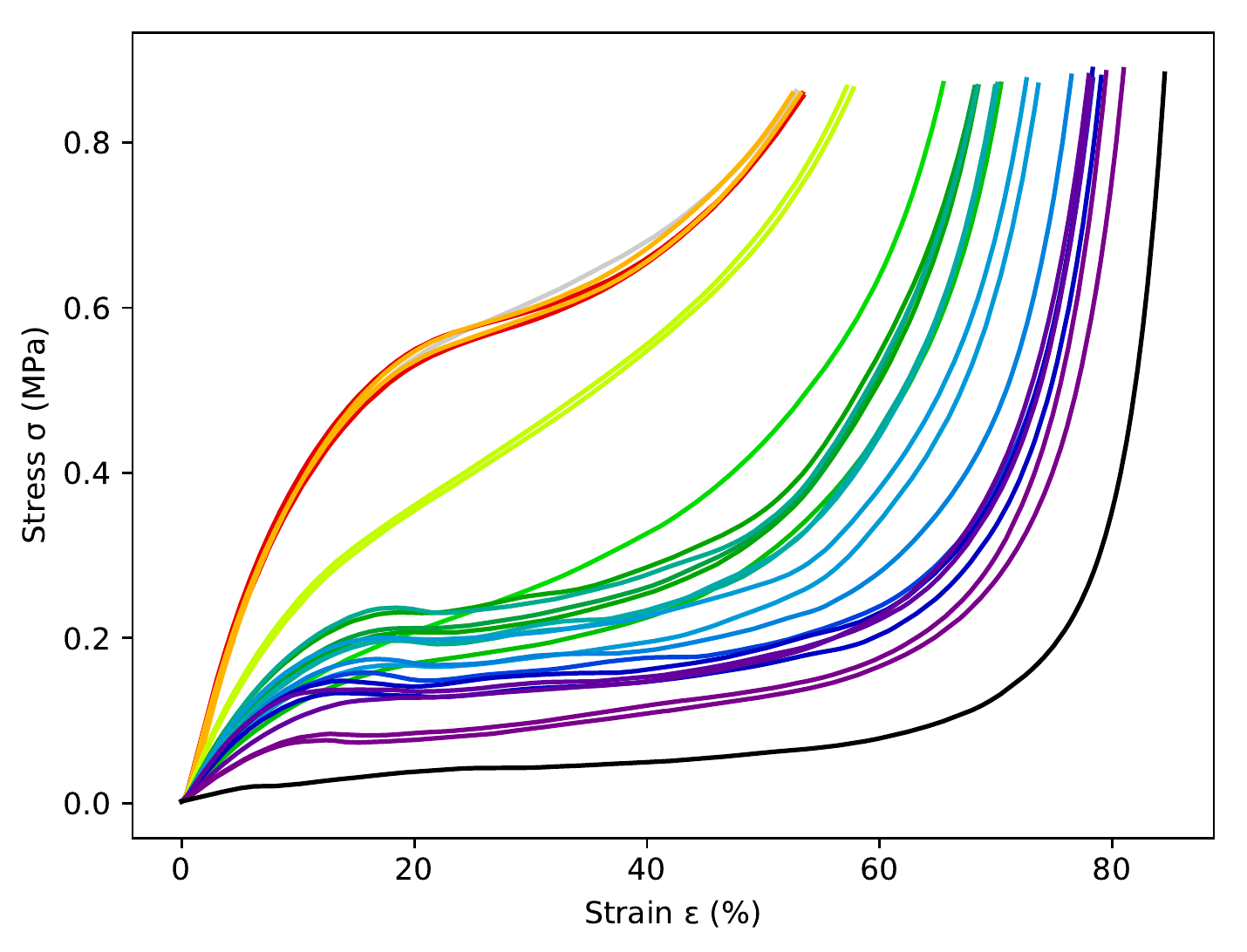}
        \caption{Horizontal compression}
        \label{fig:stress_strain_results_side}
    \end{subfigure}
    \begin{subfigure}[t]{0.9\columnwidth}
        \centering
         \includegraphics[angle=90,width=.8\columnwidth]{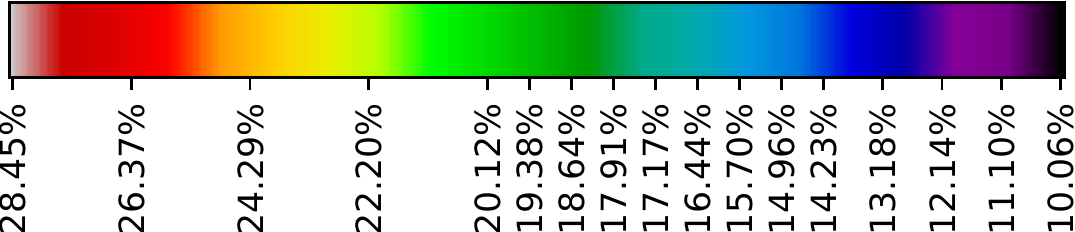}
    \end{subfigure}
    \caption{
Stress - strain - density graphs of the results of our compression experiments.
The structure is more compliant in the vertical direction than in the horizontal direction.
From these results we can conclude that dithering can provide a spectrum of material properties;
higher densities start collapsing at higher strain values and their collapse trajectory is shorter.
    }
    \label{fig:stress_strain_results}
\end{figure}

The stress-strain results are plotted in \cref{fig:stress_strain_results}.
%
One interesting observation from the data is that the stress-strain graphs are roughly horizontal for a long range of strain values.
This is caused by the structure collapsing and folding in on itself.
Such plateaus are typical of foams \citep{ashby2006properties,avalle2001characterization}, 
and they are important design variables for common applications of foams~\cite{mills2003polymer}.
\Cref{fig:partial_collapse} shows how such collapse can be localized to only a particular Z range in the case of vertical compression.

\begin{figure}
        \centering
    \begin{subfigure}[t]{0.48\columnwidth}
        \centering
        \includegraphics[width=\columnwidth]{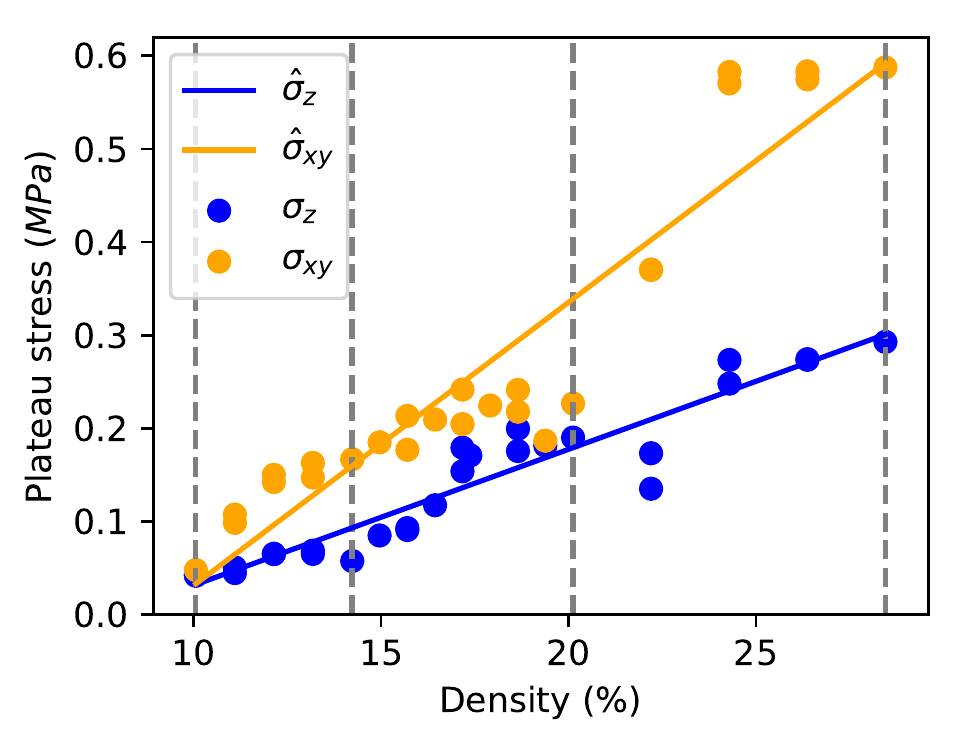}
        \includegraphics[width=\columnwidth]{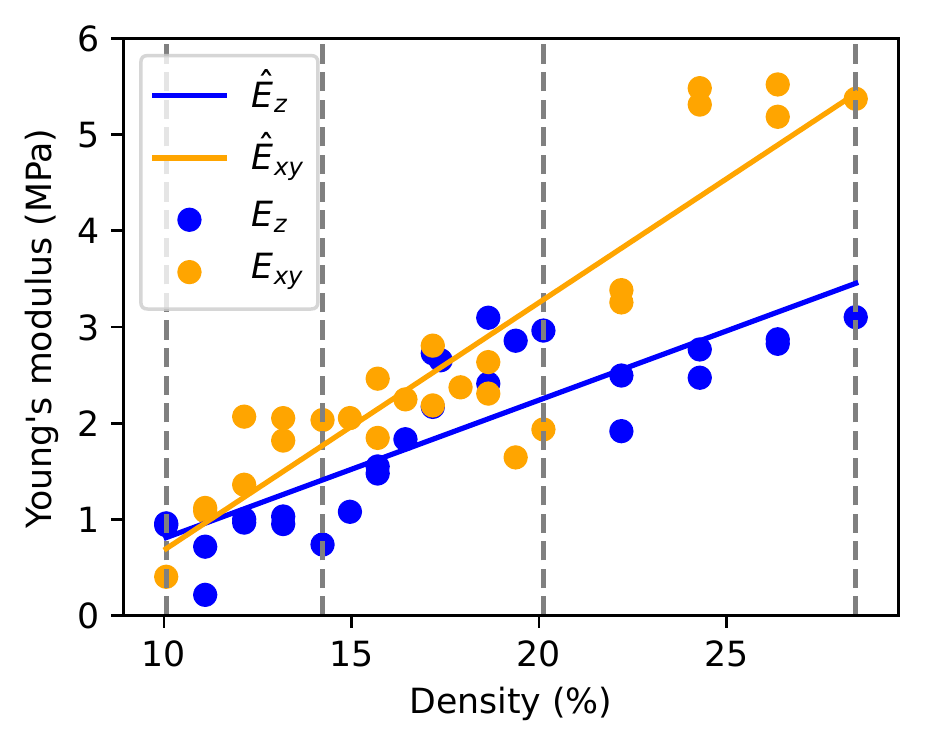}
        \caption{Dithering}
        \label{fig:plateau_stress_results}
    \end{subfigure}
    \begin{subfigure}[t]{0.48\columnwidth}
        \centering
        \includegraphics[width=\columnwidth]{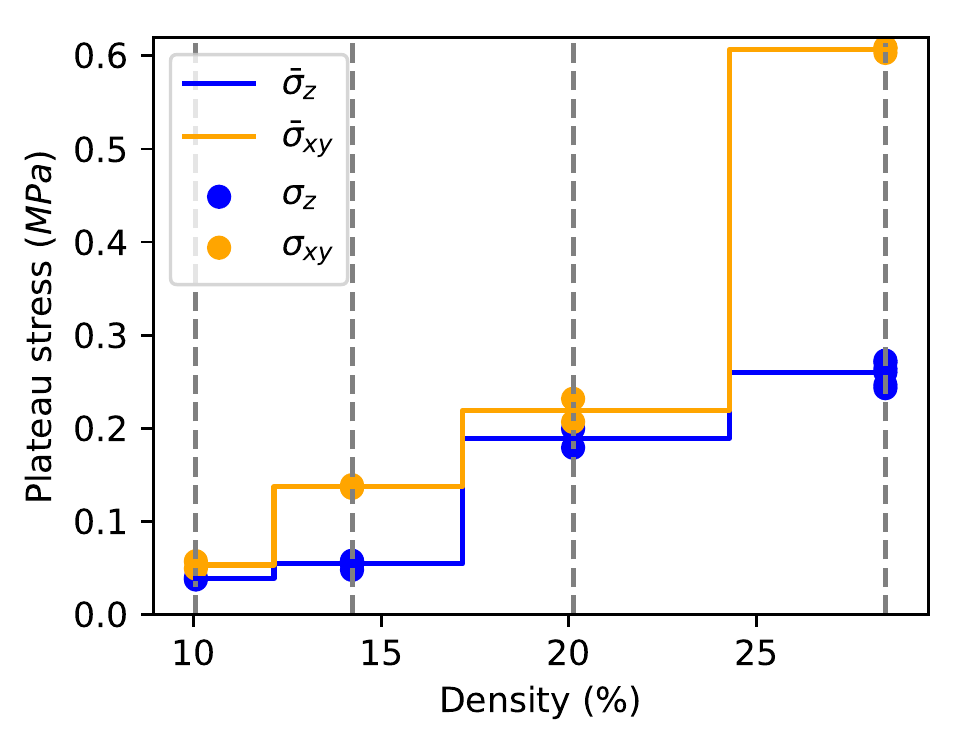}
        \includegraphics[width=\columnwidth]{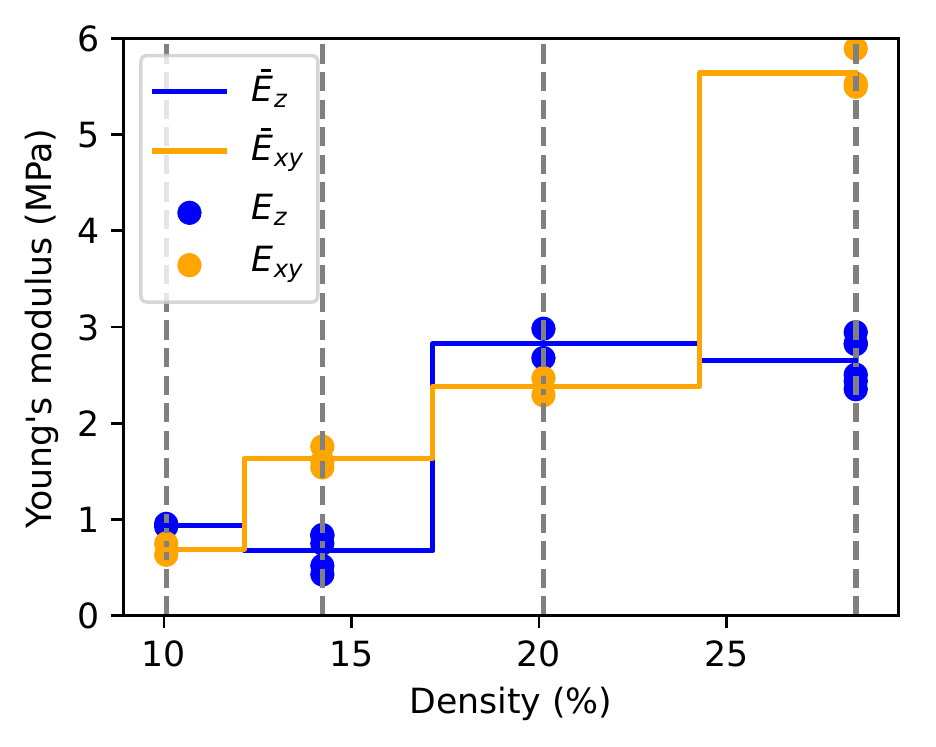}
        \caption{No dithering}
        \label{fig:youngs_modulus_results}
    \end{subfigure}
    \caption{
Plotting Young's modulus and plateau stress for different densities.
Lines represent linear regression results over dithered data and average values of not dithered data.
Dithering provides material properties in between the discrete densities.
However, they are not monotonically increasing with the density,
which makes mapping required material properties to corresponding infill densities non-trivial.
    }
    \label{fig:stiffness_results}
\end{figure}

The constant stress along that range is a fundamental characteristic of such a plateau.
We estimate it by taking \changed{the average stress along} the part of the stress-strain graph which has a local tangent modulus below \SI{0.4}{\mega\pascal}.
The resulting values have been plotted alongside the Young's modulus values in \cref{fig:youngs_modulus_results}.

\begin{figure*}
        \centering
        \includegraphics{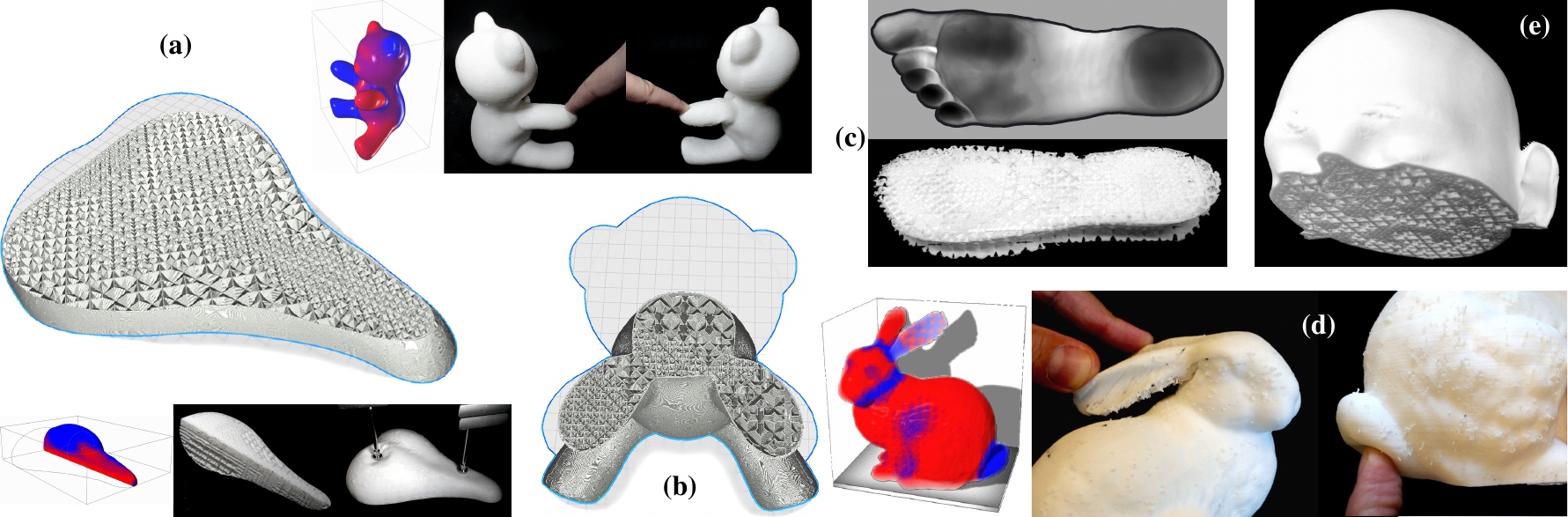}
    \caption{
    Various examples of applications of CrossFill.
(a)
A bicycle saddle with a density specification.
A weight of \SI{33}{\newton} is added on various locations to show the different response of different density infill.
(b)
A teddy bear with a density specification.
(c)
A shoe sole with densities based on a pressure map of a foot.
(d)
The Stanford bunny painted with a density specification.
(e)
A medical phantom with an example density distribution for calibrating \changed{an MRI scanning procedure}.
    }
        \label{fig:applications}
\end{figure*}

\subsection{Discussion}
\label{sec:discussion}
\paragraph{Accuracy} 
At a kernel size of $16 w$ (\SI{6.08}{\milli\meter}) the average absolute local error is low for smooth input density distributions.
Because of the constraint that neighboring linked cells can only differ by a single subdivision level the two distributions with sharp contrast edges or high frequency detail score considerably worse.
\changed{See {\cref{fig:accuracy}}, `contrast plane' and `sphere shell'.}
\changed{The relative error decreases as the resolution increases. Therefore, d}epending on the application \changed{the user} might decide that at a specific resolution the accuracy is good enough.

\subsubsection*{Computation time} \noindent
Our algorithms take up over \SI{80}{\percent} of the total gcode generation time;
there is room for improvement. 
%
It should be noted that we \changed{not optimized} the code for loading the density specification data into the subdivision tree \changed{or} the code for connecting polygons,
which currently consume the majority of the processing times.
Polygon extraction and polygon limiting times depend highly on the number of layers, which is relatively small for the shoe sole example.

\paragraph{Elastic behavior}
It should be noted that the first local maximum in the stress-strain graphs of \cref{fig:stress_strain_results} are not the yield points.
Most of the deformation applied to the structure is elastic deformation.
\changed{We observed} the material to creep back to its original shape at ever decreasing rates.
After \changed{$24$ hours} a strain of merely \SI{0.8}{\percent} remained after an initial strain above \SI{50}{\percent}, meaning that the plastic deformation is negligible at stresses with magnitudes as high as in our tests \changed{({\SI{2}{\kilo\newton}})}.

For larger densities the strain at which the structure is fully collapsed is lower.
When we compress a sample with \SI{40}{\percent} density \changed{by} \SI{60}{\percent} then the stress-strain graph should exhibit a local tangent equal to the base properties of the material used.
In our test results we see that the tangent tends toward the Young's modulus of TPU: \SI{26}{\mega\pascal} \cite{ultimaker2018tpu}.
This agrees with literature on the densification of foams \cite{ashby2006properties}.

In most aspects the structures are more compliant in the vertical direction than in the horizontal.
The plateau heights in \cref{fig:stress_strain_results_top} are lower than the corresponding ones in \subref{fig:stress_strain_results_side}
and the strain regions of the plateaus are wider as well.
Of course some aspect of this anisotropy is caused by the layerwise buildup of FDM,
but a larger part of the difference is most likely caused by the geometric structure.
The structure can collapse in the Z direction easily because of the alternating E- and C-embeddings.

The vertical Young's modulus $E_z$ is lower than the horizontal Young's modulus $E_{xy}$ for most densities.
For densities around \SI{10}{\percent} and \SI{20}{\percent}, that relationship is reversed, though.
That can be explained by the fact that around those densities the subdivision structure contains mostly Q-prism cells which are filled with CrossFill surface patches with a more vertical slope than those for H-prisms.
More vertical elements \changed{increase the} stiffness in the vertical direction.

The prisms only subdivide \changed{vertically} every two iterations: \changed{only the Q-prisms do}.
The \changed{H-}prisms of \SI{20.1}{\percent} are subdivided \changed{horizontally} compared to \SI{14.0}{\percent}, while the prisms of \SI{14.0}{\percent} and \SI{28.5}{\percent} are also subdivided \changed{vertically} compared to \SI{10.1}{\percent} and \SI{20.1}{\percent} respectively.
This irregularity can explain the nonmonotonicity we see in the Young's modulus with respect to infill densities in \cref{fig:youngs_modulus_results}:
horizontal subdivision decreases the slope of the surface patches in the structure, which decreases the thickness of these patches, which in turn decreases the overall stiffness.

Because the deeper subdivision level cells determine the ruled surface in horizontal continuity enforcement, surface patches of a dithered subdivision structure are predominantly determined by the higher density cells.
We therefore expect to see large changes after \SI{10}{\percent} and after \SI{20}{\percent}.
We can see large jumps in the vertical Young's modulus at those places in \cref{fig:youngs_modulus_results}.
Also the plateau heights for horizontal compressions show large jumps.

When comparing the dithered results in \cref{fig:stiffness_results} to the ones without dithering we can conclude that dithering indeed provides more granular control of the overall material properties of the manufactured part.
However, because of the nonmonotonicity in these results it is not trivial to define a process for the designer to choose which infill density is needed at a location in a design.

%
\subsection{Applications}
\label{section:applications}
The structures generated by our method support variable compliance, which can control the deformation of an object under certain loads.
We have designed a density specification for the Stanford bunny model using Autodesk Monolith~\cite{monolith2018}, as shown in \cref{fig:applications}d.

Because the CrossFill structures behave much like foams, it could be attractive \changed{for} "cushioning, packaging and energy absorption"~\cite{ashby2006properties}.
Foams with variable stiffness could be used for a personalized bike saddle (see \cref{fig:applications}a) or for personalized shoe soles (see \cref{fig:applications}c).

A boundary mesh of a teddy bear along with a density distribution show how different densities lead to different bending behavior (see \cref{fig:applications}b).
The difference in density distribution in the two arms causes the arms to deform in a different way under the same load.

Because our method is principally density-based, it could also be useful in situations where \changed{the relevant material properties are volumetric;
CrossFill could prove useful for imaging \emph{phantoms}}
- objects which can be used in the medical field to evaluate \changed{magnetic resonance imaging (MRI)} scan procedures.
We have printed an example of what could serve as a phantom in \cref{fig:applications}e.

\section{Conclusion and future work}
%
\label{section:conclusion}
In this paper, we have introduced a new infill structure, CrossFill, which can provide spatially graded density to match a user-specified density distribution. 
CrossFill is carefully designed so that it is self-supporting and can be fabricated from a single, continuous and overlap-free toolpath on each layer.
Algorithms for generating the lower bound subdivision levels and dithering the subdivision levels have been developed to accurately match the prescribed density distribution. 
To use CrossFill as infill structures of a given 3D model, we have presented an algorithm to connect the toolpaths of CrossFill and the toolpaths for the given model’s shell into a single continuous extrusion path. 
The performance of CrossFill has been verified on a variety of experimental tests and applications. 

The study of experimental tests shows that CrossFill acts very much like a foam although future work needs to be conducted to further explore the mapping between density and other material properties.
Another line of research is to further enhance the dithering technique, e.g. changing the weighing scheme of error diffusion.

\section*{References}
\bibliography{_mybib,_CFLproject}

\end{document}